\documentclass[11pt]{article}

\usepackage[margin=0.8in]{geometry}
\onecolumn % for one column layouts
\usepackage{graphicx}%
\usepackage{multirow}%
\usepackage{mathrsfs}%
\usepackage{amsmath,amssymb,amsfonts, amsthm}%
\usepackage{mathtools}%
\usepackage{accents}%
\usepackage{xcolor}%
\usepackage{textcomp}%
\usepackage{manyfoot}%
\usepackage{booktabs}%
\usepackage{algorithm}%
\usepackage{algorithmicx}%
\usepackage{algpseudocode}%
\usepackage{listings}%
\usepackage{natbib}%
\usepackage{prodint}%
\usepackage{makecell}%
\usepackage{nicematrix}%
\usepackage{comment}
\usepackage{lscape}
\usepackage{cancel}
\usepackage{enumitem}
\usepackage{amsthm} 
\usepackage{adjustbox}
\usepackage{hyperref}
\usepackage{authblk}
\hypersetup{
    colorlinks=true,
    linkcolor=blue,
    filecolor=blue,      
    urlcolor=blue,
    citecolor=blue,
    }
\usepackage{threeparttable}

\allowdisplaybreaks

\theoremstyle{thmstyleone}%
\newtheorem{theorem}{Theorem}%  meant for continuous numbers
\theoremstyle{thmstyletwo}%
\newtheorem{lemma}{Lemma}%
\theoremstyle{thmstylethree}%

\newcommand{\keywords}[1]{\par\addvspace\baselineskip \noindent\textbf{Keywords:}\enspace\ignorespaces#1}

\title{Addressing the Influence of Unmeasured Confounding in Observational Studies with Time-to-Event Outcomes: \\ A Semiparametric Sensitivity Analysis Approach}

\author[1]{Linda Amoafo}
\author[2]{Shiyao Xu}
\author[3]{Elizabeth Platz}
\author[2,$\ast$]{Daniel Scharfstein}

\affil[1]{Statistics - Diabetes, Eli Lilly and Company, Lilly Corporate Center, Indianapolis Indiana, 46285, USA}
\affil[2]{Department of Population Health Science, University of Utah School of Medicine, 295 Chipeta Way, Salt Lake City, Utah, 84108, USA}
\affil[3]{Department of Epidemiology, Johns Hopkins Bloomberg School of Public Health, 615 North Wolfe Street, Baltimore, Maryland, 21205, USA}

\affil[$\ast$]{Corresponding author. Email: \href{email:email-id.com}{daniel.scharfstein@hsc.utah.edu}}

\date{}

\begin{document}
\maketitle

\abstract{In this paper, we develop a semiparametric sensitivity analysis approach designed to address unmeasured confounding in observational studies with time-to-event outcomes. We target estimation of the marginal distributions of potential outcomes under competing exposures using influence function-based techniques. We derive the non-parametric influence function for uncensored data and map the uncensored data influence function to the observed data influence function. Our methodology is motivated by and applied to an observational study evaluating the effectiveness of radical prostatectomy (RP) versus external beam radiotherapy with androgen deprivation (EBRT+AD) for the treatment of prostate cancer. We also present a realistic simulation study demonstrating the finite-sample properties of our estimation procedure. %Our approach provides a practical tool for quantifying robustness of observational study findings to unmeasured confounding.}
}

\section{Introduction}

The causal effect of two competing treatments has been formalized as a contrast between the distributions of the potential outcomes (i.e., outcomes under the various treatment options)~\citep{rubin2005causal}.  In an experimental setting, treatment assignment is via an external process, and randomization probabilistically ensures that both measured and unmeasured confounders are balanced between treatment groups. Causal effects can then be estimated by simply comparing the distribution of outcomes between the treated and untreated with/without covariate adjustment. In a setting where randomized trials are impractical,  observational studies become essential. However, in the observational setting, an internal process determines treatment assignment; for example, existing guidelines may determine a patient's treatment, or each patient, in consultation with their medical team, decides their own treatment. This internal assignment can lead to systematic differences between treatment groups with respect to measured and unmeasured patient characteristics that are associated with the outcome under investigation.  In causal inference, one key, typically untestable, assumption is conditional exchangeability (or no unmeasured confounding), which states that treatment assignment is statistically independent of the potential outcomes {\em conditional} on a set of measured covariates. If this assumption holds (along with consistency, positivity, and no interference \citep{hernan2020causal}), statistical adjustments can be made to recover the true causal effect of interest.  However, what happens if conditional exchangeability fails to hold? It would be useful to have a sensitivity analysis tool that evaluates the robustness of inferences to deviations from this assumption. In this paper, we develop a sensitivity analysis tool for the observational setting with a time-to-event outcome that is subject to right censoring.  We are specifically interested in drawing inference about the difference in the marginal distributions of the time-to-event under competing treatments.

\subsection{Motivating Example}

To anchor ideas, consider the observational study conducted by \cite{ennis2018brachytherapy}. In this study, the authors analyzed the survival outcomes of patients with prostate cancer who underwent one of three treatments: (1) radical prostatectomy (RP), (2) external beam radiotherapy (EBRT) combined with androgen deprivation (AD), or (3) EBRT plus brachytherapy with or without AD.  Here, we focus on two of these treatments: RP and EBRT+AD. RP is a surgical procedure, while EBRT+AD is not.  In their Table 1, \cite{ennis2018brachytherapy} reported pre-treatment characteristics (age, prostate-specific antigen, race/ethnicity, insurance status, income, education, Charlson co-morbidity index, Gleason score, clinical T stage, and year of diagnosis) of the patients by treatment groups. Relative to those treated with EBRT+AD, patients treated with RP were younger, more likely to be white, more likely to have private insurance/managed care, have higher education, have higher income, and have lower Gleason scores. These factors are all associated with better survival and imply that an unadjusted analysis may suggest better survival for RP relative to EBRT+AD  {\em even if} there is no true casual difference.  In fact, \cite{ennis2018brachytherapy}) reported a marked difference between the unadjusted survival curves for RP versus EBRT+AD in favor of better survival with RP.  After adjustment for measured factors, the difference was attenuated but remained marked.

~\cite{chen2018challenges} wrote an editorial in response to \cite{ennis2018brachytherapy} study, saying 
\begin{quote}
``comparisons of widely differing treatments such as radical surgery and RT [radiotherapy] in prostate or any other cancer are particularly difficult to interpret because of known inherent differences in patient characteristics between the treatment groups. As a radiation oncologist, it is not uncommon for me to treat patients who my urologic colleagues feel are not ideal surgical candidates because of existing comorbidities. Although all published studies have attempted to statistically account for some measure of comorbidity burden, existing instruments, such as the Charlson score, are crude and unable to fully account for differences between patients receiving surgical treatment and those receiving RT."  %Clinically, we know that a portion of patients in the latter group are unable to undergo surgery, regardless of what their Charlson score might be."
\end{quote}
~\cite{chen2018challenges} argues that there was inadequate control for overall health status, comorbidity burden, and disease characteristics between surgical and RT patients, writing ``some urologists may preferentially select patients with relatively low-volume disease or other more favorable characteristics (such as magnetic resonance imaging findings supportive of resectability) for radical prostatectomy, whereas patients who receive RT may more commonly have disease nearer the aggressive end of the high-risk spectrum."  Thus, ~\cite{chen2018challenges} believes that there are unmeasured confounding factors that have not been adjusted for when comparing the competing treatments in this observational study.  That is, if these factors had been available, they would likely indicate that patients receiving surgery would have less aggressive disease and better health status than patients receiving radiotherapy. This suggests that the adjustment performed based on the measured characteristics in Table 1 of~\cite{ennis2018brachytherapy} may be inadequate, and the reported benefit of RT over EBRT could be too optimistic.  The sensitivity analysis tool developed in this paper will allow us to evaluate the robustness of the findings of~\cite{ennis2018brachytherapy}  that are purported to be influenced by unmeasured confounding.  

\subsection{Review of Two Sensitivity Analysis Approaches}

Two sensitivity analysis procedures have been developed that directly relate to our setting and objective. In the style of \cite{robins2000sensitivity} and \cite{brumback2004sensitivity}, \cite{klungsoyr2009sensitivity}  considered drawing inference about the discrete-time causal marginal hazard ratio and addressed unmeasured confounding by introducing a user-specified function that ``describes lack of exchangeability between baseline exposed and non-exposed within levels of measured confounders, on a hazard ratio scale". The further this function differs from one, the greater the deviation from exchangeability (i.e., the greater the influence of unmeasured confounders above and beyond measured confounders). The function is parameterized and inference by inverse weighting is conducted over a broad range of parameter values.  Their approach assumes that censoring is independent of failure given measured factors and it can be used to draw inference about the difference in marginal distributions under competing treatments. Building on the work of \cite{rosenbaum2002observational}, \cite{tan2006distributional} and \cite{zhao2019sensitivity}, \cite{lee2024sensitivity} considered drawing inference about the causal effect of treatment quantified in terms of difference in restricted mean survival time. In their approach, they assume that the ratio of the odds of treatment given measured and unmeasured covariates to the odds of treatment given measured covariates is bounded between $1/\Lambda$ and $\Lambda$, where $\Lambda \geq 1$ serves as a sensitivity analysis parameter.  The larger the value of $\Lambda$ the greater the allowable deviation from exchangeability. They develop an optimization algorithm with an inverse weighted objective function to compute lower and upper bounds on the difference in restricted mean survival time as a function of $\Lambda$. The approach assumes that censoring and failure are marginally independent and, as acknowledged by the authors, cannot be easily extended to accommodate a less restrictive assumption.  Also acknowledged is the computationally intensive nature of the their procedure as it requires two optimizations, one with parameters equal to the number of censored individuals and one with parameters equal to the number of uncensored individuals. Their approach could be extended to construct bounds on the difference in marginal distributions under competing treatments, but this would require optimizations at each time point.  Importantly, both approaches rely on correct specification of fully parametric models for the conditional distribution of treatment given measured covariates. 

Our sensitivity analysis approach is similar to that of \cite{klungsoyr2009sensitivity} with the following key exceptions. First, we introduce a user-specified function describing lack of exchangeability between baseline exposed and non-exposed within levels of measured confounders, on a proportional odds scale. Second, we develop a semiparametric inference methodology that does not require fully parametric modeling of any distribution.  We do require specification of models for three conditional distributions: (1) treatment given measured covariates, (2) time to event given treatment and measured covariates, and (3) time to censoring given treatment and measured covariates. These models can be semi-parametrically specified provided certain convergent rate conditions are satisfied, and the consistency of our procedure only requires correct specification of the conditional distribution of time to event given treatment and measured covariates.

\subsection{Outline of Paper}

The paper is organized as follows. % In Section 2, we will present a review of causal inference methods that have been proposed for  conducting sensitivity analysis of observational studies with a time-to-event outcome that is subject to right censoring.  
In Section 2, we introduce our methodology.  In Section 3, we evaluate the robustness of the findings from~\cite{ennis2018brachytherapy}. In Section 4, we present a simulation analysis to show the performance of our method. Section 5 is devoted to discussion.

\section{Methods}\label{sec4}
\subsection{Notation}

Let $X$ be pre-treatment measured covariates and $T$ be treatment received ($T=1$ for RP, $T=0$ for EBRT+AD).  Let $Y(t)$ be the time to event under treatment $t$ ($t=1$ for RP, $t=0$ for EBRT+AD).  Let $F=(Y(0),Y(1))$. We assume that $Y =  T Y(1) + (1-T)Y(0)$; these quantities are defined in a world with perfect follow-up. Our target of inference is $P[Y(t) \leq s]$ for $ s\in[0, \tau]$, for fixed $\tau < \infty$. Let $F_t(\cdot|x) = P[Y \leq \cdot |T=t,X=x]$ and $\pi_t(x) = P[T=t|X=x]$.

Let $C$ be a follow-up time that is truncated by $\tau^\dagger>\tau$; $C$ may preempt observation of $Y$. Let $G_t(\cdot|X) = P[C>\cdot |T=t,X=x]$.  We assume there is no follow-up after the occurrence of $Y$. Let $\widetilde{Y} = \min(Y,C)$ be the observed follow-up time and $\Delta = I(Y \leq C)$ indicate observation of the event time of interest.  For a random individual, we let the uncensored data be denoted by  $O = (X,T,Y)$ and the censored data by $\widetilde{O}=(X,T,\widetilde{Y},\Delta)$.  Let $P$ and $\widetilde{P}$ be the true distributions of $O$ and $\widetilde{O}$, respectively. Note that $P$ is characterized by $F_X(\cdot)$ (the marginal distribution of $X$), $\pi_1(X)$, $F_0(\cdot|X)$ and $F_1(\cdot|X)$; $\widetilde{P}$ is additionally characterized by $G_0(\cdot|X)$ and $G_1(\cdot|X)$ (under the non-informative censoring assumption discussed below).

%$F = (X,T,Y(0),Y(1))$ 
%$\widetilde{O]=(X,T,\tilde{Y},\Delta)$.

We assume that we observe $n$ i.i.d. copies of $\widetilde{O}$.Subscript $i$ will be used to denote data specific to individual $i$. %Our goal is draw inference about the marginal distribution of $Y(t)$, specifically, $S_t(s) = P[Y(t)>s]$.

\subsection{Assumptions} \label{sec32}

For $t=0,1$ and $s\in[0, \tau]$, we assume 
\begin{equation}
\resizebox{.9\hsize}{!}{$\mbox{logit} \{ P[Y(t) \leq s | T = 1-t, X=x] \} = \mbox{logit} \{ P[Y(t) \leq s | T = t, X=x]\} + \gamma_t, \;\; \mbox{for all} \; \;{\color{blue}x},$}
\label{eqn1}
\end{equation} 
where $\gamma_t \in \Gamma_t$ (compact set) is a fixed sensitivity analysis parameter that governs deviations from conditional exchangeability. This specifies a proportional odds relationship between the distribution of $Y(t)$ given $T=1-t$ and $X$ and the the distribution of $Y(t)$ given $T=t$ and $X$.
Note that $\gamma_t=0$ implies that $T$ is independent of $Y(t)$ given $X=x$ (i.e.,  conditional exchangeability); this means that, within levels of $X$, the distribution of $Y(t)$ is the same for those whose observed treatment is $1-t$ and  those whose observed treatment is $t$.  When $\gamma_t$ is greater (less) than zero, we are assuming that, within levels of $X$, the distribution of $Y(t)$ is skewed toward shorter (longer) survival times for those whose observed treatment is $1-t$ relative to those whose observed treatment is $t$; the difference increases with the absolute magnitude of $\gamma_t$. 

Using Bayes' rule, (\ref{eqn1}) can be re-written to show that
\begin{equation}
\exp(\gamma_t) = \frac{P[T=1-t|Y(t)\leq s,X=x] }{P[T=t|Y(t) \leq s, X=x]} \frac{P[T=t|Y(t) > s, X=x]}{P[T=1-t|Y(t) > s, X=x]}.
\label{eqn1a}
\end{equation}
Thus, $\gamma_t$ is the conditional (on $X=x$) log odds ratio of receiving treatment $1-t$ for individuals with $Y(t) \leq s$ versus those with $Y(t)>s$.  This result is a consequence of using the logit function in (\ref{eqn1}).

Our assumption posits that the deviation from conditional exchangeability does not depend on $s$ or the levels of $X$; while this assumption is not required for our methodology  (i.e., we can replace $\gamma_t$ by a specified function $\gamma_t(s,x)$), it greatly simplifies the sensitivity analysis. Specifically, inference about treatment effects can be visualized in two dimensions. Allowing a more complicated sensitivity analysis function that depends on $s$ and/or $x$ would make visualization of inferences unwieldy. Thus, use of $\gamma_t$ is a compromise that allows exploration along a restricted path of deviations from conditional exchangeability; specification of $\gamma_t \not = 0$ is conceptually no different than assuming $\gamma_t=0$.

We further assume that 
\begin{equation}
C \mbox{ is independent of } Y \mbox{ given }T \mbox{ and } X.    
\label{censoring}
\end{equation}  
This implies that
\begin{equation}
\lambda^{\dagger}_t(u|x) = \lambda_t(u|x),
\label{eqn2}
\end{equation}
where $\lambda^{\dagger}_t(u|x) = \lim_{h \rightarrow 0+} P[u \leq \widetilde{Y} < u+h, \Delta=0\;|\; \widetilde{Y} \geq u, T=t, X=x]/h$ is the treatment/cause-specific (conditional on $X=x$) hazard of censoring and $\lambda_t(u|x) = \lim_{h \rightarrow 0+} P[u \leq C < u+h |\; C \geq u, T=t, X=x]/h$ is the net (conditional on $X$) hazard for censoring.  %That is, we are assuming that censoring is non-informative, i.e., it is explainable by measured covariates. 
Note 
$G_t(u|x) =\exp\left(-\Lambda^{\dagger}_t(u|x) \right)$, where $\Lambda^{\dagger}_t(\cdot|x)$ is the cumulative treatment/cause-specific (conditional on $X=x$) hazard of censoring.  Assumption (\ref{censoring}) also implies that 
\begin{equation}
\upsilon^{\dagger}_t(u|x) = \upsilon_t(u|x),
\label{eqn3}
\end{equation}
where $\upsilon^{\dagger}_t(u|X) = \lim_{h \rightarrow 0+} P[u \leq \widetilde{Y} < u+h, \Delta=1\;|\; \widetilde{Y} \geq u, T=t,X=x]/h$ is the treatment/cause-specific (conditional on $X$) hazard of failure and $\upsilon_t(u|X) = \lim_{h \rightarrow 0+} P[u \leq Y < u+h |\; Y \geq u, T=t, X=x]/h$ is the net (conditional) hazard of failure. Note $F_t(u|x) = 1-\exp \left( -\Upsilon^{\dagger}_t(u|x) \right)$, where 
$\Upsilon^{\dagger}_t(\cdot|x)$ is the cumulative treatment/cause-specific (conditional on $X=x$) hazard of failure.

\subsection{Identifiability}

Under Assumptions (\ref{eqn1}) and (\ref{censoring}), we can express $P[Y(t) \leq s]$ ($s\in[0, \tau]$) as a function of the distribution of the observed data as follows:
\begin{eqnarray}
\psi_t(s;\gamma_t) %& = & \int_x \left\{ P[Y > s | T=t, X=x] P[T=t|X=x] + \right. \\
%&& \left. \frac{ P[Y > s | T=t,X=x] \exp \{ \gamma_t \}  }{ P[Y \leq  s | T=t,X=x]  + P[Y > s | T=t,X=x] \exp \{ \gamma_t \} }  \right\} P[T=1-t|X=x]  dF(x) \\
& = & \int_x \left\{ F_t(s|x) \pi_t(x) + \frac{ F_t(s|x) \exp \{ \gamma_t  \}  }{ 1- F_t(s|x)   + F_t(s|x)  \exp \{ \gamma_t  \} }  \pi_{1-t}(x) \right\}  dF_X(x). %= \psi_t(P),
\label{eqn4}
\end{eqnarray}
%where
%\[
%F_t(s|x) = 1-\Prodi_{u \leq s} \left( 1- d\Upsilon^{\dagger}_t(u|x) \right),
%\]
%$\Upsilon^{\dagger}_t(\cdot|x)$ is the cumulative treatment/cause-specific (conditional on $X=x$) hazard of failure and $
%F_X(\cdot)$ is the marginal distribution of $X$. 
For fixed $\gamma_t$, the right hand side of (\ref{eqn4}) depends on quantities that are identified from the distribution of the observed data. %\textcolor{blue}{Intuitively, since $\gamma_t$ is a user-specified sensitivity parameter rather than an identifiable quantity, it cannot be learned from observed data regardless of sample size.} 
In Appendix A of the Supplementary Material, we provide an extended discussion of identifiability and explain why (\ref{eqn1}) and (\ref{censoring}) place no restrictions on the  law of $\widetilde{O}$ and why restrictions on the law of $\widetilde{O}$ provide no information about $\gamma_t$ ($t=0,1)$.  To ease notation in the formulae that follow, we suppress on $s$ and $\gamma_t$ (e.g., we refer to $\psi_t(s;\gamma_t)$ as $\psi_t$).

\subsection{Uncensored Data Influence Function}

In Appendix B of the Supplementary Material, we derive the uncensored data non-parametric influence function for $\psi_t$ under Assumption (\ref{eqn1}):
\begin{equation}
\phi_t(P,\psi_t)(O)  = \underbrace{\phi_{t1}(P)(O)  + \phi_{t2}(P)(O)  + \phi_{t3}(P)(O)}_{\phi_t(P)(O)} - \psi_t,
\end{equation}
where
\begin{eqnarray*}
%\phi_t(O) & = & \phi_{t1}(O)  + \phi_{t2}(O)  + \phi_{t3}(O) \\
\phi_{t1}(P)(O) & =&  I(T=t) I(Y \leq s)\left\{ 1+ \frac{\pi_{1-t}(X)}{\pi_t(X)} \frac{ \exp(\gamma_t)}{\{ 1-F_t(s|X) + F_t(s|X) \exp(\gamma_t)\}^2} \right\},  \\
\phi_{t2}(P)(O) & =&  - I(T=t)  \left\{ \frac{\pi_{1-t}(X)}{\pi_t(X)} \frac{  F_t(s|X) \exp(\gamma_t)}{\{ 1-F_t(s|X) + F_t(s|X) \exp(\gamma_t)\}^2} \right\}, \\
\phi_{t3}(P)(O) & =&  I(T=1-t) \left\{ \frac{F_t(s|X) \exp(\gamma_t)}{1-F_t(s|X) + F_t(s|X) \exp(\gamma_t)} \right\}.     
\end{eqnarray*}

\subsection{Observed Data Influence Function}

The uncensored data influence function can be used to form an observed data influence function \citep{rotnitzky2005inverse} using the following steps: (1) define an indicator variable, $\xi$ that takes on the value 1 when the full data influence function is observed and 0 otherwise, (2) compute the conditional probability that $\xi=1$ given $O$, (3) multiply the the full data influence function by $\xi$ and divide by the he conditional probability that $\xi=1$ given $O$ (inverse weighted term), (4) compute negative of the projection of the inverse weighted term onto the space spanned by scores associated with the censoring mechanism (augmentation term) and (5) add the inverse weighted and augmentation terms to compute an observed data influence function. 

 In our problem, $\xi = \Delta + (1-\Delta) I( \widetilde{Y} \geq s)$ and $P[\xi=1|O] = G_{T}(\min(\widetilde{Y},s)^-| X)$. %, where $G_t(u|X) = P[ C > u | T=t,X] =  \Prodi_{u' \leq u} (1- d\Lambda^{\dagger}_t(u'|X))$, $\Lambda^{\dagger}_t(\cdot|X)$ is the cumulative treatment/cause-specific (conditional on $X$) hazard of censoring. 
So, the inverse weighted term is: 
\[
\frac{\Delta + (1-\Delta) I( \widetilde{Y} \geq s) }{G_{T}(\min(\widetilde{Y},s)^-| X)}\phi_t(P;\psi_t)(O) 
\]

The space spanned by scores associated with the censoring mechanism is 
\[
\bigg\{\int h(u,T,X)dM_c(\widetilde{P})(u,T,X): h(u,T,X)\bigg\},
\]
where $dM_c(\widetilde{P})(u,T,X)=dN_c(u)-I(\widetilde{Y}\geq u)d\Lambda^{\dagger}_T(u|X)$,  $N_c(u)=I(\widetilde{Y}\leq u, \Delta=0)$ and $dN_c(u)=N_c(u)-N_c(u^-)$.  In Appendix C of the Supplementary Material, we derive the projection of the inverse weighted term onto this space as $\int h^*(\widetilde{P},\psi_t)(u,T,X)dM_c(\widetilde{P})(u,T,X)$, where
\begin{eqnarray*}
h^*(\widetilde{P},\psi_t)(u,T,X) & = & - \frac{E[ \phi_{t1}(P)(O) | Y \geq u, T,X]}{G_T(u^-|X)} - \frac{I(u<s)}{G_T(u^-|X)}\bigg(\phi_{t2}(P)(O)+\phi_{t3}(P)(O)-\psi_t\bigg)
\end{eqnarray*}
We can write the negative of the projection as
\[
\frac{1-\Delta}{G_T(\widetilde{Y}^-|X)} l_{t}(P,\psi_t)(\widetilde{Y},T,X) - \int_0^{\widetilde{Y}} \frac{l_t(P,\psi_t)(u,T,X)}{G_T(u^-|X)}  d\Lambda^{\dagger}_T(u|X), 
\]
where
\[
l_t(P,\psi_t)(u,T,X) = \underbrace{l_{t1}(P)(u,T,X)  + l_{t2}(P)(u,T,X)  + l_{t3}(P)(u,T,X)}_{l_t(P)(u,T,X)}  - I(u < s) \psi_t,
\]
\begin{eqnarray*}
%l_t(u,T,X) & = & l_{t1}(u,T,X)  + l_{t2}(u,T,X)  + l_{t3}(u,T,X) \\
l_{t1}(P)(u,T,X) & =& E[ \phi_{t1}(P)(O) | Y \geq u, T,X] \\
& =&  I(T=t)  I(u \leq s) \frac{ F_t(s|X) - F_t(u^-|X) }{1-F_t(u^-|X)} \times \\
& & \hspace*{0.5in} \left\{1+  \frac{\pi_{1-t}(X)}{\pi_t(X)} \frac{ \exp(\gamma_t)}{\{ 1-F_t(s|X) + F_t(s|X) \exp(\gamma_t)\}^2} \right\},  \\
l_{t2}(P)(u,T,X) & = &  %- I(T=t)  \left\{ \frac{\pi_{1-t}(X)}{\pi_t(X)} \frac{  F(u|t,X) \exp(\gamma_t)}{\{ 1-F(u|t,X) + F(u|t,X) \exp(\gamma_t)\}^2} \right\} \\%
%- I(T=t') \left\{ \frac{\pi_{1-t}(X)}{\pi_t(X)} \frac{  F_t(s|X) \exp(\gamma_t)}{\{ 1-F_t(s|X) + F_t(s|X) \exp(\gamma_t)\}^2} \right\} \\ %
I(u<s) \phi_{t2}(P)(O), \\
l_{t3}(P)(u,T,X) & = &  %I(T=1-t) \left\{ \frac{F(u|t,X) \exp(\gamma_t)}{1-F(u|t,X) + F(u|t,X) \exp(\gamma_t)} \right\} %
%I(t=1-t') \left\{ \frac{F_t(s|X) \exp(\gamma_t)}{1-F_t(s|X) + F_t(s|X) \exp(\gamma_t)} \right\}   %
I(u<s) \phi_{t3}(P)(O). 
\end{eqnarray*}

Thus, the observed data influence function takes the form:
\begin{align*}
& \widetilde{\phi}_t(\widetilde{P},\psi_t)(\widetilde{O}) \\
& = \frac{\Delta + (1-\Delta) I( \widetilde{Y} \geq s) }{G_{T}(\min(\widetilde{Y},s)^-| X)}\phi_t(P,\psi_t)(O) + \frac{(1-\Delta)}{G_T(\widetilde{Y}^-| X)} l_{t}(P,\psi_t)(\widetilde{Y},T,X) -  \int_0^{\widetilde{Y}} \frac{l_{t}(P,\psi_t)(u,T,X)}{G_T(u^-|X)} d\Lambda^{\dagger}_T(u|X) \\
%& = & \frac{I(T=t) \Delta}{G_{t}(\widetilde{Y}-| X)}\phi_t(O;F_t(s)) + \frac{I(T=t) (1-\Delta)}{G_t(\widetilde{Y}-| X)} l_{t}(\widetilde{Y},t,X;F_t(s)) - I(T=t) \int_0^{\widetilde{Y}} \frac{l_{t}(u,t,X;F_t(s))}{G_t(u-|X)} d\Lambda^{\dagger}_C(u|T=t,X) \\
%& = & \frac{I(T=1-t) \Delta}{G_{1-t}(\widetilde{Y}-| X)}\phi_t(O;F_t(s)) + \frac{I(T=1-t) (1-\Delta)}{G_{1-t}(\widetilde{Y}-| X)} l_{t}(\widetilde{Y},1-t,X;F_t(s)) - I(T=1-t) \int_0^{\widetilde{Y}} \frac{l_{t}(u,1-t,X;F_t(s))}{G_{1-t}(u-|X)} d\Lambda^{\dagger}_C(u|T=1-t,X) 
& = \underbrace{\frac{\Delta + (1-\Delta) I( \widetilde{Y} \geq s) }{G_{T}(\min(\widetilde{Y},s)^-| X)}\phi_t(P)(O) + \frac{(1-\Delta)}{G_T(\widetilde{Y}^-| X)} l_{t}(P)(\widetilde{Y},T,X) - \int_0^{\widetilde{Y}} \frac{l_{t}(P)(u,T,X)}{G_T(u^-|X)} d\Lambda^{\dagger}_T(u|X)}_{g_t(\widetilde{P})(\widetilde{O})} - \\
 & \; \; \;  \underbrace{\left\{ \frac{\Delta + (1-\Delta) I( \widetilde{Y} \geq s) }{G_{T}(\min(\widetilde{Y},s)^-| X)} + \frac{(1-\Delta) I(\widetilde{Y} < s)}{G_T(\widetilde{Y}^-| X)}  -  \int_0^{\widetilde{Y}} \frac{I(u < s)}{G_T(u^-|X)} d\Lambda^{\dagger}_T(u|X) \right\}}_{h_t(\widetilde{P})(\widetilde{O})} \psi_t.
\end{align*}

In Lemma 3 of Appendix D, we prove a result that is critical for establishing the asymptotic distribution of our estimator for $\psi_t$. Specifically, we show that $\vline \; E \left[  g_t(\widetilde{P}^*)(\widetilde{O}) - h_t(\widetilde{P}^*)(\widetilde{O})  \psi_t \right] \; \vline$ is bounded above by:
%the sum of eight terms, each of which is a second order product of differences between a component of $\widetilde{P}^*$ and the corresponding component of $\widetilde{P}$: 
\begin{align}
& \sup_{u\in[0, \tau]}\left|\left|F^\ast_t(u|X)-F_t(u|X)\right|\right|_{L_2} \times \sup_{u\in[0, \tau]}\left|\left|G^\ast_{t}(u| X)-G_{t}(u| X)\right|\right|_{L_2} \nonumber \\
& + \sup_{u\in[0, \tau]}\left|\left|F^\ast_t(u|X)-F_t(u|X)\right|\right|_{L_2} \times \sup_{u\in[0, \tau]}\left|\left|G^\ast_{1-t}(u| X)-G_{1-t}(u| X)\right|\right|_{L_2} \nonumber \\
& + \sup_{u\in[0, \tau]}\left|\left|F^\ast_t(u|X)-F_t(u|X)\right|\right|_{L_2} \times \left|\left|\pi^\ast_t(X)-\pi_t(X)\right|\right|_{L_2} \nonumber  \\
& + \left\{ \sup_{u\in[0, \tau]}\left|\left|F^\ast_t(u|X)-F_t(u|X)\right|\right|_{L_2} \right\}^2 + E\left[\int_0^{\tau}(G^\ast_{t}(u|X)-G_{t}(u|X))^2du\right] \nonumber  \\
& + E\left[\int_0^{\tau}\left(F^\ast_t(u|X)-F_t(u|X)\right)^2du\right] + E\left[\int_0^{\tau}\left(\upsilon^{\dagger *}_{t}(u|X)-\upsilon^{\dagger}_{t}(u|X)\right)^2du\right], \label{bound}
\end{align}
%\begin{enumerate}
%    \item  $\sup_{u\in[0, \tau]}\left|\left|F^\ast_t(u|X)-F_t(u|X)\right|\right|_{L_2} \times \sup_{u\in[0, \tau]}\left|\left|G^\ast_{t}(u| X)-G_{t}(u| X)\right|\right|_{L_2}$, 
%    \item $\sup_{u\in[0, \tau]}\left|\left|F^\ast_t(u|X)-F_t(u|X)\right|\right|_{L_2} \times \sup_{u\in[0, \tau]}\left|\left|G^\ast_{1-t}(u| X)-G_{1-t}(u| X)\right|\right|_{L_2}$
%    \item $\sup_{u\in[0, \tau]}\left|\left|F^\ast_t(u|X)-F_t(u|X)\right|\right|_{L_2} \times \left|\left|\pi^\ast_t(X)-\pi_t(X)\right|\right|_{L_2}$
%    \item $\left\{ \sup_{u\in[0, \tau]}\left|\left|F^\ast_t(u|X)-F_t(u|X)\right|\right|_{L_2} \right\}^2$
%    \item $E\left[\int_0^{\tau}(G^\ast_{t}(u|X)-G_{t}(u|X))^2du\right]$
%    \item $E\left[\int_0^{\tau}(G^\ast_{1-t}(u|X)-G_{1-t}(u|X))^2du\right]$
%    \item $E\left[\int_0^{\tau}\left(F^\ast_t(u|X)-F_t(u|X)\right)^2du\right]$
%    \item $E\left[\int_0^{\tau}\left(\upsilon^{\dagger *}_{t}(u|X)-\upsilon^{\dagger}_{t}(u|X)\right)^2du\right]$,
%\end{enumerate}
where $\widetilde{P}^*$ is any distribution for $\widetilde{O}$ and $\left| \left| c(X) \right| \right|_{L_2} = \sqrt{ \int c(x)^2 dF_X(x)}$. 

\subsection{Additional Modeling} \label{section3.6}

The observed data influence function depends on $\widetilde{P}$, specifically through $d\Lambda^{\dagger}_t(\cdot|X)$, $d\Lambda^{\dagger}_{1-t}(\cdot|X)$, $d\Upsilon^{\dagger}_t(\cdot|X)$ and $\pi_1(X)$ ($\pi_0(X) = 1-\pi_1(X)$). %\textcolor{cyan}{
%In Appendix D of the Supplementary Material, we show that the observed data influence function $\widetilde{\phi}_t(\widetilde{O};\psi_t(s))$ is robust to mis-specification of $d\Lambda^{\dagger}_t(\cdot|X)$, $d\Lambda^{\dagger}_{1-t}(\cdot|X)$ and $\pi_1(X)$. } 
In order to use the influence function as the basis of inference for $\psi_t$, we need to estimate these functions of $X$.  Unfortunately, 
%where $\Upsilon^{\dagger}_t(\cdot|X)$ is cumulative treatment/cause-specific (conditional on $X$) hazard of failure.  
the rates of convergence of non-parametric estimators of these quantities decay with the dimension of $X$ \citep{stone1982optimal}. One way to ensure that our influence based estimator of $\psi_t$  converges at a root-$n$ rate is for estimators of these functions of $X$ to converge at rates faster than $n^{1/4}$ \citep{newey1990semiparametric,bickel1993efficient}. This can be accomplished by imposing modeling restrictions on these functions. We specify proportional hazards regression models ~\citep{cox1972regression} for $d\Lambda^{\dagger}_t(\cdot|X)$ and $d\Upsilon^{\dagger}_t(\cdot|X)$, i.e.,
\begin{equation}
d\Lambda^{\dagger}_t(\cdot|X) = d\Lambda^{\dagger}_t(\cdot) \exp \{ \beta_t' X \},
\label{eqn6}
\end{equation}
\begin{equation}
d\Upsilon^{\dagger}_t(\cdot|X) = d\Upsilon^{\dagger}_t(\cdot) \exp \{ \alpha_t' X \},
\label{eqn7}
\end{equation}
where $d\Lambda^{\dagger}_t(\cdot)$ and  $d\Upsilon^{\dagger}_t(\cdot)$ are unspecified treatment/cause-specific baseline hazard functions and $\beta_t$ and $\alpha_t$ are treatment-specific regression parameters. We posit a generalized additive logistic regression model~\citep{hastie2017generalized} for $\pi_1(X)$.  As discussed in the next section, estimators of the parameters of these three models have been shown to converge at rates faster than $n^{1/4}$. 
%}\textcolor{cyan}{and the rate conditions in Theorem~\ref{rate} are satisfied. }

\subsection{Estimation}

\subsubsection*{Estimation of $\widetilde{P}$}

In (\ref{eqn6}) and (\ref{eqn7}), the regression parameters can be estimated using partial likelihood~\citep{cox1975partiallikelihood} and the baseline hazards can be estimated by kernel smoothing the Breslow's estimator of the cumulative baseline hazards~\citep{breslow1972discussion}, using an Epanechnikov kernel and a bandwidth of $b_n$ ~\citep{ramlau1983smooth}. For $t=0,1$, denote these estimators by $\widehat{\beta}_t$, $\widehat{\alpha}_t$, $d\widehat{\Lambda}^{\dagger}_t(u)$ and $d\widehat{\Upsilon}^{\dagger}_t(u)$; We can then estimate $F_t(u|X)$ by $\widehat{F}_t(u|X)= 1-\exp\left( -\widehat\Upsilon^{\dagger}_t(u)\exp \{ \widehat{\alpha}_t' X\} \right)$ and $G_t(u|X)$ by $\widehat{G}_t(u|X)= \exp\left(- \widehat{\Lambda}^{\dagger}_t(u) \exp \{ \widehat{\beta}_t' X \}\right)$.
\citet{wells1994smooth} %and \citet{smith2024smooth} 
shows that if the chosen bandwidth $b_n$ satisfies $\lim_{n\rightarrow\infty}n b_n^5=d<\infty$, then $\Big|\Big|\widehat{\upsilon}^{\dagger}_{t}(u)-\upsilon^{\dagger}_{t}(u)\Big|\Big|_{L_2}$ and $\Big|\Big|\widehat{\lambda}^{\dagger}_{t}(u)-\lambda^{\dagger}_{t}(u)\Big|\Big|_{L_2}$ are $O_P(n^{-2/5})$. In addition, $\widehat{\beta}_t$ and $\widehat{\alpha}_t$ converge to $\beta_t$ and $\alpha_t$, respectively, at root-$n$ rates~\citep{fleming2013counting}. %In Appendix E of the Supplementary Material, Lemma 3 and Lemma 4 
It can then be shown that $\sup_{u\in[0, \tau]}\Big|\Big|\widehat{F}_t(u|X)-F_t(u|X)\Big|\Big|_{L_2}=O_P(n^{-2/5})$, $\sup_{u\in[0, \tau]}\Big|\Big|\widehat{G}_t(u|X)-G_t(u|X)\Big|\Big|_{L_2}=O_P(n^{-2/5})$, $E\left[\int_0^{\tau}\left(\widehat{\upsilon}^{\dagger}_{t}(u|X)-\upsilon^{\dagger}_{t}(u|X)\right)^2du\right]=O_P(n^{-4/5})$,
% By the Continuous Mapping Theorem, we can further show that 
$E\left[\int_0^{\tau}\left(\widehat{F}_t(u|X)-F_t(u|X)\right)^2du\right]=O_p(n^{-4/5})$ and $E\left[\int_0^{\tau}\left(\widehat{G}_t(u|X)-G_t(u|X)\right)^2du\right]=O_P(n^{-4/5})$.
%By the continuous mapping theorem, we can show that the convergence rate for $\int_0^{\tau} \left(\widehat{\upsilon}^{\dagger}_{t}(u)-\upsilon^{\dagger}_{t}(u)\right)^2 du $ and $\int_0^{\tau} \left(\widehat{\lambda}^{\dagger}_{t}(u)-\lambda^{\dagger}_{t}(u)\right)^2 du$ are faster than $n^{1/2}$. Then, $E\left[\int_0^{\tau}\left(\widehat{\lambda}^{\dagger}_{t}(u)\exp\{\widehat{\beta}_{t}'X\}-\lambda^{\dagger}_{t}(u)\exp\{\beta_{t}'X\}\right)^2du\right]$ and $E\left[\int_0^{\tau} \left(\widehat{\upsilon}^{\dagger}_{t}(u)\exp\{\widehat{\alpha}_{t}'X\}-\upsilon^{\dagger}_{t}(u)\exp\{\alpha_{t}'X\}\right)^2du\right]$ are $O_p(n^{-1})$. Since $\widehat{F}_t(u|X)$ and $\widehat{G}_t(u|X)$ are estimated with $\widehat{\beta}_t$, $\widehat{\alpha}_t$, $d\widehat{\Lambda}^{\dagger}_t(u)$ and $d\widehat{\Upsilon}^{\dagger}_t(u)$, we know that the rate conditions for $\widehat{F}_t(u|X)$ and $\widehat{G}_t(u|X)$ in Lemma~\ref{rate} are satisfied as well. 
%note that the estimators of the baseline hazards jump at the observed censoring and failure times, respectively. 
Estimation of the generalized additive logistic regression model for $\pi_1(X)$ uses a back-fitting algorithm~\citep{hastie2017generalized}.  Denote the estimator for $\pi_t(X)$ by $\widehat{\pi}_t(X)$. \cite{horowitz2004nonparametric} showed that
$
\Big|\Big| \ \widehat{\pi}_t(X) - \pi_t(X) \ \Big|\Big|_{L_2} =O_P(n^{-2/5}).
$

Putting these results together with bound (\ref{bound}) on $\vline \; E \left[  g_t(\widetilde{P}^*)(\widetilde{O}) - h_t(\widetilde{P}^*)(\widetilde{O})  \psi_t \right] \; \vline$ shows that  $E \left[  g_t(\widehat{\widetilde{P}})(\widetilde{O}) - h_t(\widehat{\widetilde{P}})(\widetilde{O})  \psi_t \right]$ is $o_P(n^{-1/2})$, where $\widehat{\widetilde{P}}$ is an estimator of $\widetilde{P}$ using the modeling assumptions and estimation procedures described above.

\subsubsection*{Estimation of $\psi_t$}

To estimate $\psi_t$, we use an influence function-based split sampling procedure.  Specifically, we randomly split the $n$ observations into $K$ disjoint sets, where $S_i$ denotes the split membership of the $i$th individual.  The size of the $k$th disjoint set is denoted by $n_k=n/K$ (i.e., $n_k = O (n )$ ). Let $\widehat{\widetilde{P}}^{(-k)}$ be an application of the above estimation procedure based on all individuals except those in the $k$th split.

Our split-sample estimator of $\psi_t$ is the solution $\widehat{\psi}_t$, to
\begin{equation}
\sum_{k=1}^K \sum_{i \in S_k} \left\{ g_t\left(\widehat{\widetilde{P}}^{(-k)}\right)(\widetilde{O}_i) - h_t\left(\widehat{\widetilde{P}}^{(-k)}\right)(\widetilde{O}_i) \psi_t \right\} = 0,
\end{equation}
which takes the form:
\begin{equation}
\widehat{\psi}_t = \frac{\sum_{k=1}^K \sum_{i \in S_k} g_t\left(\widehat{\widetilde{P}}^{(-k)}\right)(\widetilde{O}_i) }{\sum_{k=1}^K \sum_{i \in S_k} h_t\left(\widehat{\widetilde{P}}^{(-k)}\right)(\widetilde{O}_i)}.
\end{equation}
In Theorem 1 of Appendix D, we show that $\widehat{\psi}_t$ is robust to mis-specification of $G_0(\cdot|X)$, $G_1(\cdot|X)$ and $\pi_t(X)$. That is, consistency of $\widehat{\psi}_t$ only relies on the correct specification of $F_t(\cdot|X)$, not on $G_0(\cdot|X)$, $G_1(\cdot|X)$ and $\pi_t(X)$. In Theorem 2 of Appendix D, we show that 
\begin{align}
& \frac{1}{\sqrt{n}} \sum_{k=1}^K \sum_{i \in S_k} \left\{ g_t\left(\widehat{\widetilde{P}}^{(-k)}\right)(\widetilde{O}_i) - h_t\left(\widehat{\widetilde{P}}^{(-k)}\right)(\widetilde{O}_i) \psi_t \right\} \nonumber \\
& = \underbrace{\frac{1}{\sqrt{n}} \sum_{k=1}^K \sum_{i \in S_k} \left\{ g_t\left(\widetilde{P}\right)(\widetilde{O}_i) - h_t\left(\widetilde{P}\right)(\widetilde{O}_i) \psi_t \right\}}_{\stackrel{D}{\rightarrow} N(0,E[\widetilde{\phi}_t(\widetilde{P},\psi_t)(\widetilde{O}) ^2]) \mbox{ by Central Limit Theorem}} + o_P(1),
\end{align}
where bound (\ref{bound}) on $\vline \; E \left[  g_t(\widetilde{P}^*)(\widetilde{O}) - h_t(\widetilde{P}^*)(\widetilde{O})  \psi_t \right] \; \vline$ was critical showing that the remainder term is $o_P(1)$. In Theorem 3 of Appendix D, we use the mean value theorem to show that
\begin{equation}
\sqrt{n} (\widehat{\psi}_t - \psi_t) =   \frac{ \frac{1}{\sqrt{n}}\sum_{k=1}^K \sum_{i \in S_k} \left\{ g_t\left(\widehat{\widetilde{P}}^{(-k)}\right)(\widetilde{O}_i) - h_t\left(\widehat{\widetilde{P}}^{(-k)}\right)(\widetilde{O}_i) \psi_t \right\} }{\frac{1}{n} \sum_{k=1}^K \sum_{i \in S_k} h_t\left(\widehat{\widetilde{P}}^{(-k)}\right)(\widetilde{O}_i)}.
\end{equation}
The numerator converges in distribution to $N(0,E[\widetilde{\phi}_t(\widetilde{P},\psi_t)(\widetilde{O}) ^2])$
and the deominator converges in probability to $E[ h_t(\widetilde{P}(\widetilde{O})]$. By Slutsky's theorem, $\sqrt{n} (\widehat{\psi}_t - \psi_t)$ converges in distribution to \\$N(0,E[ h_t(\widetilde{P}(\widetilde{O})]^{-2} E[\widetilde{\phi}_t(\widetilde{P},\psi_t)(\widetilde{O}) ^2])$. The variance of $\widehat{\psi}_t$ can be estimated by
$$
\widehat{\sigma}^2_t =   \left\{\displaystyle\sum_{k=1}^K\sum_{i\in\mathcal{S}_k}h_t\left(\widehat{\widetilde{P}}^{(-k)}\right)(\widetilde{O}_i)\right\}^{-2}\left\{\displaystyle\sum_{k=1}^K\sum_{i\in\mathcal{S}_k}\left(g_t\left(\widehat{\widetilde{P}}^{(-k)}\right)(\widetilde{O}_i) - h_t\left(\widehat{\widetilde{P}}^{(-k)}\right)(\widetilde{O}_i) \widehat{\psi}_t\right)^2\right\}.
$$

\subsubsection*{Logistics}

The influence function $\widetilde{\phi}_t(\widetilde{P},\psi_t)(\widetilde{O})$ depends on 
$\int_0^{\widetilde{Y}} \frac{l_{t}(P)(u,T,X)}{G_T(u^-|X)} d\Lambda^{\dagger}_T(u|X)$ and $\int_0^{\widetilde{Y}} \frac{I(u < s)}{G_T(u^-|X)} d\Lambda^{\dagger}_T(u|X)$.  We approximate these integrals numerically using the midpoint rule, where we use a common set ${\mathcal U}$ of $M$ equally spaced midpoints that cover $(0,\tau^{\dagger})$.

Our estimation procedure for $\psi_t$ depends on $g_t\left(\widehat{\widetilde{P}}^{(-k)}\right)(\widetilde{O}_i)$ and $h_t\left(\widehat{\widetilde{P}}^{(-k)}\right)(\widetilde{O}_i)$ ($i \in S_k$, $k=1,\ldots, K$), which in turn depend on estimated $1/\pi_t(X)$'s, $1/G_t(\widetilde{Y}|X)$'s, $1/G_{1-t}(\widetilde{Y}|X)$'s, $1/G_t(s|X)$'s, $1/G_{1-t}(s|X)$'s, $1/G_t(u_1|X)$'s, $1/G_{1-t}(u_1|X)$'s, ...,  $1/G_t(u_M|X)$'s, $1/G_{1-t}(u_M|X)$'s (for $u_1,\ldots,u_M \in {\mathcal U})$. These inverse weights can become excessively large leading to extreme values of $g_t\left(\widehat{\widetilde{P}}^{(-k)}\right)(\widetilde{O}_i)$ and $h_t\left(\widehat{\widetilde{P}}^{(-k)}\right)(\widetilde{O}_i)$ for some $i \in S_k$, $k = 1,\ldots,K$.  To avoid this, we perform $99.5$th percentile truncation separately for each inverse weight across split samples. 

%For the estimated $1/G_t(u|X)$'s and $1/G_{1-t}(u|X)$'s, we considered a fine grid of points in $(0, \tau^{\dagger})$. 

 While our estimator of $\psi_t$ will be asymptotically monotone in $s$, it may not be in finite samples.  To address this issue, we employ the pool adjacent violators algorithm (PAVA)~\citep{leeuw2009isotone} to ensure that the estimates are monotonic. PAVA works by iteratively adjusting the values of adjacent data points until they are non-decreasing (in the case of increasing curves) or non-increasing (in the case of decreasing curves).   PAVA does not change the asymptotic distribution of our estimator.

%$\hat{\sigma}^2_{\gamma_t}$, where 
%\[ 
%\left \{ \frac{1}{n} \sum_{i = 1}^n \widehat{h}_t(O_i))\right\}^{-2} \left \{ \frac{1}{n^2} \sum_{i = 1}^n \left\{\widehat{g}_t(O_i) - \widehat{h}_t(O_i)\hat{F}_t(s) \right\}^2\right\}.
%\]
After applying trunaction and PAVA, we construct a logit transformed $95\%$ Wald confidence interval for $\psi_t$. Using the Delta method, we estimate the variance of $\log(\frac{\widehat{\psi}_t}{1-\widehat{\psi}_t})$ by $\left\{\widehat{\psi}_t(1-\widehat{\psi}_t)\right\}^{-2} \widehat{\sigma}^2_t$.
%$$
%    \left\{\widehat{\psi}_t(1-\widehat{\psi}_t)\right\}^{-2}\left\{\displaystyle\sum_{k=1}^K\sum_{i\in\mathcal{S}_k}h_t(\widehat{\widetilde{P}}^{(-k)})(\widetilde{O}_i)\right\}^{-2}\left\{\displaystyle\sum_{k=1}^K\sum_{i\in\mathcal{S}_k}\left(g_t(\widehat{\widetilde{P}}^{(-k)})(\widetilde{O}_i) - h_t(\widehat{\widetilde{P}}^{(-k)})(\widetilde{O}_i) \widehat{\psi}_t\right)^2\right\}
%$$
We then calculate a Wald-based confidence interval for $\log(\frac{\psi_t}{1-\psi_t})$, then back transform the lower and upper bound to the probability scale. We also consider parametric bootstrap~\citep{efron1993introduction} to estimate the variance, $\widehat{\sigma}_{boot,t}^2$, of $\widehat{\psi}_t$.  We then use the above procedure with $\widehat{\sigma}_{boot,t}^2$ replacing $\widehat{\sigma}_{t}^2$ to construct a confidence interval for $\psi_t$.
%construct a logit transformed $95\%$ bootstrap confidence interval for $\psi_t$. Specifically, we use the bootstrap estimate of the variance for $\widehat{\psi}_t$, multiplied by $\left\{\widehat{\psi}_t(1-\widehat{\psi}_t)\right\}^{-2}$, as the bootstrap variance for $\log(\frac{\widehat{\psi}_t}{1-\widehat{\psi}_t})$. Then we apply the same procedure as above to construct the logit transformed confidence interval.  }

\subsection{Interpretation and Calibration of Sensitivity Analysis Parameters}

Determining plausible ranges for $\gamma_t$ is a challenge that requires expert judgment. Here, we review its interpretation and provide two ideas for calibration.

\subsubsection*{Interpretation}
%To help researchers quantify deviations from the assumption of no unmeasured confounding, 
As seen in Section \ref{sec32}, $\gamma_t$ has two interpretations. First, it represents the conditional (on $X=x$) log odds ratio of $Y(t)\leq s$ for individuals who receive treatment $T=1-t$ versus those who receive treatment $T=t$. Second, it represents  the conditional (on $X=x$) log odds ratio of receiving treatment $1-t$ for individuals with $Y(t) \leq s$ versus those with $Y(t)>s$.

\subsubsection*{Calibration}

External data sources can be used to calibrate the value of $\gamma_t$.  Suppose there exists an external data source that has treatment and outcome data collected in a similar fashion as the study cohort and it can be argued that the external data source has  a higher (or lower) risk profile than patients in the cohort.  Then the survival curve from the external data source can be used to bound the survival curve in the study cohort.  This can then be used to bound $\gamma_t$.

 Another approach to evaluate the plausibility of different values of $\gamma_t$ is to compute an induced counterfactual survival for $Y(t)$ given $T=1-t$. These curves are then compared to observed survival data or to survival curves from other studies. Subject matter experts can judge the plausibility of the implied survival distributions to determine reasonable bounds for $\gamma_t$.

\section{Data Analysis}\label{sec5}

We illustrate our proposed method using National Cancer Database (NCDB) data. NCDB is a nationwide comprehensive oncology outcomes database that records $\approx$ 72\% of newly diagnosed cancer cases in the United States annually~\citep{NCDB}. ~\cite{ennis2018brachytherapy} studied prostate cancer patients who underwent one of three treatments:  a surgical procedure radical prostatectomy (RP), treated by one of two therapeutic procedures, external beam radiotherapy (EBRT) combined with androgen deprivation (AD), or  EBRT  plus brachytherapy with or without  AD. In our analysis, we focus on patients with a diagnosis between 2004 and 2010, and similar to the approach taken by~\cite{ennis2018brachytherapy}, we include patients with (1) adenocarcinoma of the prostate and with non-metastatic stages (no N+, no M+) and (2) clinical T stage cT3 or higher, biopsy Gleason score ranging from 8 to 10, or PSA $\geq 20$ ng/dL. We further excluded low risk patients (i.e., Gleason score $\leq 6$ and T-stage = cT1) who at the time of their diagnosis would have been very likely to receive RP. While we increased the comparability of the clinical features of the two treatment groups through this latter restriction, concerns about confounding due to unmeasured factors remain.

 We are interested in comparing treatment with RP versus treatment with EBRT + AD with respect to time from treatment initiation until death. For the purpose of this analysis, we set $\tau=130$ months and $\tau^\dagger=150$ months. Baseline covariates ($X$) included age, PSA, clinical T stage, Charlson-Deyo score, biopsy Gleason score, insurance status, income (divided into quartiles based on zip code of residence), education (divided into quartiles based on the proportion of residents in the patient’s zip code who did not graduate high school), and race.  For our cohort, Table \ref{Table1} shows the descriptive statistics of covariates overall and by treatment received. Patients who underwent the RP procedure were, on average, seven years younger than those treated with EBRT+AD (62.5 vs. 69.4). There were dramatic differences in type of insurance: a higher percentage of RP patients had private insurance/managed care (56.2\% vs. 28.5\%), while a greater percentage of EBRT+AD patients had Medicare (60.6\% vs. 36.2\%).

 Our regression models for treatment, censoring, and failure have additive effects of covariates.  The generalized additive model for treatment naturally allows for non-linear effects of age and PSA.  In the censoring and failure models, we model the effects of these latter covariates using natural cubic splines.  We additionally apply an Epanechnikov kernel with bandwidth of 1 month to smooth the estimated cumulative baseline hazards in the censoring and failure models. 
 
 %To avoid extreme weights, we compute the $99.5$th percentile of elements in the set $\left\{ 1/\widehat{\pi}_t^{(-1)}(X_{i_1}), \ldots, 1/\widehat{\pi}_t^{(-K)}(X_{i_K}): i_1 \in S_1, \ldots i_K \in S_K \right\}$ and then truncate the elements by the computed percentile.
 %$\left\{1/\widehat{G}_t^{(-1)}(u|X_{i_1}), \ldots, 1/\widehat{G}_t^{(-K)}(u|X_{i_K}): i_1 \in S_1, \ldots i_K \in S_K, u\in U\right\}$,\\
 %$\left\{1/\widehat{G}_{1-t}^{(-1)}(u|X_{i_1}), \ldots, 1/\widehat{G}_{1-t}^{(-K)}(u|X_{i_K}): i_1 \in S_1, \ldots i_K \in S_K, u\in U\right\}$, \\
 %$\left\{1/\widehat{G}_{t}^{(-1)}(\widetilde{Y}_{i_1}|X_{i_1}), \ldots, 1/\widehat{G}_{t}^{(-K)}(\widetilde{Y}_{i_K}|X_{i_K}): i_1 \in S_1, \ldots i_K \in S_K\right\}$, and\\
 %$\left\{1/\widehat{G}_{1-t}^{(-1)}(\widetilde{Y}_{i_1}|X_{i_1}), \ldots, 1/\widehat{G}_{1-t}^{(-K)}(\widetilde{Y}_{i_K}|X_{i_K}): i_1 \in S_1, \ldots i_K \in S_K\right\}$, respectively, and perform weight truncation with these values. $U$ are a set of time points ranging from 0 to 150. Note that the inverse censoring weights in the observed data influence function are evaluated at different time points (i.e. at observed event times $\widetilde{Y}$, $s$ or time points between $0$ and $\widetilde{Y}$). For 
 %The $99.5$th percentile is computed and applied to each time point separately.
 
\begin{figure}[hbt!]  
\includegraphics[width=\textwidth,height=\textheight,keepaspectratio]{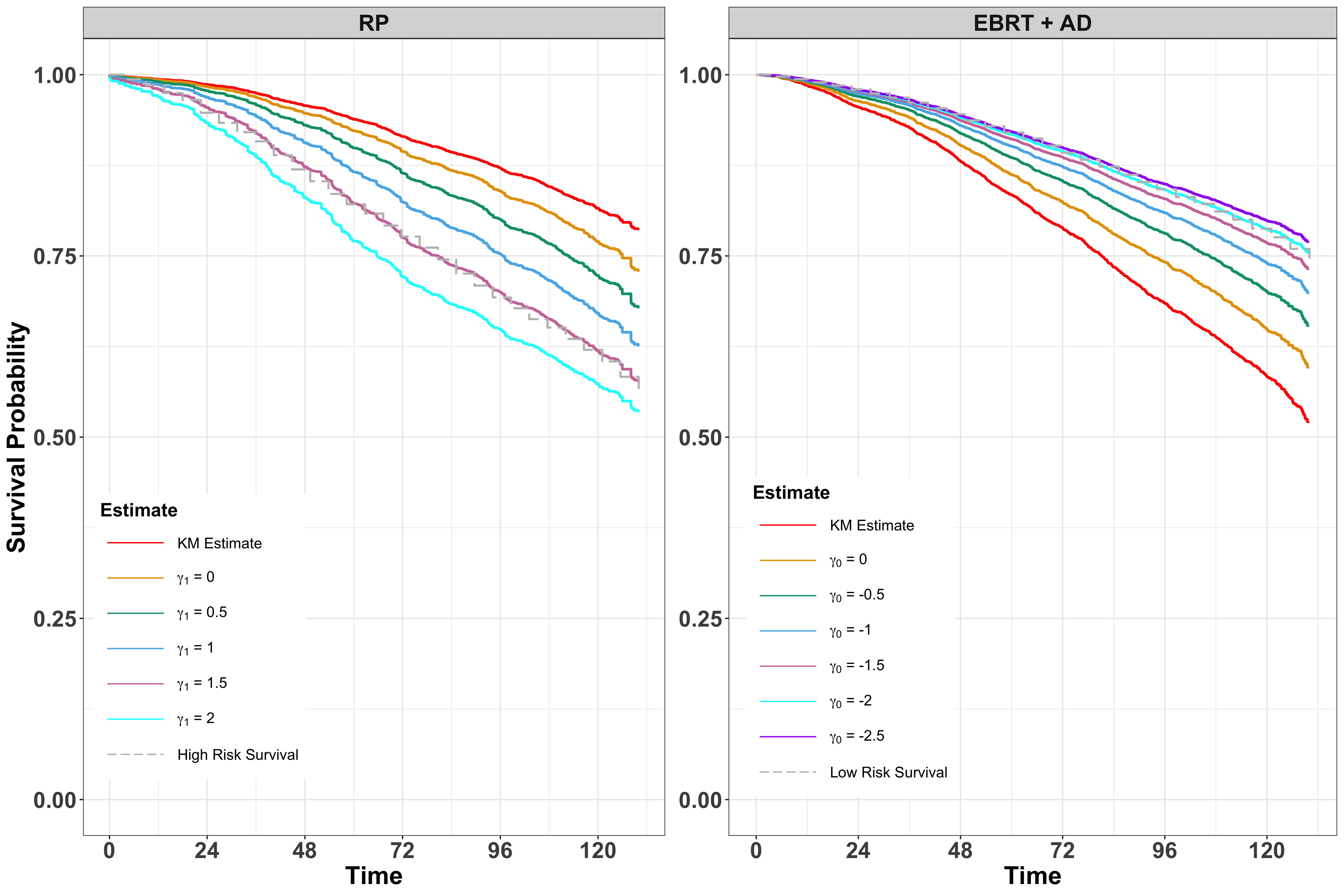}
\caption{Estimated survival curves for undergoing RP (left panel) and EBRT+AD (right panel) treatments, at specified sensitivity parameter $\gamma_t$ values.}
\label{fig2}
\end{figure}

\cite{chen2018challenges} argue that, within levels of measured covariates, patients who receive RP are likely to be healthier than those receiving EBRT+AD.  This implies that $\gamma_1$ should be positive and $\gamma_0$ should be negative. Figure (\ref{fig2}) shows, for each treatment (RP: left panel; EBRT+AD: right panel), estimated survival curves for various sensitivity parameter values: $0.0\leq \gamma_1 \leq 2$ and $-2.5 \leq \gamma_0 \leq 0.0$. For reference, the figure also presents treatment-specific Kaplan-Meier curves (in red). Notice how the RP (EBRT+AD) Kaplan-Meier curve is an over (under)- estimate of the survival experience had everyone received RP (EBRT+AD). 

In Figure \ref{fig2}, the grey curve in the RP (EBRT+AD) panel is an attempt at an under (over)-estimate of survival, thus allowing us to bound the value of the sensitivity analysis parameters. For RP, we include an estimated survival curve  for a cohort of higher-risk patients (from the NCDB prostate cancer database) who were diagnosed between 2004 and 2010 and identified with biopsy Gleason scores ranging from 8 to 10 and PSA $\geq$20 ng/dL or T stage $\geq$ cT3.  Table \ref{Table1} shows the descriptive statistics of this higher-risk cohort. The survival of this higher-risk cohort should be worse than if everyone in our main cohort had received RP, i.e., $0 \leq \gamma_1 \leq 1.5$.  For EBRT+AD treatment, we include an estimated survival curve  for a lower-risk group of patients (from the NCDB prostate cancer database) who were diagnosed between 2004 and 2010 
and identified by T stage cT2 or lower (excluding those with T stage $\leq$ cT2a, biopsy Gleason scores $\leq$ 6, and PSA levels $<$ 10 ng/dL). NCCN guidelines ~\citep{richardnccn} recommend active surveillance as the initial therapy for these patients. Table \ref{Table1} shows the descriptive statistics of this lower-risk cohort. The survival of this cohort should be better than if everyone in our cohort had received EBRT+AD, i.e., $-2.0 \leq \gamma_0 \leq 0$.

\begin{figure}[hbt!]
\includegraphics[width=\textwidth,height=\textheight,keepaspectratio]{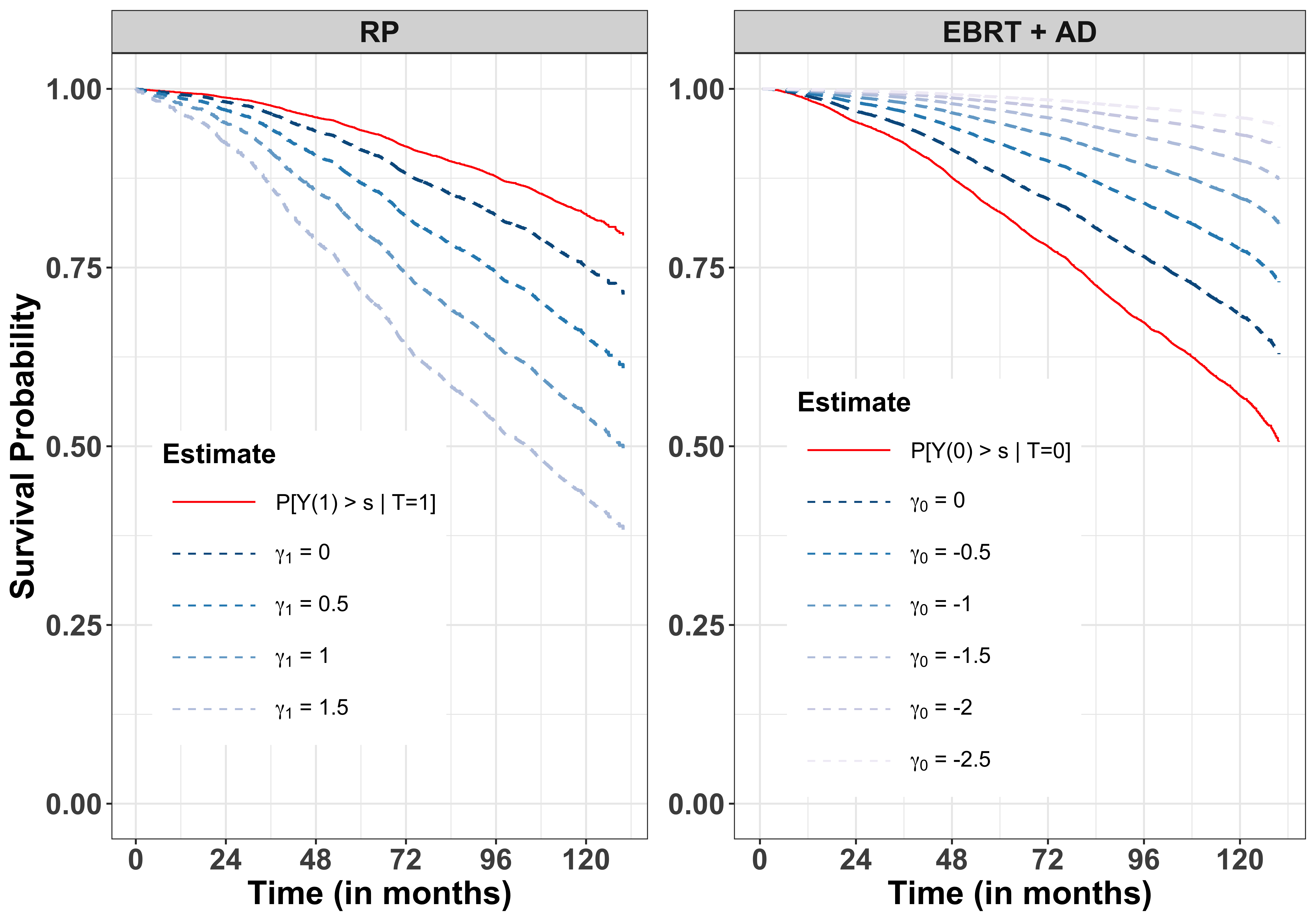}
\caption{Induced survival curves for undergoing RP for people who actually underwent EBRT+AD (left panel) and EBRT+AD for people who actually underwent RP (right panel) treatments, at specified values of $\gamma_t$.}
\label{fig3}
\end{figure}

Another way to assess the plausibility of various values of $\gamma_t$ is to compute induced estimates of the survival curve of $Y(t)$ given $T = 1-t$ as a function of $\gamma_t$ (see Figure \ref{fig3}). For fixed $\gamma_t$, $P[Y(t) >  s | T= 1-t]$ is equal to $(P[Y(t) > s] - P[Y(t) > s | T= t] P[T=t])/P[T=1-t]$, where $P[Y(t) > s | T= t]$ is estimated using the proportional hazards model along with the empirical distribution of $X$ among those who received treatment $t$, $P[T=t']$ is estimated by the observed proportion of individuals with who received treatment $t'$ and $P[Y(t) > s]$ is estimated using (\ref{eqn4}). In the left (right) panel of  Figure \ref{fig3}, we show the induced survival curves for survival under RP (EBRT+AD) for those who actually took EBRT+AD (RP). For example, the estimated 5-year survival under RP (EBRT+AD) for patients who actually received EBRT+AD (RP) would be 82\% (93\%) when $\gamma_1=1.0$ ($\gamma_0=-1.0$) versus 93\% (86\%)  when $\gamma_1=0$ ($\gamma_0=0$). Subject matter experts can use such calculations to judge the plausibility of specific choices of the sensitivity parameters.

\begin{figure}[hbt!]
\includegraphics[width=\textwidth,height=\textheight,keepaspectratio]{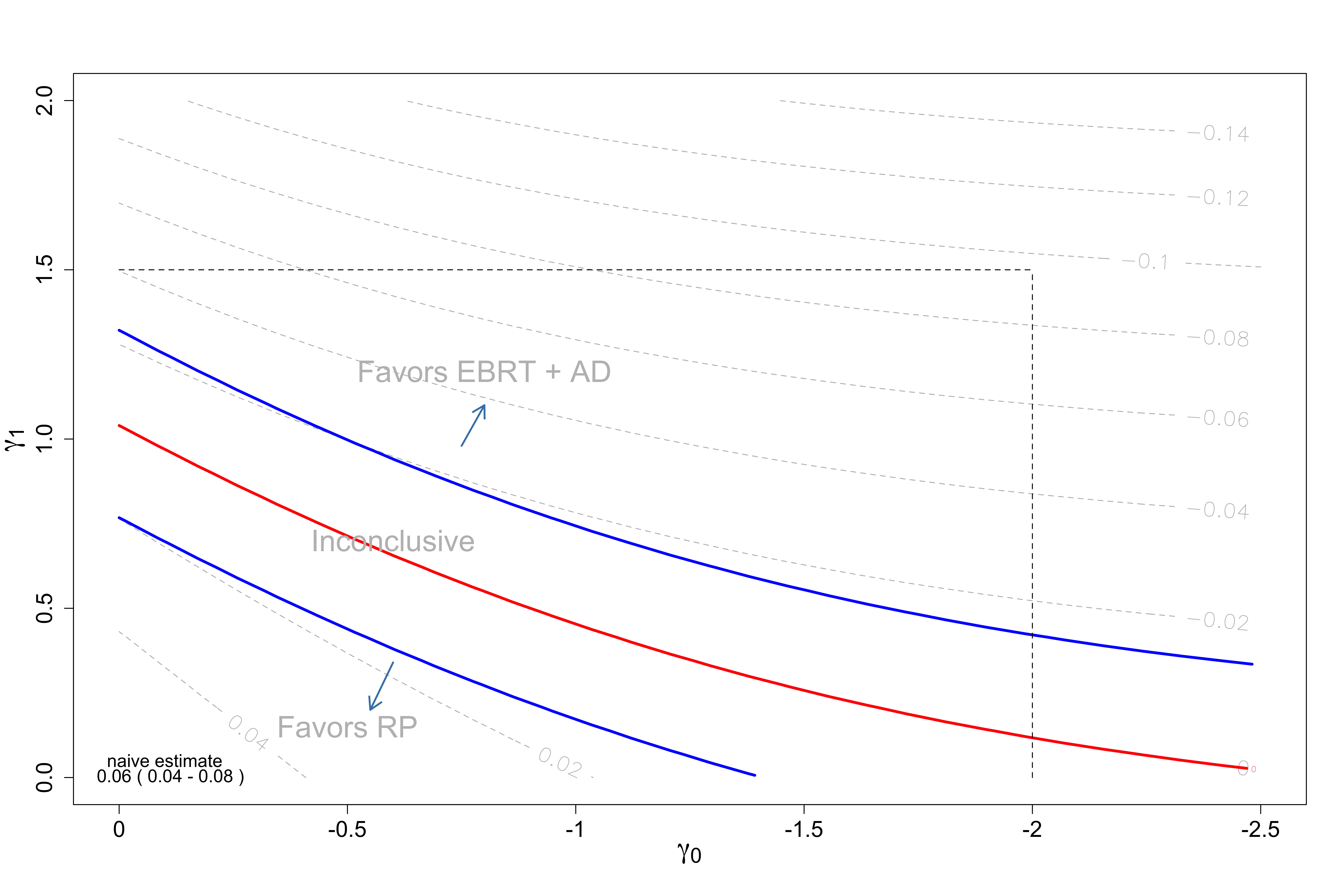}
\caption{Contour plot of estimates of the 5-year survival benefit of undergoing RP versus EBRT + AD for the different combinations of sensitivity parameters $\gamma_1$ and $\gamma_0$. The blue curves reflect the area of inconclusive results (not favoring either RP or EBRT + AD)}
\label{fig4}
\end{figure}

Figure \ref{fig4} displays a contour plot of estimates of the 5-year survival benefit of undergoing RP versus EBRT + AD for the different combinations of sensitivity parameters $\gamma_1$ and $\gamma_0$, respectively. The figure includes the estimate of the 5-year survival benefit of RP over EBRT + AD when $\gamma_1=\gamma_0 = 0$ (i.e., no unmeasured confounding). This estimate suggests that the 5-year survival benefit of undergoing RP is $7\%$ higher than undergoing EBRT + AD (95\% CI: 6\% to 8\%). The red curve represents the contour along which the 5-year survival benefit is estimated to be zero. 
We also include the region of sensitivity parameters (bracketed by the blue curves) that would lead to inconclusive results about the relative effect of RP versus EBRT+AD (i.e., 95\% confidence interval includes 0). The region to the left (right) of the lower (upper) blue lines indicates combinations of sensitivity analysis parameters that yield evidence in favor of RP (EBRT+AD).  These blue curves show the combination of sensitivity parameter values that lead to the same 95\% lower and upper confidence interval bounds for the 5-year survival benefit of zero. 
For reference, we include horizontal and vertical lines at $\gamma_1 = 1.5$ and $\gamma_0 = -2.0$ to indicate the bounds of the sensitivity parameters that we derived from the grey curves depicted in Figure \ref{fig2}.  Our analysis suggests that there are values of sensitivity analysis parameters that would suggest that EBRT+AD is more effective than RP.  This would happen if, for example, $\gamma_1=1.0$ and $\gamma_0=-1.0$; the plausibility of these values can be judged by reviewing the induced survival curves presented in Figure \ref{fig3}.

\section{Simulation Study}\label{sec8}

\subsection{Data Generation}

We conducted a simulation study to assess the performance of our method in recovering the true survival curve for a chosen $\gamma_t$ value. To build a realistic simulation study, we used data from our cohort to build an observed data-generating mechanism. Specifically, we used the empirical distribution of $X$, the estimated parameters from Weibull regression models for (\ref{eqn6}) and (\ref{eqn7}), and the estimated GAM model for treatment as the true observed data generating mechanisms. We considered choices of $\gamma_1 = (0,1,2,3)$ $\gamma_0=(0,-1,-2,-3)$.  For each choice of $\gamma_t$ along with the true distribution of the observed data, we compute the true survival curve using (\ref{eqn4}). We considered sample sizes of 1000, 3000, and 5000. For each individual $i$, we simulated data as follows: (1) randomly sampled covariate vector $(X_i)$ from the original dataset, (2)  using $X_i$, draw treatment assignment ($T_i$) using the GAM model, (3) using $X_i$, draw survival time $(Y_i)$ from the Weibull regression model (\ref{eqn6}) with $t=T_i$  and (4) using $X_i$, draw censoring time $(C_i)$ from the Weibull regression model (\ref{eqn7}) with $t=T_i$, and (5) set $\widetilde{Y}_i = \min \{ Y_i, C_i \}$. For each sample size, we simulated 2000 datasets;  for each simulated dataset, we estimated the treatment-specific survival curves using our proposed method. We evaluated estimation bias and $95\%$ confidence interval coverage.  We considered the two types of Wald-based confidence intervals discussed in Section 2.7: one with influence-based standard errors and one with parametric bootstrap-based standard errors. % confidence intervals (Use the bootstrap (sampling with replacement) to compute the absolute value of a data-driven $T$ distribution. Then, the appropriate quantile (95\%) from the $T$ distribution is plugged into the standard confidence interval equation to obtain a symmetric bootstrap$-t$ confidence interval) 
%\citep{wilcox2011introduction} 
%We compute confidence intervals on the logit scale and then transform them back to the probability scale. 
Coverage is measured as the proportion of simulated samples whose 95\% confidence intervals contain the truth.

\subsection{Results}

\begin{figure}[hbt!]
\includegraphics[width=\textwidth,height=\textheight,keepaspectratio]{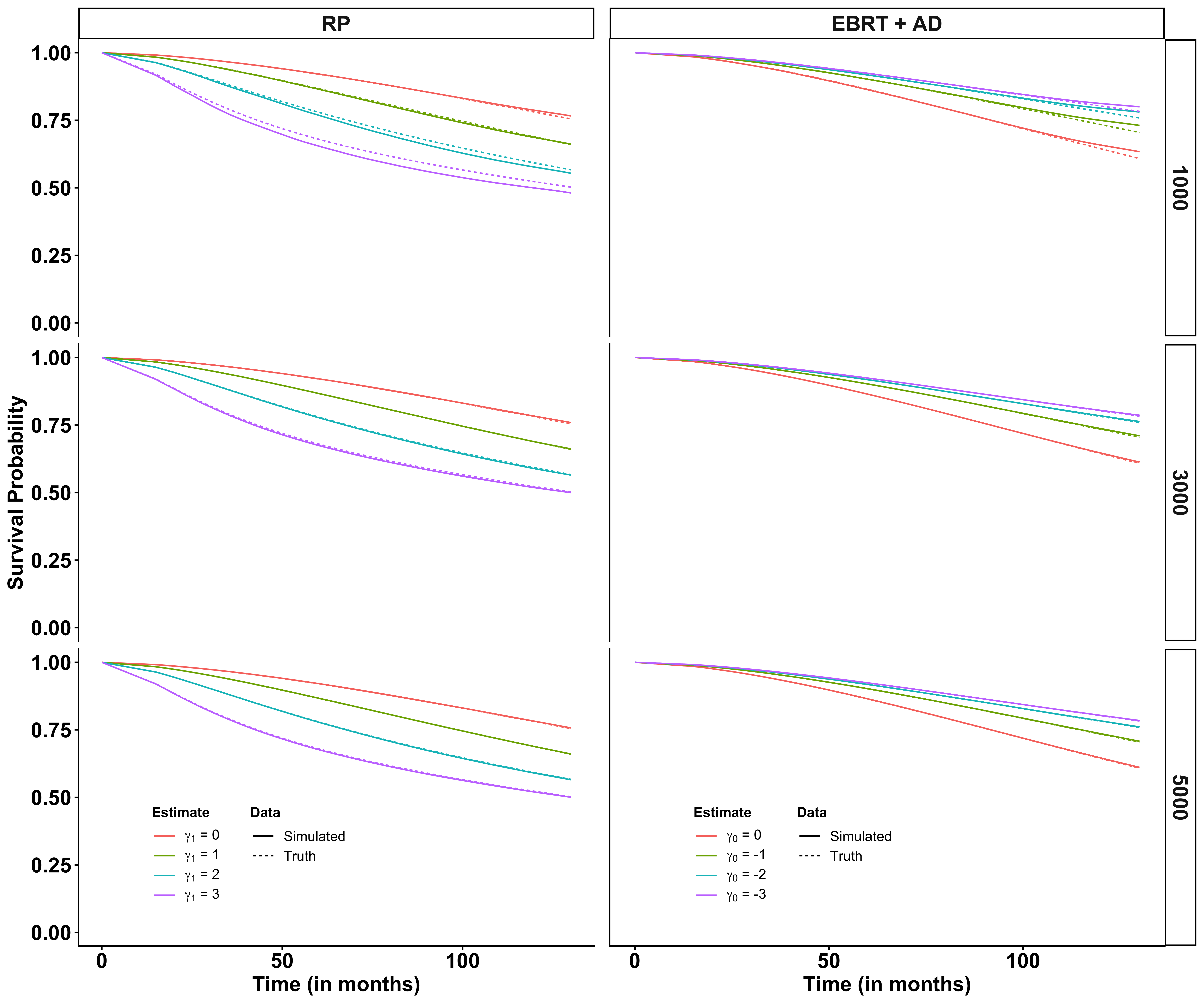}
\caption{Simulation study results with sample sizes of 1000 (1st row), 3000 (2nd row), and 5000 (3rd row), for RP (1st panel), and EBRT $+$ AD (2nd panel) treatments}
\label{fig5}
\end{figure}

Figure \ref{fig5} presents the results of our simulation study; the left (right) panel corresponds to RP (EBRT+AD) and the
rows represent sample sizes 1000, 3000 and 5000, respectively. For each $\gamma_t$, the dashed lines represent the true survival curves and solid lines represent averages of estimates over 2000 simulations. The figure shows that, with correct specifications of $d\Lambda^{\dagger}_t(\cdot|X)$, $d\Lambda^{\dagger}_{1-t}(\cdot|X)$, $d\Upsilon^{\dagger}_t(\cdot|X)$ and $\pi_1(X)$, bias goes to zero with increasing sample size. 

%Dashed lines represent the treatment-specific survival curve for a sensitivity parameter $\gamma_t$ averaged across 2000 simulations with sample sizes of 1000, 3000, and 5000, while solid lines correspond to the truth. The figure reveals that the method yields estimates closer to the truth under the specific sensitivity parameter values, demonstrating little to no bias with increased sample size. 
Table \ref{coverage} shows bias and $95\%$ confidence interval coverage for different $\gamma_t$ values at $2, 5, \text{and } 10$ years. Regardless of the value of $\gamma_t$, bias is low and decreases with sample size. Wald confidence intervals with influence-based standard errors have coverage close to the nominal level, with the exception of RP at $2$ years.  When parametric bootstrap standard errors are used in the construction of confidence intervals, coverage rates are at or slightly higher than the nominal level.
%\textcolor{cyan}{Regardless of sample size or value of $\gamma_t$, , Wald confidence intervals with influence-based standard errors have coverage close to the nominal level, with the exception of RP at $2$ years. Wald confidence intervals with parametric bootstrap standard errors tend to have better,, and gets close to the nominal level with increased sample size.}

\section{Discussion}

%The limitation of an adjusted survival curve is that it might still be biased due to unmeasured confounding. 
In this manuscript, we developed a semiparametric sensitivity analysis approach to address unmeasured confounding in observational studies with time-to-event outcomes subject to right censoring. Our approach allows researchers to quantitatively explore how survival curves and associated treatment effects change under different assumptions about unmeasured confounding.

Equation (\ref{eqn1}) with $\gamma_t$ serving as a non-identified, treatment-specific sensitivity analysis parameter is the core of our approach. The choice of the logit function allowed mapping from the unit interval to the real line and  re-expression of (\ref{eqn1}), using Bayes' rule, into a restriction on the treatment selection mechanism in the form of Equation (\ref{eqn1a}). Our approach can be modified to handle alternatives to the logit function, such as the inverse of the cumulative distribution function or survival function of any continuous random variable. However, the restriction on the treatment selection mechanism may be more complicated. For example, if we replace logit$(x)$ by $g(x) = \log( -\log(1-x))$ in (\ref{eqn1}), then (\ref{eqn1}) implies that the conditional (on $X=x$) hazard function of $Y(t)$ for those who received treatment $1-t$ is equal to the conditional (on $X=x$) hazard function of $Y(t)$ for those who received treatment $t$ times $\exp(\gamma_t)$. In addition, our approach can be easily extended to accommodate alternative models for the $F_t(\cdot|X)$, $G_t(\cdot|X)$ and $\pi_t(X)$, provide the model parameters can be estimated at rates faster than $n^{1/4}$.

Our approach assumes that censoring is independent of failure within levels of measured factors. Like exchangeability, this assumption is untestable.  If there is concern that unmeasured factors related to the outcome are associated with censoring, an additional layer of sensitivity analysis can be added (see, for example, \cite{scharfstein2002estimation}).

Our methodology has limitations. As discussed in Section 2.2, we assume, for the sake of simplicity, that deviations from conditional exchangeability do not depend on $s$ and the levels of $X$. While this assumption can be relaxed, visualization of the results will be more complication. As discussed in Section 2.8, determining plausible ranges for $\gamma_t$ is challenging and requires expert judgment. While our estimator does enjoy some robustness properties, it does requite correct specification of a model for $F_t(\cdot,X)$.

Sensitivity analyses like ours are rarely reported in the scientific literature. Rather, the discussion sections of scientific articles typically mention unmeasured confounding as a potential limitation. This is not surprising as guidelines that provide recommendations on the reporting of results of observational studies do not specifically call for sensitivity analyses.  STROBE~\citep{von2007strengthening} recommends discussing ``limitations of the study, taking into account sources of potential bias" and ``direction and magnitude of any potential bias". ROBINS-I \citep{sterne2016robins,sterne2016guidance}  recommends that subject matter experts categorize risk of ``bias due to confounding" into one of four categories: ``Low risk", ``Moderate risk", ``Serious risk" and ``Critical risk".  In making the categorization, the tool asks experts to assess whether ``the true effect estimate [can] be predicted to be greater or less than the estimated effect in the study because one or more of the important confounding domains was not controlled for".  

The reporting of such qualitative assessments of bias in the limitations sections of scientific papers would be greatly strengthened by quantitative sensitivity analyses. Subject matter experts could then review the results of the sensitivity analyses to judge the robustness of the study findings to various degrees of unmeasured confounding. This will allow a more reliable and nuanced interpretation of study findings. 

R code for implementing the approach considered in this manuscript can be found at\\https://github.com/LindaAmoafo/SemiparSens.

\section*{Supplementary Material}
Appendix includes all of the technical details.

\section*{Competing interests}
No competing interest is declared.

\section*{Author contributions statement}

L.A., S.X. and D.S. developed and implemented the methodology.  L.A. and S.X. contributed equally.  Using her expertise in cancer epidemiology, E.P. provided critical insights into the data analysis. All authors contributed to writing the manuscript.

\section*{Acknowledgments}

The authors would like to thank Lingyuan Hu who provided guidance on access to the National Cancer Database.
%{\color{red} The authors thank the anonymous reviewers for their valuable suggestions. This work is supported in part by funds from the National Science Foundation (NSF: \# 1636933 and \# 1920920).}

%%%%%%%%%%%%%%
\newpage
\bibliographystyle{plainnat}
\bibliography{reference}

%%%%%%%%%%%%%%

\newpage

\begin{landscape}
\begin{table}[hbt!]
%\centering
%\small
\setlength\tabcolsep{2.5pt}
\caption{Descriptive Statistics for covariates by treatment}
%\begin{adjustwidth}{-10cm}{}
%\resizebox{\textwidth}{!}{
\begin{tabular}{|cccccc|}
\toprule
& \multicolumn{3}{c}{\makecell{\large Our Cohort}} & \multicolumn{2}{c|}{\makecell{\large Comparable Cohort}}\\
\midrule
\textbf{Covariates} & \textbf{\makecell{RP\footnotemark[1] \\ (N=5005)}} & \textbf{ \makecell{EBRT Plus AD\footnotemark[1] \\(N=3888)}} & \textbf{\makecell{Overall \\ (N=8893)}} & \textbf{\makecell{Higher Risk\footnotemark[2] \\ (N=3433)}} & \textbf{ \makecell{Lower Risk\footnotemark[3] \\(N=41476)}}\\
\midrule
\multicolumn{1}{|l}{\textbf{Age (in years):} Mean (SD)} & 62.5 (7.1) & 69.4 (8.5) & 65.5 (8.4) & 66.9 (9.7) & 65.1 (8.5) \\
\textbf{Prostate-Specific Antigen:} Mean (SD) & 19.6 (21.8) & 17.5 (20.0) & 22.5 (23.6) & 31.6 (27.6) & 12.5 (16.2)\\

\multicolumn{6}{|l|}{\textbf{Race}} \\
White  & 4117 (82.3\%) & 3040 (78.2\%) & 7157 (80.5\%) & 2634 (76.7\%) & 33346 (80.4\%) \\
Non-White  &	888 (17.7\%) & 848 (21.8\%) &	1736 (19.5\%) & 799 (23.3\%) & 8130 (19.6\%)\\
%Black & 714 (14.3\%) &	732 (18.8\%) &	1446 (16.3\%)  & 2623 (18.2\%) & 6851 (16.5\%)\\
%Other  &	174 (3.5\%) & 116 (3\%)&	290 (3.3\%) & 478 (3.3\%) & 1279 (3.1\%)\\

\multicolumn{6}{|l|}{\textbf{Insurance Status}} \\
Private Insurance/Managed Care  & 2811 (56.2\%) & 1110 (28.5\%)& 3921 (44.1\%) & 1332 (38.8\%) & 19211 (46.3\%) \\
Medicare  & 1814 (36.2\%) & 2357 (60.6\%)& 4171 (46.9\%) & 1693 (49.3\%) & 18716 (45.1\%)\\
Other  & 380 (7.6\%) & 421 (10.8\%)& 801 (9\%) & 408 (11.9\%) & 3549 (8.6\%)\\

\multicolumn{6}{|l|}{\textbf{Income level(\$)}} \\
$<30,000$  & 598 (11.9\%) & 601 (15.5\%)& 1199 (13.5\%) & 529 (15.4\%) & 5433 (13.1\%)\\
$30,000-34,999$  & 808 (16.1\%) & 689 (17.7\%)& 1497 (16.8\%)& 617 (18\%) & 7011 (16.9\%) \\
$35,000-49,999$ & 1326 (26.5\%) & 1097 (28.2\%)& 2423 (27.2\%)  & 928 (27\%) & 11177 (26.9\%)\\
$>45,000$ & 2273 (45.4\%) & 1501 (38.6\%)& 3774 (42.4\%)  & 1359 (39.6\%) & 17855 (43\%)\\

\multicolumn{6}{|l|}{\textbf{Education level\footnotemark[4]}} \\
$<14$  & 742 (14.8\%) & 705 (18.1\%)& 1447 (16.3\%) & 658 (19.2\%) & 6759 (16.3\%)\\
$14-19.99$  & 983 (19.6\%) & 912 (23.5\%)& 1895 (21.3\%) & 769 (22.4\%) & 9032 (21.8\%)\\
$20-28.99$  & 1202 (24\%) & 948 (24.4\%)& 2150 (24.2\%) & 832 (24.2\%) & 9595 (23.1\%)\\
$>29$  & 2078 (41.5\%) & 1323 (34\%)& 3401 (38.2\%) & 1174 (34.2\%) & 16090 (38.8\%)\\

\multicolumn{6}{|l|}{\textbf{Charlson Comorbidity Index}} \\
0  & 4064 (81.2\%) & 3292 (84.7\%)& 7356 (82.7\%) & 2904 (84.6\%) & 34404 (82.9\%)\\
$\geq1$  & 941 (18.8\%) & 596 (15.3\%)& 1537 (17.3\%) & 529 (15.4\%) & 7072 (17.1\%)\\

\multicolumn{6}{|l|}{\textbf{Gleason score}} \\
$\leq7$  & 1488 (29.7\%) & 905 (23.3\%)& 2393 (26.9\%) & 1013 (29.5\%) & 33079 (79.8\%)\\
$8$  & 2273 (45.4\%) & 1647 (42.4\%)& 3920 (44.1\%) & 1235 (36\%) & 5247 (12.7\%)\\
$\geq9$  & 1244 (24.9\%) & 1336 (34.4\%) & 2580 (29\%) & 1185 (34.5\%) & 3150 (7.6\%)\\

\multicolumn{6}{|l|}{\textbf{Clinical T stage}} \\
$\leq$ cT2  & 4238 (84.7\%) & 3227 (83\%)& 7465 (83.9\%) & 1517 (44.2\%) & 41476 (100\%)\\
$\geq$ cT3  & 767 (15.3\%) & 661 (17\%)& 1428 (16.1\%) & 1916 (55.8\%) & 0 (0\%)\\
%\midrule
%\multicolumn{4}{|l|}{\thead[l]{$^a$ }}\\
%\multicolumn{4}{|l|}{\thead[l]{$^b$ }}\\
%\multicolumn{4}{|l|}{\thead[l]{$^c$ }}\\
\bottomrule
\end{tabular}
%\footnotetext[1]{\tiny Abbreviations: AD, androgen deprivation; EBRT, external beam radiation; RP, radical prostatectomy.}
%\footnotetext[2]{\tiny Patients classified as higher-risk have biopsy Gleason scores ranging from 8 to 10, and PSA$\geq$20 ng/dL or T stage cT3 or higher.}
%\footnotetext[3]{\tiny Patients in the lower-risk category are defined by T stage cT2 or lower, excluding those with T stage cT2a or lower, biopsy Gleason scores $\leq$ 6, and PSA levels $<$ 10 ng/dL.}
%\footnotetext[4]{\tiny Percentiles of adults in the patient’s zip code who did not graduate from high school.}
%\footnotetext[3]{Myocardial infarction, congestive heart failure, peripheral vascular disease, cerebrovascular disease, dementia, chronic pulmonary disease, rheumatology disease, peptic ulcer disease, mild liver disease, diabetes, diabetes with chronic complications, hemiplegia or paraplegia, renal disease.}
\begin{tablenotes}%
\item[$^{1}$] Abbreviations: AD, androgen deprivation; EBRT, external beam radiation; RP, radical prostatectomy.
\item[$^{2}$] Patients classified as higher-risk have biopsy Gleason scores ranging from 8 to 10, and PSA$\geq$20 ng/dL or T stage cT3 or higher.
\item[$^{3}$] Patients in the lower-risk category are defined by T stage cT2 or lower, excluding those with T stage cT2a or lower, biopsy Gleason \\scores $\leq$ 6, and PSA levels $<$ 10 ng/dL.
\item[$^{4}$] Percentiles of adults in the patient’s zip code who did not graduate from high school.
\end{tablenotes}
\label{Table1}
%\end{adjustwidth}
\end{table}
\end{landscape}

\begin{landscape}
\begin{table}[ht!]
%\renewcommand{\theadfont}{\normalsize}
%\centering
\caption{Simulation results averaged over simulation replicates. Bias and $95\%$ confidence interval coverage for different $\gamma_t$ values at $2, 5, \text{and } 10$ years. For RP, $\gamma_t$'s are positive. For EBRT+AD, $\gamma_t$'s are negative.}
\setlength{\tabcolsep}{6pt}
%\begin{tabular}{lllccccclccccc}
%\begin{tabular}{|ll|cccccc|cccccc|}
\begin{adjustbox}{max width=\linewidth}
%\fontsize{100}{50}\selectfont
%\resizebox{\textwidth}{!}{
\begin{tabular}{|ll|cccccc|cccccc|cccccc|}

\toprule

 & &  \multicolumn{6}{c|}{Sample Size = 1000} & \multicolumn{6}{c|}{Sample Size = 3000} & \multicolumn{6}{c|}{Sample Size = 5000} \\
 \midrule
 & & \multicolumn{2}{c}{Bias} & \multicolumn{2}{c}{\makecell{$95\%$ Wald\\ Confidence Interval \\Coverage Rate}} & \multicolumn{2}{c|}{\makecell{$95\%$ Bootstrap \\Confidence Interval \\Coverage Rate}} & \multicolumn{2}{c}{Bias} & \multicolumn{2}{c}{\makecell{$95\%$ Wald\\ Confidence Interval \\Coverage Rate}} & \multicolumn{2}{c|}{\makecell{$95\%$ Bootstrap\\Confidence Interval \\Coverage Rate}} & \multicolumn{2}{c}{Bias} & \multicolumn{2}{c}{\makecell{$95\%$ WALD\\ Confidence Interval \\Coverage Rate}} & \multicolumn{2}{c|}{\makecell{$95\%$ Bootstrap \\Confidence Interval \\Coverage Rate}}\\
\midrule
\textbar $\gamma_t$\textbar & Years & RP & EBRT $+$ AD & RP & EBRT $+$ AD & RP & EBRT $+$ AD & RP & EBRT $+$ AD & RP & EBRT $+$ AD & RP & EBRT $+$ AD & RP & EBRT $+$ AD & RP & EBRT $+$ AD & RP & EBRT $+$ AD\\
\midrule
$0$ & 2 & 0.00 & 0.00 & 0.92 & 0.93 & 0.94 & 0.94 & 0.00 & 0.00& 0.92 & 0.94& 0.96 & 0.95& 0.00 & 0.00 & 0.93 & 0.95 & 0.95 & 0.96\\
& 5 & 0.00 & 0.00 & 0.94 & 0.95 & 0.96 & 0.96 & 0.00 & 0.00& 0.94 & 0.95& 0.96 & 0.97& 0.00 & 0.00 & 0.94 & 0.95 & 0.97 & 0.97 \\
& 10 & 0.01 & 0.02 & 0.93 & 0.93 & 0.97 & 0.97 & 0.00 & 0.00& 0.93 & 0.94& 0.96 & 0.97& 0.00 & 0.00 & 0.93 & 0.95 & 0.97& 0.98\\
\midrule
$1$ & 2 & 0.00 & 0.00 & 0.89 & 0.94 & 0.93 & 0.94 & 0.00& 0.00& 0.90& 0.95& 0.95 &0.96& 0.00 & 0.00 & 0.92 & 0.95 & 0.95 & 0.95\\
 & 5 & 0.00  & 0.00 & 0.93 & 0.96 & 0.96 & 0.96 & 0.00& 0.00& 0.94 & 0.95&0.96 &0.97 & 0.00 & 0.00 & 0.94 & 0.94 & 0.96 & 0.97\\
 & 10 & 0.00 & 0.02 &  0.95 & 0.92 & 0.97& 0.96 & 0.00 & 0.00& 0.94 & 0.95& 0.96&0.97& 0.00 & 0.00 & 0.94 & 0.94 & 0.97 & 0.97\\
 \midrule
 $2$ & 2 & 0.00 & 0.00 & 0.87 & 0.94 & 0.92 & 0.94 & 0.00& 0.00& 0.90& 0.96& 0.95 & 0.96& 0.00 & 0.00 & 0.91 & 0.95 & 0.95 & 0.96\\
 & 5 & -0.01 & 0.00 & 0.94 & 0.96 & 0.94 & 0.96 & 0.00&0.00 &  0.94 & 0.96& 0.95& 0.97& 0.00 & 0.00 & 0.95 & 0.94 & 0.96 & 0.96\\
 & 10 & -0.01 & 0.01 & 0.95 & 0.92 & 0.96 & 0.97 &0.00 & 0.00&0.94 &0.95 &0.97 &0.98& 0.00 & 0.00 & 0.94 & 0.94 & 0.97 &0.98 \\
 \midrule
 $3$ & 2 & -0.01 & 0.00 & 0.88 & 0.94 & 0.91 & 0.94 & 0.00& 0.00&0.91 & 0.95&0.95 &0.96& 0.00 & 0.00 & 0.92 & 0.95 & 0.95 & 0.96\\
 & 5 & -0.03 & 0.00 & 0.95 & 0.95 & 0.94 & 0.96 & -0.01& 0.00&0.95 &0.96 & 0.95&0.97& 0.00 & 0.00 & 0.95 & 0.94 & 0.95 & 0.96\\
 & 10 & -0.02 & 0.01 & 0.94 & 0.92 & 0.96 & 0.98 & 0.00& 0.00&0.94 & 0.94& 0.97&0.98& 0.00 & 0.00 & 0.94 & 0.95 & 0.96 & 0.97\\
\bottomrule
\end{tabular}
%}
\end{adjustbox}
%\end{tabular}

\label{coverage}
\end{table}
\end{landscape}

\newpage

\newpage

\newpage

%%%%%%%%%%%%%%
\newpage
\appendix
\noindent {\bf \Large Appendix} \vspace{0.25cm}

\section{Identification Mappings}

 Let ${\mathcal{P}}_{T,X}$,${\mathcal{P}}_{Y|T,X}$ and  ${\mathcal{P}}_{C|T,X}$ be the space of distributions of $T$ and $X$, $Y$ given $T$ and $X$, and $C$ given $T$ and $X$, respectively. 
 Let 
  ${\mathcal{P}}_{Y,C,T,X}$ be the space of distributions for $(Y,C,T,X)$ that satisfies (3) in the main manuscript, i.e., ${\mathcal{P}}_{Y,C,T,X}= \{ p_{Y,C,T,X}(\cdot,\cdot,\cdot,\cdot) = p_Y(\cdot|T,X) p_C(\cdot|T,X) p_{T,X}(\cdot,\cdot): p_Y(\cdot|T,X) \in {\mathcal{P}}_{Y|T,X}, p_C(\cdot|T,X) \in {\mathcal{P}}_{C|T,X}, p_{T,X}(\cdot,\cdot) \in {\mathcal{P}}_{T,X} \}$.
 %Let ${\mathcal P}^*$ be the space of distributions of $(Y,C,T,X)$ and $\widetilde{O}$, respectively.
 Let $\widetilde{{\mathcal P}}$ be the space of distributions of $\widetilde{O}$.%that satisfy the additional modeling assumptions described in Section \ref{section3.6}. 

\cite{tsiatis1975nonidentifiability} showed that under (3) in the main manuscript, there exists a unique function $\widetilde{g}$ that maps from  $\widetilde{{\mathcal P}}$ to ${\mathcal P}_{Y,C,T,X}$. Letting $\widetilde{h}$ be the function that marginalizes a distribution for $(Y,T,C,X)$ to a distribution for $\widetilde{O}$, \cite{tsiatis1975nonidentifiability} also showed that $\widetilde{g}(\widetilde{h}(P^\star))=P^\star$ for all $P^\star \in {\mathcal P}_{Y,C,T,X}$.  and $\widetilde{h}(\widetilde{g}(\widetilde{P}))=\widetilde{P}$ for all $\widetilde{P} \in \widetilde{{\mathcal P}}$.  %This means there is a 1-1 mapping between $\widetilde{{\mathcal P}}_{Y,C,T,X}$ and ${\mathcal P}^*_{Y,C,T,X}$.

%Note that the range of $\widetilde{g}$, $\mathcal{R}(\widetilde{g})$, is a proper subset of ${\mathcal{P}}_{Y,C,T,X}$. 
%for each   $\widetilde{P} \in \widetilde{{\mathcal P}}$ there exists a mapping to  $P_{Y,C,T,X} \in {\mathcal P}_{Y,C,T,X}$ which marginalizes back to $\widetilde{P}$.  That is, there is a 1-1 mapping, $g$, between a $\widetilde{{\mathcal P}}$ to a ${\mathcal P}_{Y,C,T,X}$. 

 In what follows, consider a fixed $\boldsymbol{\gamma} \in \boldsymbol{\Gamma}$. Let $\mathcal{P}^{\boldsymbol{\gamma}}_{F,C,T,X}$ be the set of distributions for $F,C,T,X$ satisfying (1) for $t=0,1$ in the main manuscript and (3) in the main manuscript. That is,
\begin{align*}
& \mathcal{P}^{\boldsymbol{\gamma}}_{F,C,T,X} = \bigg\{ p_{F,C,T,X}(\cdot,\cdot,\cdot,\cdot,\cdot)  = p_{F,C}(\cdot,\cdot,\cdot|T,X) p_{T,X}(\cdot,\cdot): p_{T,X} \in \mathcal{P}_{T,X}, \\
& \hspace*{0.5in}  p_{F,C}(\cdot,\cdot,\cdot|T,X) \mbox{ is restricted so that }\\
&  \hspace*{1in}  p_{Y(0),C}(\cdot,\cdot|T=0,X) = p_{Y}(\cdot|T=0,X)  p_{C}(\cdot|T=0,X), \\
&   \hspace*{1in} p_{Y(1),C}(\cdot,\cdot|T=1,X) = p_{Y}(\cdot|T=1,X)  p_{C}(\cdot|T=1,X), \\
&  \hspace*{1in}  p_{Y(0)}(\cdot|T=1,X) \mbox{ is linked to } p_{Y}(\cdot|T=0,X) \mbox{ via } (1) \mbox{ in main} , \\
& \hspace*{1in}   p_{Y(1)}(\cdot|T=0,X) \mbox{ is linked to } p_{Y}(\cdot|T=1,X) \mbox{ via } (1) \mbox{ in main}, \mbox{ where} \\
&\hspace*{1in}  p_Y(\cdot|T,X)\in \mathcal{P}_{Y|T,X},  p_C(\cdot|T,X)\in \mathcal{P}_{C|T,X}\bigg\}
\end{align*}
Let $\mathcal{Q}^{\boldsymbol{\gamma}}_{F,C,T,X}$ be be a subset of $\mathcal{P}^{\boldsymbol{\gamma}}_{F,C,T,X}$ that satisfy two additional assumptions: $Y(0),C$ independent of $Y(1)$ given $T=0$ and $X$, and  $Y(1),C$ independent of $Y(0)$ given $T=1$ and $X$. 

There exists a unique function $g_{\boldsymbol{\gamma}}$ that maps from  $\mathcal {P}_{Y,C,T,X}$ to $\mathcal{Q}^{\boldsymbol{\gamma}}_{F,C,T,X}$.
Letting $h$ be the function that marginalizes a distribution for $F,C,T,X$ to a distribution for $Y,C,T,X$, it can be shown that $g_{\boldsymbol{\gamma}}(h(Q^{\boldsymbol{\gamma}}))=Q^{\boldsymbol{\gamma}}$ for all $Q^{\boldsymbol{\gamma}}\in \mathcal{Q}^{\boldsymbol{\gamma}}_{F,C,T,X}$ and  $h(g_{\boldsymbol{\gamma}}(P^\star))=P^\star$ for all $P^\star\in \mathcal {P}_{Y,C,T,X}$. 
%There are many distributions in $\mathcal{P}^{\boldsymbol{\gamma}}_{F,C,T,X}$ that map to a given $P^*$ under $h(\cdot)$. The set of such distributions form equivalence classes around each $Q^{\boldsymbol{\gamma}}\in \mathcal{Q}^{\boldsymbol{\gamma}}_{F,C,T,X}$.  
For $Q^{\boldsymbol{\gamma}} \in \mathcal{Q}^{\boldsymbol{\gamma}}_{F,C,T,X}$, let $\mathcal{P}^{\boldsymbol{\gamma}}_{F,C,T,X}(Q^{\boldsymbol{\gamma}})$ be the set of distributions in $\mathcal{P}^{\boldsymbol{\gamma}}_{F,C,T,X}$ that agree with $Q^{\boldsymbol{\gamma}}$ with respect to (a) distribution of $T,X$, (b) distribution of $Y(0),C$ given $T=0$ and $X$, (c) distribution of $Y(1),C$ given $T=1$ and $X$,
(d) distribution of $Y(0)$ given $T=1$ and $X$, and (e) distribution of $Y(1)$ given $T=0$ and $X$.  The distributions in $\mathcal{P}^{\boldsymbol{\gamma}}_{F,C,T,X}(Q^{\boldsymbol{\gamma}})$ form an equivalence class. {\em Importantly, each distribution in the equivalence class has the same 
conditional distributions of $Y(0)$ given $X$, $Y(1)$ given $X$, and the same marginal distributions of $Y(0)$ and  $Y(1)$; they differ with respect to the conditional distributions of $Y(0),Y(1),C$ given $T=0$ and $X$ and $Y(0),Y(1),C$ given $T=1$ and $X$.} If $P^{\boldsymbol{\gamma}} \in \mathcal{P}^{\boldsymbol{\gamma}}_{F,C,T,X}(Q^{\boldsymbol{\gamma}})$, we write $P^{\boldsymbol{\gamma}} \sim Q^{\boldsymbol{\gamma}}$. We use the  notation $\mathcal{P}^{\boldsymbol{\gamma}}_{F,C,T,X} / \sim$ to indicate the set of all equivalence classes formed by ranging over all $Q^{\boldsymbol{\gamma}} \in \mathcal{Q}^{\boldsymbol{\gamma}}_{F,C,T,X}$.  The equivalence classes in $\mathcal{P}^{\boldsymbol{\gamma}}_{F,C,T,X} / \sim$ are disjoint and their union is equal to  $\mathcal{P}^{\boldsymbol{\gamma}}_{F,C,T,X}$. For each $P^{\boldsymbol{\gamma}} \in \mathcal{P}^{\boldsymbol{\gamma}}_{F,C,T,X}$, there exists  a unique $Q^{\boldsymbol{\gamma}} \in \mathcal{Q}^{\boldsymbol{\gamma}}_{F,C,T,X}$ such that $P^{\boldsymbol{\gamma}} \sim Q^{\boldsymbol{\gamma}}$, $h(P^{\boldsymbol{\gamma}})=h(Q^{\boldsymbol{\gamma}})$ and $g_{\boldsymbol{\gamma}}(h(P^{\boldsymbol{\gamma}}))=Q^{\boldsymbol{\gamma}}$.  For each $P^\star\in \mathcal {P}_{Y,C,T,X}$, we define $g_{\boldsymbol{\gamma}}^*(P^\star) = [g_{\boldsymbol{\gamma}}(P^\star)]\sim$ to be a function that maps $P^\star$ to the equivalence class associated with $g_{\boldsymbol{\gamma}}(P^\star)$.  For each $P^\star\in \mathcal {P}_{Y,C,T,X}$, we know that $h([g_{\boldsymbol{\gamma}}(P^\star)]\sim) = P^\star$.

\begin{figure}[hbt!]  
\includegraphics[width=\textwidth,height=\textheight,keepaspectratio]{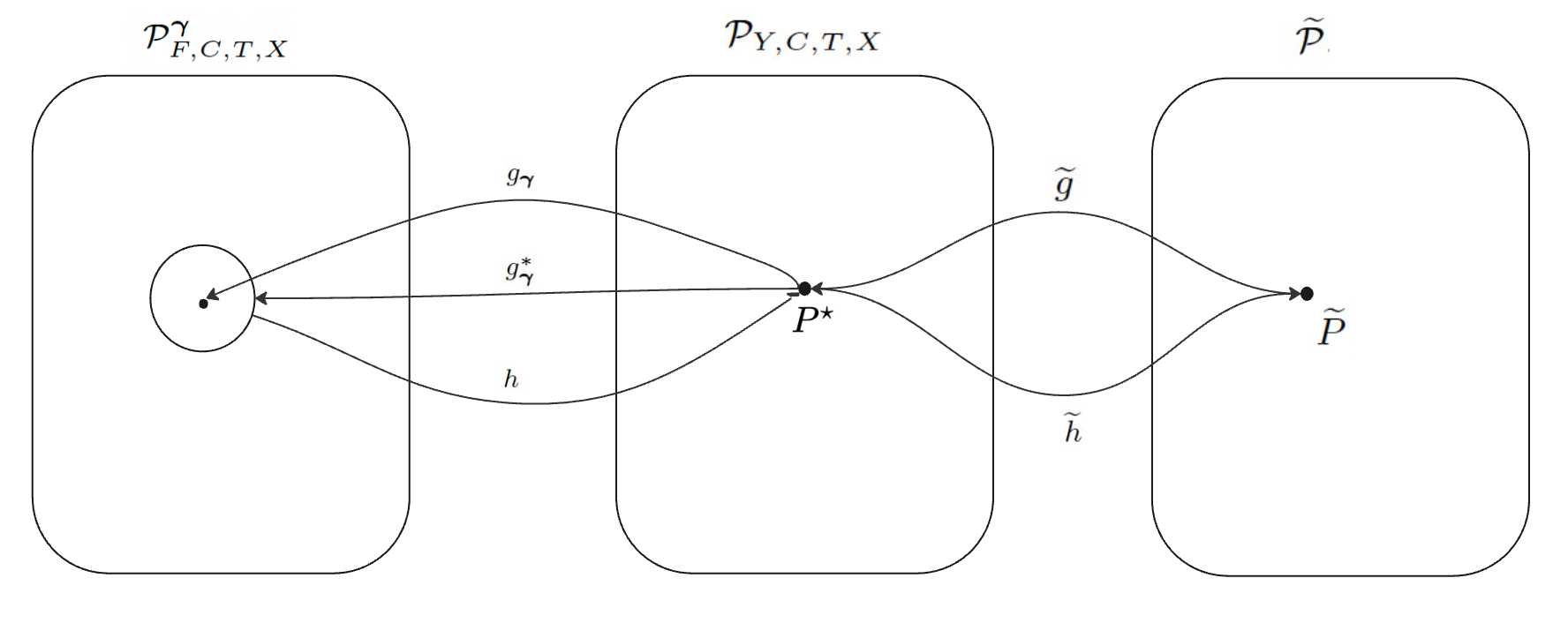}
\caption{Identification Mappings}
\label{fig1}
\vspace{-15mm}
  {\footnotesize \textit{Note.} In the left box, the point represents an element in $\mathcal{Q}^{\boldsymbol{\gamma}}_{F,C,T,X}$ and the circle around the point denotes the equivalence class associated with it.}
\end{figure}

Figure \ref{fig1} presents a graphical depiction of these mappings.  Putting these results together, we now consider identification of the marginal distribution of $Y(t)$ from $\widetilde{P} \in \widetilde{\mathcal{P}}$. We need to show that if $P^{\boldsymbol{\gamma}}_1$ and $P^{\boldsymbol{\gamma}}_2$ are two distinct distributions in $\mathcal{P}^{\boldsymbol{\gamma}}_{F,C,T,X}$ that marginalize to the same $\widetilde{P} \in \widetilde{\mathcal{P}}$, then the marginal distributions of $Y(t)$ derived from $P^{\boldsymbol{\gamma}}_1$ and $P^{\boldsymbol{\gamma}}_2$ must be the same. Associated with $P^{\boldsymbol{\gamma}}_j$ we know that there exists $Q^{\boldsymbol{\gamma}}_j \in \mathcal{Q}^{\boldsymbol{\gamma}}_{F,C,T,X}$ such that $P^{\boldsymbol{\gamma}}_j \sim Q^{\boldsymbol{\gamma}}_j$, $h(P^{\boldsymbol{\gamma}}_j)=h(Q^{\boldsymbol{\gamma}}_j)$ and $g_{\boldsymbol{\gamma}}(h(P^{\boldsymbol{\gamma}}_j))=Q^{\boldsymbol{\gamma}}_j$ ($j=1,2$).   %Associated with $P^{\boldsymbol{\gamma}}_2$ we know that there exists $Q^{\boldsymbol{\gamma}}_2 \in \mathcal{Q}^{\boldsymbol{\gamma}}_{F,C,T,X}$ such that $P^{\boldsymbol{\gamma}}_2 \sim Q^{\boldsymbol{\gamma}}_2$, $h(P^{\boldsymbol{\gamma}}_2)=h(Q^{\boldsymbol{\gamma}}_2)$ and $g_{\boldsymbol{\gamma}}(h(P^{\boldsymbol{\gamma}}_2))=Q^{\boldsymbol{\gamma}}_2$. 
We also know that $\widetilde{h}(h(P^{\boldsymbol{\gamma}}_1))= \widetilde{h}(h(P^{\boldsymbol{\gamma}}_2))$.  
This implies that $\widetilde{g}(\widetilde{h}(h(P^{\boldsymbol{\gamma}}_1)))= \widetilde{g}(\widetilde{h}(h(P^{\boldsymbol{\gamma}}_2)))$, which implies that $h(P^{\boldsymbol{\gamma}}_1) = h(P^{\boldsymbol{\gamma}}_2)$.  So, $g_{\boldsymbol{\gamma}}(h(P^{\boldsymbol{\gamma}}_1)) = g_{\boldsymbol{\gamma}}(h(P^{\boldsymbol{\gamma}}_2))$, which implies that $Q_1^{\boldsymbol{\gamma}} = Q_2^{\boldsymbol{\gamma}}$.  Thus, $P_1^{\boldsymbol{\gamma}} \sim P_2^{\boldsymbol{\gamma}}$.  That is, they share the same marginal distribution of $Y(t)$.

These results hold for all $\boldsymbol{\gamma} \in \boldsymbol{\Gamma}$ and whatever be
$\widetilde{P} \in \widetilde{\mathcal{P}}$.  In particular, it holds for any subset of  $\widetilde{\mathcal{P}}$, i.e., the one induced by the additional modeling restrictions introduced in Section 2.6.   Therefore, there is no information in the observed data about $\boldsymbol{\gamma}$ and any restrictions of $\widetilde{\mathcal{P}}$ do not restrict the range of $\boldsymbol{\gamma}$.

\section{Uncensored data non-parametric influence function for \texorpdfstring{$\psi_t$}{TEXT} %
}

This appendix contains details of the derivation of the uncensored data non-parametric influence function for $\psi_t=P[Y(t) \leq  s] $ under Assumption (1) of the main manuscript.\\

Under Assumption (1) in the main manuscript, 
\[
\mbox{logit} \{ P[Y(t) \leq s | T = 1-t, X=x] \} = \mbox{logit} \{ P[Y(t) \leq s | T = t, X=x\} + \gamma_t,
\]
and 
\begin{eqnarray*}
 \psi_t & = & \int_x \left\{ P[Y \leq s | T=t, X=x] P[T=t|X=x] + \right. \\
&& \left. \frac{ P[Y \leq s | T=t,X=x] \exp \{ \gamma_t \}  }{ P[Y >  s | T=t,X=x]  + P[Y \leq s | T=t,X=x] \exp \{ \gamma_t \} }  P[T=1-t|X=x]  \right\} dF(x), \\
& = & \int_x \left\{ F_t(s|x) \pi_t(x) + \frac{ F_t(s|x) \exp \{ \gamma_t \}  }{ S_t(s|x)   + F_t(s|x)  \exp \{ \gamma_t \} }  \pi_{1-t}(x)   \right\} dF(x) \equiv \psi_t(P).
\end{eqnarray*}

Consider a statistical model $\mathcal{M}$ composed of distributions $P^\ast$, with $P$ denoting the true distribution. A distribution $P^\ast \in \mathcal{M}$ is characterized by $F^\ast_t(y \mid x) = P^\ast(Y\leq y \mid T=t, X=x)$, $\pi^\ast_{t}(x) = P^\ast(T=t \mid X=x)$, and $F^\ast(x) = P^\ast(X \leq x).$ Let $\{ P^\ast_{\theta}: P^\ast_{\theta} \in \mathcal{M}\}$. %Let $s(O)$ be the score for $\theta$ evaluated at $\theta=0$. 
We consider parametric submodels of the following form:
\begin{align*}
dF^\ast_{\theta}(x) & =  dF(x) \{ 1 + \epsilon h(x) \} \\
dF^\ast_{t,\theta}(y|x) & = dF_{t}(y\mid x) \{ 1 + \eta_t k_t(y,x) \} \\
\pi^\ast_{t,\theta}(x) & = \frac{\{ \pi_1(x) \exp \{ \delta l(x) \} \}^t \pi_0(x)^{1-t}  }{ \pi_1(x) \exp \{ \delta l(x) \} + \pi_0(x)}
\end{align*}
where $\theta = (\epsilon,\eta_0,\eta_1,\delta)$, $\mathbb{E}[h(X)]=0$, $\mathbb{E}[k_t(Y,X) \mid T=t,X]=0$ and $l(X)$ is any function of $X$. The associated score functions are $h(X)$, $T k_1(Y,X) + (1-T) k_0(Y,X)$, and $\{ T - \pi_1(X) \} l(X).$ 

\noindent 
The target parameter as a function of $P^\ast_{\theta}$, $\psi_t(P^\ast_{\theta})$, is
\begin{eqnarray*}
 \psi_t(P^\ast_{\theta}) & = & \int_x \int_y I(y \leq s) dF^\ast_{t,\theta}(y \mid x) \pi^\ast_{t,\theta}(x)  dF^\ast_{\theta}(x) + . \\
 && \; \;  \; \; \;. \int_x  \frac{ \int_y I(y \leq s) \exp(\gamma_t) dF^\ast_{t,\theta}(y \mid x) }{ \int_y I(y > s) dF^\ast_{t,\theta}(y \mid x) +  \int_y I(y \leq  s) dF^\ast_{t,\theta}(y \mid x) \exp(\gamma_t)  } \pi^\ast_{1-t,\theta}(x) dF^\ast_{\theta}(x).
\end{eqnarray*}
The derivative of $\psi_t(P^\ast_{\theta})$ with respect to $\epsilon$ evaluated at $\theta=0$ is
\[
    \int_x \left\{ F_t(s|x) \pi_t(x) + \frac{F_t(s|x) \exp(\gamma_t)}{S_t(s|x) + F_t(s|x) \exp(\gamma_t)} \pi_{1-t}(x) \right\} h(x) dF(x) . 
\]
The derivative of $\psi_t(P^\ast_{\theta})$ with respect to $\eta_t$ evaluated at $\theta=0$ is
\begin{align*}
& \int_x  \int_y I(y \leq s) k_t(y,x)dF_t(y|x) \pi_t(x) dF(x) + \\
& \int_x \int_y \ \frac{ \left\{ I(y \leq s) - F_t(s|x)  \right\} \exp(\gamma_t)  }{\{ S_t(s|x) + F_t(s|x) \exp(\gamma_t)\}^2} \frac{\pi_{1-t}(x)}{\pi_t(x)}  k_t(y,x)  dF_t(y|x)  \pi_t(x) dF(x)  
\end{align*}
The derivative of $\psi_t(P^\ast_{\theta})$ with respect to $\eta_{1-t}$ evaluated at $\theta=0$ is 0. \\
The derivative of $\psi_t(P^\ast_{\theta})$ with respect to $\delta$ evaluated at $\theta=0$ is
\[
    \int_x   (-1)^{t+1}  \left\{F_t(s|x) - \frac{ F_t(s|x) \exp(\gamma_t)}{S_t(s|x) + F_t(s|x) \exp(\gamma_t)} \right\} \pi_1(x) \ \pi_0(x) \ l(x) \ dF(x).
\]
Any mean zero observed data random variable can be expressed as
\[
d(O)  = a(X) + I(T=1) \{ b_1(Y,X)+ \pi_0(X) c(X) \} + I(T=0) \{ b_0(Y,X) - \pi_1(X) c(X) \} 
\]
\[
d(O)  = a(X) + \sum_{t'=0}^1 I(T=t') \{ b_{t'}(Y,X)+ (-1)^{t'+1} \pi_{1-t'}(X) c(X) \} 
\]
where $\mathbb{E}[a(X)]=0$, $\mathbb{E}[b_t(Y,X)|T=t,X]=0$ and $c(X)$ is an unspecified function of $X$.  The set of all $d(O)$ is the non-parametric tangent space.  To find the non-parametric efficient influence function, we need to find
choices of $a(X)$, $b_t(Y,X)$ and $c(X)$ such that 
$\mathbb{E}[a(X)h(X)] = \partial \psi_t(P^\ast_{\theta})/ \partial \epsilon \; \vline_{\; \theta=0}$, $\mathbb{E}[I(T=t) b_t(Y,X)k_t(Y,X)] = \partial \psi_t(P^\ast_{\theta})/ \partial \eta_t \; \vline_{\; \theta=0}$ and $\mathbb{E}[(T-\pi_1(X))^2 c(X) l(X)]= {\color{blue}\partial} \psi_t(P^\ast_{\theta})/ \partial \delta \; \vline_{\; \theta=0}$. It can be shown that
\begin{align*}
&a(X) = F_t(s|X) \pi_t(X) + \frac{F_t(s|X) \exp(\gamma_t)}{S_t(s|X) + F_t(s|X) \exp(\gamma_t)} \pi_{1-t}(X) - \psi_t(P) 
\\
&b_t(Y,X) = \left\{ I(Y \leq s) - F_t(s|X) \right\} \left\{ 1  + \frac{\pi_{1-t}(X)}{\pi_t(X)} \frac{ \exp(\gamma_t)}{ \{ S_t(s|X) + F_t(s|X) \exp(\gamma_t)\}^2} \right\}
\\
&b_{1-t}(Y,X) = 0 
\\
&c(X) =  (-1)^{t+1}  \left\{F_t(s|X) - \frac{ F_t(s|X) \exp(\gamma_t)}{S_t(s|X) + F_t(s|X) \exp(\gamma_t)} \right\}. 
\end{align*}
Hence, the non-parametric efficient influence function that corresponds to $\psi_t$ is as follows: 
\begin{align*}
\phi_t(P, \psi_t)(O) =& \  I(T=t) \left\{ I(Y\leq s) + \left\{ I(Y \leq s) - F_t(s|X) \right\} \left\{ \frac{\pi_{1-t}(X)}{\pi_t(X)} \frac{ \exp(\gamma_t)}{\{ S_t(s|X) + F_t(s|X) \exp(\gamma_t)\}^2} \right\} \right\}  \nonumber \\
& + I(T=1-t) \frac{F_t(s|X) \exp(\gamma_t)}{S_t(s|X) + F_t(s|X) \exp(\gamma_t)}  - \psi_t(P) 
\end{align*}
\[
\phi_t(P, \psi_t)(O)  = \phi_{t}(P)(O)  - \psi_t(P)
\]
%\end{appendix}

\section{Observed data non-parametric influence function for \texorpdfstring{$\psi_t$}{TEXT} %
}

This appendix shows details on deriving the observed data non-parametric influence function for \texorpdfstring{$\psi_t$}{TEXT}, denoted $\widetilde{\phi}_t(\widetilde{P}, \psi_t)(\widetilde{O})$. \citet{rotnitzky2005inverse} and \citet{rotnitzky1992inverse} show that 
\begin{eqnarray*}
\widetilde{\phi}_t(\widetilde{P}, \psi_t)(\widetilde{O})=\frac{\Delta + (1-\Delta) I( \widetilde{Y} \geq s) }{G_{T}(\min(\widetilde{Y},s)^-| X)}\phi_t(P, \psi_t)(O)-\int h^*(u, T, X)dM_c(u,T,X)
\end{eqnarray*}
where $\int h^*(u, T, X)dM_c(u,T,X)$ is the projection of the inverse weighted term $\frac{\Delta + (1-\Delta) I( \widetilde{Y} \geq s) }{G_{T}(\min(\widetilde{Y},s)^-| X)}\phi_t(P, \psi_t)(O)$ onto the space spanned by scores associated with the censoring mechanism, 
\[
\bigg\{\int h(u,T,X)dM_c(u,T,X): h(u,T,X)\bigg\},
\]
where $dM_c(u,T,X)=dN_c(u)-I(\widetilde{Y}\geq u)d\Lambda^{\dagger}_T(u|X)$,  $N_c(u)=I(\widetilde{Y}\leq u, \Delta=0)$ and $dN_c(u)=N_c(u)-N_c(u^-)$. We further adopt the notations for survival: $dM(u,T,X))=dN(u)-I(\widetilde{Y}\geq u)d\Upsilon^{\dagger}_t(u)$, $N(u)=I(\widetilde{Y}\leq u, \Delta=1)$ and $dN(u)=N(u)-N(u^-)$. 

Since the observed data influence function $\widetilde{\phi}_t(\widetilde{P}, \psi_t)(\widetilde{O})$ is orthogonal to the space, we know that for any $h(u, T, X)$, 
\begin{eqnarray*}
E\bigg[\bigg(\frac{\Delta + (1-\Delta) I( \widetilde{Y} \geq s) }{G_{T}(\min(\widetilde{Y},s)^-| X)}\phi_t(P, \psi_t)(O)-\int h^*(u, T, X)dM_c(u,T,X)\bigg)\int h(u, T, X)dM_c(u,T,X)\bigg]=0
\end{eqnarray*}

We can further organize the inverse weighted term as 
\begin{align*}
&\frac{\Delta+(1-\Delta)I(\widetilde{Y}\geq s)}{G_{T}(\min(\widetilde{Y}, s)^-| X)}\phi_t(P, \psi_t)(O)\\
&\hspace{1cm}=\bigg(\frac{\Delta}{G_{T}(\min(\widetilde{Y}, s)^-| X)}+\frac{(1-\Delta)I(\widetilde{Y}\geq s)}{G_{T}(s^-| X)}\bigg)\bigg(\phi_{t1}(P)(O)+\phi_{t2}(P)(O)+\phi_{t3}(P)(O)-\psi_t\bigg)\\
&\hspace{1cm}=\frac{\Delta}{G_{T}(\min(\widetilde{Y}, s)^-| X)}\phi_t(P, \psi_t)(O)+\frac{(1-\Delta)I(\widetilde{Y}\geq s)}{G_{T}(s^-| X)}\underbrace{\bigg(\phi_{t2}(P)(O)+\phi_{t3}(P)(O)-\psi_t\bigg)}_{\phi_{t_{-1}}(P)(T, X; \psi_t)}\\
&\hspace{1cm}=\frac{\Delta}{G_{T}(\widetilde{Y}^-| X)}\phi_{t1}(P)(\widetilde{Y}, T, X)+\frac{\Delta}{G_{T}(\min(\widetilde{Y}, s)^-| X)}\phi_{t_{-1}}(P, \psi_t)(T, X)+\frac{(1-\Delta)I(\widetilde{Y}\geq s)}{G_{T}(s^-| X)}\phi_{t_{-1}}(P, \psi_t)(T, X)\\
&\hspace{1cm}=\int\frac{\phi_{t1}(P)(u, T, X)}{G_{T}(u^-| X)}dN(u)+\int\frac{\phi_{t_{-1}}(P, \psi_t)(T, X)}{G_{T}(\min(u, s)^-| X)}dN(u)+\int\frac{I(u\geq s)\phi_{t_{-1}}(P, \psi_t)(T, X)}{G_{T}(s^-| X)}dN_c(u)
\end{align*}

So we need to derive $h^*(u, T, X)$ that can satisfy the below equation for any $h(u, T, X)$:
\begin{align*}
E\bigg[&\int\frac{\phi_{t1}(P)(u, T, X)}{G_{T}(u^-| X)}dN(u)\int h(u, T, X)dM_c(u,T,X)\\
&+\int\frac{\phi_{t_{-1}}(P, \psi_t)(T, X)}{G_{T}(\min(u, s)^-| X)}dN(u)\int h(u, T, X)dM_c(u,T,X)\\
&+\int\frac{I(u\geq s)\phi_{t_{-1}}(P, \psi_t)(T, X)}{G_{T}(s^-| X)}dN_c(u)\int h(u, T, X)dM_c(u,T,X)\\
&-\int h^*(u, T, X)dM_c(u,T,X)\int h(u, T, X)dM_c(u,T,X)\bigg]=0
\end{align*}
Let $N_1+N_2+N_3-N_4=0$, where
\begin{align*}
N_1&=E\bigg[\int\frac{\phi_{t1}(P)(u, T, X)}{G_{T}(u^-| X)}dN(u)\int h(u, T, X)dM_c(u,T,X)\bigg]\\
&=E\bigg[\int_{u'}\int_{u}\frac{\phi_{t1}(P)(u, T, X)}{G_{T}(u^-| X)}dN(u)h(u', T, X)dN_c(u')-\int_{u'}\int_{u}\frac{\phi_{t1}(P)(u, T, X)}{G_{T}(u^-| X)}dN(u)h(u', T, X)I(\widetilde{Y}\geq u)d\Lambda^{\dagger}_T(u'|X)\bigg]\\
&=-E\bigg[\int_{u'}\int_{u}\frac{\phi_{t1}(P)(u, T, X)}{G_{T}(u^-| X)}dN(u)h(u', T, X)I(\widetilde{Y}\geq u)d\Lambda^{\dagger}_T(u'|X)\bigg]\\
&=-E\bigg[\int_{u'}\frac{\Delta\phi_{t1}(P)(\widetilde{Y}, T, X)}{G_{T}(\widetilde{Y}^-| X)}h(u', T, X)I(\widetilde{Y}\geq u)d\Lambda^{\dagger}_T(u'|X)\bigg]\\
&=-E\bigg[\int_{u'}E\bigg(\frac{\Delta\phi_{t1}(P)(\widetilde{Y}, T, X)}{G_{T}(\widetilde{Y}^-| X)}\bigg|\widetilde{Y}\geq u', T, X\bigg)h(u', T, X)I(\widetilde{Y}\geq u)d\Lambda^{\dagger}_T(u'|X)\bigg]\\
&=-E\bigg[\int_{u'}E\bigg(\frac{I(C\geq Y)I(C\geq u')I(Y\geq u')\phi_{t1}(P)(\widetilde{Y}, T, X)}{G_{T}(\widetilde{Y}^-| X)G_{T}(u'^-| X)S_{T}(u'^-| X)}\bigg|T, X\bigg)h(u', T, X)I(\widetilde{Y}\geq u)d\Lambda^{\dagger}_T(u'|X)\bigg]\\
&=-E\bigg[\int_{u'}E\bigg(\frac{I(C\geq Y)I(Y\geq u')\phi_{t1}(P)(\widetilde{Y}, T, X)}{G_{T}(\widetilde{Y}^-| X)G_{T}(u'^-| X)S_{T}(u'^-| X)}\bigg|T, X\bigg)h(u', T, X)I(\widetilde{Y}\geq u)d\Lambda^{\dagger}_T(u'|X)\bigg]\\
&=-E\bigg[\int_{u'}E\bigg(\frac{I(Y\geq u')E(I(C\geq Y)|Y, T, X)\phi_{t1}(P)(\widetilde{Y}, T, X)}{G_{T}(\widetilde{Y}^-| X)G_{T}(u'^-| X)S_{T}(u'^-| X)}\bigg|T, X\bigg)h(u', T, X)I(\widetilde{Y}\geq u)d\Lambda^{\dagger}_T(u'|X)\bigg]\\
&=-E\bigg[\int_{u'}E\bigg(\frac{I(Y\geq u')\phi_{t1}(P)(\widetilde{Y}, T, X)}{G_{T}(u'^-| X)S_{T}(u'^-| X)}\bigg|T, X\bigg)h(u', T, X)I(\widetilde{Y}\geq u)d\Lambda^{\dagger}_T(u'|X)\bigg]\\
&=-E\bigg[\int_{u'}\frac{E(\phi_{t1}(P)(\widetilde{Y}, T, X)|Y\geq u', T, X)}{G_{T}(u'^-| X)}h(u', T, X)I(\widetilde{Y}\geq u)d\Lambda^{\dagger}_T(u'|X)\bigg]
\end{align*}
Let $l_{t1}(P)(u',T,X)=E(\phi_{t1}(P)(\widetilde{Y}, T, X)|Y\geq u', T, X)$, 
\begin{align*}
    l_{t1}(P)(u',T,X) &= E\left[I(T=t) I(Y \leq s)\left\{ 1+ \frac{\pi_{1-t}(X)}{\pi_t(X)} \frac{ \exp(\gamma_t)}{\{ 1-F_t(s|X) + F_t(s|X) \exp(\gamma_t)\}^2} \right\}\bigg|Y\geq u', T, X\right]\\
    &=I(T=t)\left\{ 1+ \frac{\pi_{1-t}(X)}{\pi_t(X)} \frac{ \exp(\gamma_t)}{\{ 1-F_t(s|X) + F_t(s|X) \exp(\gamma_t)\}^2} \right\}E\left[I(Y \leq s)| Y\geq u', T, X\right]\\
    &=I(T=t)\left\{ 1+ \frac{\pi_{1-t}(X)}{\pi_t(X)} \frac{ \exp(\gamma_t)}{\{ 1-F_t(s|X) + F_t(s|X) \exp(\gamma_t)\}^2} \right\}E\left[\frac{I(Y \leq s)I(Y\geq u')}{S_{T}(u'^-| X)}\bigg|T, X\right]\\
    &=I(T=t)I(u'\leq s)\frac{F_{T}(s| X)-F_{T}(u^-| X)}{1-F_{T}(u'^-| X)}\left\{ 1+ \frac{\pi_{1-t}(X)}{\pi_t(X)} \frac{ \exp(\gamma_t)}{\{ 1-F_t(s|X) + F_t(s|X) \exp(\gamma_t)\}^2} \right\}\\
    &=I(T=t)I(u'\leq s)\frac{F_{t}(s| X)-F_{t}(u^-| X)}{1-F_{t}(u'^-| X)}\left\{ 1+ \frac{\pi_{1-t}(X)}{\pi_t(X)} \frac{ \exp(\gamma_t)}{\{ 1-F_t(s|X) + F_t(s|X) \exp(\gamma_t)\}^2} \right\}
\end{align*}

\begin{align*}
N_2&=E\bigg[\int_{u'}\int_{u}\frac{\phi_{t_{-1}}(P, \psi_t)(T, X)}{G_{T}(\min(u, s)^-| X)}dN(u)h(u', T, X)dN_c(u')\\
&\hspace{1cm}-\int_{u'}\int_{u}\frac{\phi_{t_{-1}}(P, \psi_t)(T, X)}{G_{T}(\min(u, s)^-| X)}dN(u)h(u', T, X)I(\widetilde{Y}\geq u)d\Lambda^{\dagger}_T(u'|X)\bigg]\\
&=-E\bigg[\int_{u'}\int_{u}\frac{\phi_{t_{-1}}(P, \psi_t)(T, X)}{G_{T}(\min(u, s)^-| X)}dN(u)h(u', T, X)I(\widetilde{Y}\geq u)d\Lambda^{\dagger}_T(u'|X)\bigg]\\
&=-E\bigg[\int_{u'}\frac{\Delta\phi_{t_{-1}}(P, \psi_t)(T, X)}{G_{T}(\min(\widetilde{Y}, s)^-| X)}h(u', T, X)I(\widetilde{Y}\geq u)d\Lambda^{\dagger}_T(u'|X)\bigg]\\
&=-E\bigg[\int_{u'}E\bigg(\frac{\Delta}{G_{T}(\min(\widetilde{Y}, s)^-| X)}\bigg|\widetilde{Y}\geq u', T, X\bigg)\phi_{t_{-1}}(P, \psi_t)(T, X)h(u', T, X)I(\widetilde{Y}\geq u)d\Lambda^{\dagger}_T(u'|X)\bigg]
\end{align*}
We can show that
\begin{small}
\begin{align*}
    & E\bigg(\frac{\Delta}{G_{T}(\min(\widetilde{Y}, s)^-| X)}\bigg|\widetilde{Y}\geq u', T, X\bigg) \\
    & =  \frac{ E\bigg(\frac{I(Y \leq C,Y \geq u', C\geq u')}{G_{T}(\min(\widetilde{Y}, s)^-| X)}\bigg|T, X\bigg) }{P[Y \geq u', C \geq u'|T,X]} \\
    &= \frac{E\bigg(\frac{I(Y \geq u')E(I(Y \leq C,  C\geq u')|Y, T, X)}{G_{T}(\min(\widetilde{Y}, s)^-| X)}\bigg|T, X\bigg)}{S_T(u'^-|X) G_T(u'^-|X)}\\
    &= \frac{E\bigg(I(Y \geq u')\bigg\{I(Y<s)\frac{E(I(Y \leq C)|Y, T, X)}{G_{T}(Y^-| X)}+I(Y\geq s)\frac{E(I(Y \leq C)|Y, T, X)}{G_{T}(s^-| X)}\bigg\}\bigg|T, X\bigg)}{S_T(u'^-|X) G_T(u'^-|X)}\\
    &= \frac{E\bigg(I(Y \geq u')\bigg\{I(Y<s)+I(Y\geq s)\frac{G_{T}(Y^-| X)}{G_{T}(s^-| X)}\bigg\}\bigg|T, X\bigg)}{S_T(u'^-|X) G_T(u'^-|X)}\\
    & =  I( u' > s)\frac{E\bigg(I(Y \geq u')\frac{G_{T}(Y^-| X)}{G_{T}(s^-| X)}\bigg|T, X\bigg)}{S_T(u'^-|X) G_T(u'^-|X)}  + I(u' \leq s)\frac{E\bigg(I(u'\leq Y<s)+I(Y\geq s)\frac{G_{T}(Y^-| X)}{G_{T}(s^-| X)}\bigg|T, X\bigg)}{S_T(u'^-|X) G_T(u'^-|X)}  \\
    &= I( u' > s)\frac{E\bigg(I(Y \geq u')G_{T}(Y^-| X)\bigg|T, X\bigg)}{S_T(u'^-|X) G_T(u'^-|X)G_{T}(s^-| X)}  +  I(u' \leq s)\bigg(\frac{S_T(u'^-|X)-S_T(s^-|X)}{S_T(u'^-|X) G_T(u'^-|X)}+\frac{E\bigg(I(Y \geq s)G_{T}(Y^-| X)\bigg|T, X\bigg)}{S_T(u'^-|X) G_T(u'^-|X)G_{T}(s^-| X)}\bigg)
\end{align*}
\end{small}

\begin{align*}
N_3&=E\bigg[\int_{u'}\int_{u}\frac{I(u\geq s)\phi_{t_{-1}}(P, \psi_t)(T, X)}{G_{T}(s^-| X)}dN_c(u)h(u', T, X)dN_c(u')\\
&\hspace{1cm}-\int_{u'}\int_{u}\frac{I(u\geq s)\phi_{t_{-1}}(P, \psi_t)(T, X)}{G_{T}(s^-| X)}dN_c(u)h(u', T, X)I(\widetilde{Y}\geq u)d\Lambda^{\dagger}_T(u'|X)\bigg]\\
&=E\bigg[\int_{u'}\int_{u>u'}\frac{I(u\geq s)\phi_{t_{-1}}(P, \psi_t)(T, X)}{G_{T}(s^-| X)}dN_c(u)h(u', T, X)dN_c(u')\\
&\hspace{1cm}+\int_{u'}\int_{u<u'}\frac{I(u\geq s)\phi_{t_{-1}}(P, \psi_t)(T, X)}{G_{T}(s^-| X)}dN_c(u)h(u', T, X)dN_c(u')\\
&\hspace{1cm}+\int_{u'}\int_{u=u'}\frac{I(u\geq s)\phi_{t_{-1}}(P, \psi_t)(T, X)}{G_{T}(s^-| X)}dN_c(u)h(u', T, X)dN_c(u')\\
&\hspace{1cm}-\int_{u'}\int_{u}\frac{I(u\geq s)\phi_{t_{-1}}(P, \psi_t)(T, X)}{G_{T}(s^-| X)}dN_c(u)h(u', T, X)I(\widetilde{Y}\geq u)d\Lambda^{\dagger}_T(u'|X)\bigg]\\
&=E\bigg[\int_{u'}\frac{I(u'\geq s)\phi_{t_{-1}}(P, \psi_t)(T, X)}{G_{T}(s^-| X)}h(u', T, X)dN_c(u')\\
&\hspace{1cm}-\int_{u'}\int_{u}\frac{I(u\geq s)\phi_{t_{-1}}(P, \psi_t)(T, X)}{G_{T}(s^-| X)}dN_c(u)h(u', T, X)I(\widetilde{Y}\geq u)d\Lambda^{\dagger}_T(u'|X)\bigg]\\
&=E\bigg[\int_{u'}\frac{I(u'\geq s)\phi_{t_{-1}}(P, \psi_t)(T, X)}{G_{T}(s^-| X)}h(u', T, X)I(\widetilde{Y}\geq u)d\Lambda^{\dagger}_T(u'|X)\bigg]\\
&\hspace{1cm}-E\bigg[\int_{u'}\frac{E((1-\Delta)I(\widetilde{Y}\geq s)|\widetilde{Y}\geq u', T, X)\phi_{t_{-1}}(P, \psi_t)(T, X)}{G_{T}(s^-| X)}h(u', T, X)I(\widetilde{Y}\geq u)d\Lambda^{\dagger}_T(u'|X)\bigg]
\end{align*}
and 
\begin{align*}
& E((1-\Delta)I(\widetilde{Y}\geq s)|\widetilde{Y}\geq u', T, X) \\
& =\frac{P[C < Y, Y \geq s, C \geq s, Y \geq u', C \geq u'|T,X] }{P[Y \geq u', C \geq u'|T,X] } \\
& =\frac{P[C < Y, Y \geq s, C \geq s, Y \geq u', C \geq u'|T,X] }{S_T(u'^-|X) G_T(u'^-|X) } \\
 & = \frac{P[C < Y, Y \geq \max\{s,u'\}, C \geq \max\{ s,u'\}|T,X] }{S_T(u'^-|X) G_T(u'^-|X) } \\
& = \frac{E(P[C < Y, Y \geq \max\{s,u'\}, C \geq \max\{ s,u'\}|Y, T,X] |T,X)}{S_T(u'^-|X) G_T(u'^-|X) } \\
& =\frac{E(I(Y \geq \max\{s,u'\})P[C < Y, C \geq \max\{ s,u'\}|Y, T,X] |T,X)}{S_T(u'^-|X) G_T(u'^-|X) } \\
& =I(u'>s)\frac{E(I(Y \geq u')P[C < Y, C \geq u'|Y, T,X] |T,X)}{S_T(u'^-|X) G_T(u'^-|X) } +I(u'\leq s) \frac{E(I(Y \geq s)P[C < Y, C \geq s|Y, T,X] |T,X)}{S_T(u'^-|X) G_T(u'^-|X) }\\
& = I(u'>s)\frac{E(I(Y \geq u')(G_T(u'^-|X)-G_T(Y^-|X)) |T,X)}{S_T(u'^-|X) G_T(u'^-|X) } +I(u'\leq s)\frac{E(I(Y \geq s)(G_T(s^-|X)-G_T(Y^-|X)) |T,X)}{S_T(u'^-|X) G_T(u'^-|X) }\\
& = I(u'>s)\frac{S_T(u'^-|X)G_T(u'^-|X)-E(I(Y \geq u')G_T(Y^-|X) |T,X)}{S_T(u'^-|X) G_T(u'^-|X)}  \\
& \hspace*{0.2in} + I(u'\leq s)\frac{S_T(s^-|X)G_T(s^-|X)-E(I(Y\geq s)G_T(Y^-|X) |T,X)}{S_T(u'^-|X) G_T(u'^-|X) }
\end{align*}

\begin{eqnarray*}
    N_4&=&E\bigg[\int h^*(u, T, X)dM_c(u)\int h(u, T, X)dM_c(u)\bigg]\\
    &=&E\bigg[\int_{u'}h^*(u', T, X)h(u', T, X)I(\widetilde{Y}\geq u)d\Lambda^{\dagger}_T(u'|X)\bigg]
\end{eqnarray*}

Putting everything together, for any $h(u', T, X)$,
\begin{eqnarray*}
    -E\bigg[&\int_{u'}&\frac{l_{t1}(P)(u',T,X)}{G_{T}(u'^-| X)}h(u', T, X)I(\widetilde{Y}\geq u)d\Lambda^{\dagger}_T(u'|X)\bigg]\\
    -E\bigg[&\int_{u'}&\bigg\{I( u' > s)\frac{E\bigg(I(Y \geq u')G_{T}(Y^-| X)\bigg|T, X\bigg)}{S_T(u'^-|X) G_T(u'^-|X)G_{T}(s^-| X)}+I(u' \leq s)\bigg(\frac{S_T(u'^-|X)-S_T(s^-|X)}{S_T(u'^-|X) G_T(u'^-|X)}\\
    &&+\frac{E\bigg(I(Y \geq s)G_{T}(Y^-| X)\bigg|T, X\bigg)}{S_T(u'^-|X) G_T(u'^-|X)G_{T}(s^-| X)}\bigg)\bigg\}\phi_{t_{-1}}(P, \psi_t)(T, X)h(u', T, X)I(\widetilde{Y}\geq u)d\Lambda^{\dagger}_T(u'|X)\bigg]\\
    +E\bigg[&\int_{u'}&\frac{I(u'\geq s)}{G_{T}(s^-| X)}\phi_{t_{-1}}(P, \psi_t)(T, X)h(u', T, X)I(\widetilde{Y}\geq u)d\Lambda^{\dagger}_T(u'|X)\bigg]\\
-E\bigg[&\int_{u'}&\bigg\{I(u'>s)\frac{S_T(u'^-|X)G_T(u'^-|X)-E(I(Y \geq u')G_T(Y^-|X) |T,X)}{S_T(u'^-|X) G_T(u'^-|X)G_T(s^-|X)}\\
&&+I(u'\leq s)\frac{S_T(s^-|X)G_T(s^-|X)-E(I(Y\geq s)G_T(Y^-|X) |T,X)}{S_T(u'^-|X) G_T(u'^-|X)G_T(s^-|X)}\bigg\}\phi_{t_{-1}}(P, \psi_t)(T, X)\\
&&\hspace{2cm}\times h(u', T, X)I(\widetilde{Y}\geq u)d\Lambda^{\dagger}_T(u'|X)\bigg]\\
-E\bigg[&\int_{u'}&h^*(u', T, X)h(u', T, X)I(\widetilde{Y}\geq u)d\Lambda^{\dagger}_T(u'|X)\bigg]=0
\end{eqnarray*}

So for any $h(u', T, X)$,
\begin{eqnarray*}
    E\bigg[\int_{u'}\Bigg\{&&\frac{l_{t1}(P)(u',T,X)}{G_{T}(u'^-| X)}+\bigg\{I( u' > s)\frac{E\bigg(I(Y \geq u')G_{T}(Y^-| X)\bigg|T, X\bigg)}{S_T(u'^-|X) G_T(u'^-|X)G_{T}(s^-| X)}\\
    &&+I(u' \leq s)\bigg(\frac{S_T(u'^-|X)-S_T(s^-|X)}{S_T(u'^-|X) G_T(u'^-|X)}+\frac{E\bigg(I(Y \geq s)G_{T}(Y^-| X)\bigg|T, X\bigg)}{S_T(u'^-|X) G_T(u'^-|X)G_{T}(s^-| X)}\bigg)\bigg\}\phi_{t_{-1}}(P, \psi_t)(T, X)\\
    &&-\frac{I(u'\geq s)}{G_{T}(s^-| X)}\phi_{t_{-1}}(P, \psi_t)(T, X)\\
    &&+\bigg\{I(u'>s)\frac{S_T(u'^-|X)G_T(u'^-|X)-E(I(Y \geq u')G_T(Y^-|X) |T,X)}{S_T(u'^-|X) G_T(u'^-|X)G_T(s^-|X)}\\
&&+I(u'\leq s)\frac{S_T(s^-|X)G_T(s^-|X)-E(I(Y\geq s)G_T(Y^-|X) |T,X)}{S_T(u'^-|X) G_T(u'^-|X)G_T(s^-|X)}\bigg\}\phi_{t_{-1}}(P, \psi_t)(T, X)\\
&&+h^*(u', T, X)\Bigg\}h(u', T, X)I(\widetilde{Y}\geq u)d\Lambda^{\dagger}_T(u'|X)\bigg]=0
\end{eqnarray*}
After canceling out terms, we are left with
\begin{align*}
    E\bigg[\int_{u'}\Bigg\{\frac{l_{t1}(P)(u',T,X)}{G_{T}(u'^-| X)}&+\bigg\{\frac{I(u' \leq s)}{ G_T(u'^-|X)}-\frac{I(u'\geq s)}{G_{T}(s^-| X)}+\frac{I(u'>s)}{G_T(s^-|X)}\bigg\}\phi_{t_{-1}}(P, \psi_t)(T, X)\\
&+h^*(u', T, X)\Bigg\}h(u', T, X)I(\widetilde{Y}\geq u)d\Lambda^{\dagger}_T(u'|X)\bigg]=0
\end{align*}
which can also be written as
\begin{align*}
    E\bigg[\int_{u'}\Bigg\{\frac{l_{t1}(P)(u',T,X)}{G_{T}(u'^-| X)}&+\bigg\{\frac{I(u' = s)}{ G_T(s^-|X)}+\frac{I(u' < s)}{ G_T(u'^-|X)}-\frac{I(u'> s)}{G_{T}(s^-| X)}-\frac{I(u'= s)}{G_{T}(s^-| X)}+\frac{I(u'>s)}{G_T(s^-|X)}\bigg\}\phi_{t_{-1}}(P, \psi_t)(T, X)\\
&+h^*(u', T, X)\Bigg\}h(u', T, X)I(\widetilde{Y}\geq u)d\Lambda^{\dagger}_T(u'|X)\bigg]=0
\end{align*}
and is equivalent to 
\begin{eqnarray*}
    E\bigg[\int_{u'}\Bigg\{\frac{l_{t1}(P)(u',T,X)}{G_{T}(u'^-| X)}+\frac{I(u'<s)}{G_T(u'^-|X)}\phi_{t_{-1}}(P, \psi_t)(T, X)+h^*(u', T, X)\Bigg\}h(u', T, X)I(\widetilde{Y}\geq u)d\Lambda^{\dagger}_T(u'|X)\bigg]=0
\end{eqnarray*}

We derive $h^*(u, T, X)$ as
\begin{eqnarray*}
    h^*(u, T, X)&=&-\frac{l_{t1}(P)(u,T,X)}{G_{T}(u^-| X)}-\frac{I(u<s)}{G_T(u^-|X)}\phi_{t_{-1}}(P, \psi_t)(T, X)\\
    &=&-\frac{l_{t1}(P)(u,T,X)}{G_{T}(u^-| X)}-\frac{I(u<s)}{G_T(u^-|X)}\bigg(\phi_{t2}(P)(O)+\phi_{t3}(P)(O)-\psi_t\bigg)
\end{eqnarray*}

Thus, the projection is
{\small
\begin{align*}
    \int h^*(u, T, X)&dM_c(u,T,X) \\
    &= \int_{u}\bigg\{-\frac{l_{t1}(P)(u,T,X)}{G_{T}(u^-| X)}-\frac{I(u<s)}{G_T(u^-|X)}\bigg(\phi_{t2}(P)(O)+\phi_{t3}(P)(O)-\psi_t\bigg)\bigg\}dM_c(u,T,X)\\
    &=\int_{u}\bigg\{-\frac{l_{t1}(P)(u,T,X)}{G_{T}(u^-| X)}-\frac{I(u<s)}{G_T(u^-|X)}\bigg(\phi_{t2}(P)(O)+\phi_{t3}(P)(O)-\psi_t\bigg)\bigg\}dN_c(u)\\
   &\hspace{1cm}+\int_{u}\bigg\{\frac{l_{t1}(P)(u,T,X)}{G_{T}(u^-| X)}+\frac{I(u<s)}{G_T(u^-|X)}\bigg(\phi_{t2}(P)(O)+\phi_{t3}(P)(O)-\psi_t\bigg)\bigg\}I(\widetilde{Y}\geq u)d\Lambda^{\dagger}_T(u'|X)\\
   &= -\frac{(1-\Delta)}{G_{T}(\widetilde{Y}^-| X)}l_{t1}(P)(\widetilde{Y},T,X)-\frac{(1-\Delta)I(\widetilde{Y}<s)}{G_{T}(\widetilde{Y}^-| X)}\bigg(\phi_{t2}(P)(O)+\phi_{t3}(P)(O)-\psi_t\bigg)\\
   &\hspace{1cm}+\int_{u}\bigg\{\frac{l_{t1}(P)(u,T,X)}{G_{T}(u^-| X)}+\frac{I(u<s)}{G_T(u^-|X)}\bigg(\phi_{t2}(P)(O)+\phi_{t3}(P)(O)-\psi_t\bigg)\bigg\}I(\widetilde{Y}\geq u)d\Lambda^{\dagger}_T(u'|X)\\
   &= -\frac{(1-\Delta)}{G_{T}(\widetilde{Y}^-| X)}l_{t1}(P)(\widetilde{Y},T,X)-\frac{(1-\Delta)I(\widetilde{Y}<s)}{G_{T}(\widetilde{Y}^-| X)}\bigg(\phi_{t2}(P)(O)+\phi_{t3}(P)(O)-\psi_t\bigg)\\
   &\hspace{1cm}+\int_{0}^{\widetilde{Y}}\bigg\{\frac{l_{t1}(P)(u,T,X)}{G_{T}(u^-| X)}+\frac{I(u<s)}{G_T(u^-|X)}\bigg(\phi_{t2}(P)(O)+\phi_{t3}(P)(O)-\psi_t\bigg)\bigg\}d\Lambda^{\dagger}_T(u'|X)
\end{align*}
}
And the observed data influence function is 
\begin{align*}
\widetilde{\phi}_t(\widetilde{P}, \psi_t)(\widetilde{O})&=\frac{\Delta + (1-\Delta) I( \widetilde{Y} \geq s) }{G_{T}(\min(\widetilde{Y},s)^-| X)}\phi_t(P, \psi)(O)-\int h^*(u, T, X)dM_c(u,T,X)\\
&= \frac{\Delta + (1-\Delta) I( \widetilde{Y} \geq s) }{G_{T}(\min(\widetilde{Y},s)^-| X)}\phi_t(P, \psi_t)(O)\\
&\hspace{1cm}+\frac{(1-\Delta)}{G_{T}(\widetilde{Y}^-| X)}l_{t1}(P)(\widetilde{Y},T,X)+\frac{(1-\Delta)I(\widetilde{Y}<s)}{G_{T}(\widetilde{Y}^-| X)}\bigg(\phi_{t2}(P)(O)+\phi_{t3}(P)(O)-\psi_t\bigg)\\
&\hspace{1cm}-\int_{0}^{\widetilde{Y}}\bigg\{\frac{l_{t1}(P)(u,T,X)}{G_{T}(u^-| X)}+\frac{I(u<s)}{G_T(u^-|X)}\bigg(\phi_{t2}(P)(O)+\phi_{t3}(P)(O)-\psi_t\bigg)\bigg\}d\Lambda^{\dagger}_T(u'|X)\\
&= \frac{\Delta + (1-\Delta) I( \widetilde{Y} \geq s) }{G_{T}(\min(\widetilde{Y},s)^-| X)}\phi_t(P, \psi_t)(O)\\
&\hspace{1cm}+\frac{(1-\Delta)}{G_{T}(\widetilde{Y}^-| X)}l_{t1}(P)(\widetilde{Y},T,X)+\frac{(1-\Delta)I(\widetilde{Y}<s)}{G_{T}(\widetilde{Y}^-| X)}\bigg(\phi_{t2}(P)(O)+\phi_{t3}(P)(O)-\psi_t\bigg)\\
&\hspace{1cm}-\int_{0}^{\widetilde{Y}}\bigg\{\frac{l_{t1}(P)(u,T,X)}{G_{T}(u^-| X)}+\frac{I(u<s)}{G_T(u^-|X)}\bigg(\phi_{t2}(P)(O)+\phi_{t3}(P)(O)-\psi_t\bigg)\bigg\}d\Lambda^{\dagger}_T(u'|X)
\end{align*}
where 
\begin{align*}
l_{t1}(P)(u,T,X) = I(T=t)  I(u \leq s) \frac{ F_t(s|X) - F_t(u^-|X) }{1-F_t(u^-|X)} \times \left\{1+  \frac{\pi_{1-t}(X)}{\pi_t(X)} \frac{ \exp(\gamma_t)}{\{ 1-F_t(s|X) + F_t(s|X) \exp(\gamma_t)\}^2} \right\}
\end{align*}

\section{Proof of robustness property and asymptotic results}

\begin{lemma}\label{inverse_weight}
    Assume $C$ is independent of $Y$ given $T$ and $X$.  Then,
    $$E \left[ \frac{\Delta + (1-\Delta)I( \widetilde{Y}\geq s)}{G^\ast_{T}(\min(\widetilde{Y}, s)^-| X)} \right]=E \left[  \frac{G_{T}(Y^-| X)  I(Y \leq s)}{G^\ast_{T}(Y^-| X)} + \frac{I( Y> s)G_{T}(s^-| X)}{G^\ast_{T}(s^-| X)} \right]$$
    for any $G^\ast_{\cdot}(\cdot| X)$.% and $G^\ast_{1}(\cdot| X)$. %$G^\ast_{T}(u^-| X)$. 
\end{lemma}

\textbf{Proof of Lemma~\ref{inverse_weight}}
\begin{eqnarray*}
E \left[ \frac{\Delta + (1-\Delta)I( \widetilde{Y}\geq s)}{G^\ast_{T}(\min(\widetilde{Y}, s)^-| X)} \right]&=& E \left[ \frac{\Delta I(\widetilde{Y} \geq s) +\Delta I(\widetilde{Y} < s) + (1-\Delta)I( \widetilde{Y}\geq s)}{G^\ast_{T}(\min(\widetilde{Y}, s)^-| X)} \right] \\
& = & E \left[ \frac{\Delta I(\widetilde{Y} \geq s)}{G^\ast_{T}(s^-| X)} + \frac{\Delta I(\widetilde{Y} < s)}{G^\ast_{T}(\widetilde{Y}^-| X)} + \frac{(1-\Delta)I( \widetilde{Y}\geq s)}{G^\ast_{T}(s^-| X)} \right] \\
& = & E \left[  \frac{\Delta I(\widetilde{Y} < s)}{G^\ast_{T}(\widetilde{Y}^-| X)} + \frac{I( \widetilde{Y}\geq s)}{G^\ast_{T}(s^-| X)} \right] \\
& = & E \left[  \frac{I( Y \leq C) I(Y < s)}{G^\ast_{T}(Y^-| X)} + \frac{I( Y\geq s,C \geq s)}{G^\ast_{T}(s^-| X)} \right] \\
& = & E \left[  \frac{G_{T}(Y^-| X)  I(Y < s)}{G^\ast_{T}(Y^-| X)} + \frac{I( Y\geq s)G_{T}(s^-| X)}{G^\ast_{T}(s^-| X)} \right] \\
& = & E \left[  \frac{G_{T}(Y^-| X)  I(Y \leq s)}{G^\ast_{T}(Y^-| X)} + \frac{I( Y> s)G_{T}(s^-| X)}{G^\ast_{T}(s^-| X)} \right]
\end{eqnarray*}

\begin{lemma}\label{robust}
    $E\left[g_t(G^\ast_{\cdot}(\cdot|X), F_{t}(\cdot|X), \pi^\ast_{t}(X))(\widetilde{O})\right]=E\left[g_t(\widetilde{P})(\widetilde{O})\right]$ and $E\left[h_t(G^\ast_{\cdot}(\cdot|X))(\widetilde{O})\right]=E\left[h_t(\widetilde{P})(\widetilde{O})\right]$ for any $G^\ast_{\cdot}(\cdot|X)$ and $\pi^\ast_{t}(X)$, where $G^\ast_\cdot(\cdot|X)=(G^\ast_t(\cdot|X), G^\ast_{1-t}(\cdot|X))$ and $G^\ast_\cdot(\cdot|X)=\exp\left(- \Lambda^{\ast\dagger}_\cdot(\cdot) \exp \{ {\beta^{\ast'}_\cdot} X \}\right)$,  provided that $F_{t}(\cdot|X)$ is the true conditional distribution of $Y$ given $X$ and $T=t$. 
\end{lemma}
\textbf{Proof of Lemma~\ref{robust}} For simplicity, we will sometimes abbreviate the nuisance functions by $G^*_{\cdot}\coloneq G^*_{\cdot}(\cdot|X)$, $F_{t}\coloneq F_{t}(\cdot|X)$ and $\pi^*_{t}\coloneq \pi^*_{t}(X)$. Since

\begin{align*}
&E\left[g_t(G^\ast_{\cdot}, F_{t}, \pi^
\ast_{t})(\widetilde{O})\right]\\
    &=E\left[\frac{\Delta + (1-\Delta) I( \widetilde{Y} \geq s) }{G^\ast_{T}(\min(\widetilde{Y},s)^-| X)}\phi_{t}(F_{t}, \pi^
\ast_{t})(O)\right] + E\left[\frac{(1-\Delta)}{G^\ast_{T}(\widetilde{Y}^-| X)}l_t(F_{t}, \pi^
\ast_{t})(\widetilde{Y},T,X)\right]\\
    &\hspace{1cm}-E\left[\int_0^{\widetilde{Y}} \frac{l_t(F_{t}, \pi^
\ast_{t})(u,T,X)}{G^\ast_{T}(u^-|X)} \exp\{\beta^{\ast'}_{T} X \}d\Lambda^{\ast\dagger}_{T}(u)\right]\\
    &=E\left[\frac{\Delta + (1-\Delta) I( \widetilde{Y} \geq s) }{G^\ast_{T}(\min(\widetilde{Y},s)^-| X)}\phi_{t}(F_{t}, \pi^
\ast_{t})(O)\right]+ E\left[\int\frac{l_t(F_{t}, \pi^
\ast_{t})(u,T,X)}{G^\ast_{T}(u^-| X)}dN_c(u)\right]\\
    &\hspace{1cm}-E\left[\int_0^{\widetilde{Y}} \frac{l_t(F_{t}, \pi^
\ast_{t})(u,T,X)}{G^\ast_{T}(u^-|X)} \exp\{\beta^{\ast'}_{T} X \}d\Lambda^{\ast\dagger}_{T}(u)\right]\\
    &= E\left[\frac{\Delta + (1-\Delta)I( \widetilde{Y}\geq s)}{G^\ast_{T}(\min(\widetilde{Y}, s)^-| X)}\phi_{t}(F_{t}, \pi^
\ast_{t})(O)\right]+ E\left[\int_0^{\widetilde{Y}}\frac{l_t(F_{t}, \pi^
\ast_{t})(u,T,X)}{G^\ast_{T}(u^-|X)}\exp\{\beta_{T}' X \}d\Lambda^{\dagger}_{T}(u)\right] \\
    &\hspace{1cm}-E\left[\int_0^{\widetilde{Y}} \frac{l_t(F_{t}, \pi^\ast_{t})(u,T,X)}{G^\ast_{T}(u^-|X)} \exp\{\beta^{\ast'}_{T} X \}d\Lambda^{\ast\dagger}_{T}(u)\right]\\
    &= E\left[\frac{\Delta + (1-\Delta)I( \widetilde{Y}\geq s)}{G^\ast_{T}(\min(\widetilde{Y}, s)^-| X)} \left\{\phi_{t}(F_{t}, \pi^
\ast_{t})(O)\right\}\right]\\
    &\hspace{1cm}+E\left[\int_0^{\widetilde{Y}}\frac{l_t(F_{t}, \pi^
\ast_{t})(u,T,X)}{G^\ast_{T}(u^-|X)}\left\{\exp\{\beta_{T}' X \}d\Lambda^{\dagger}_{T}(u)-\exp\{\beta^{\ast'}_{T}X \}d\Lambda^{\ast\dagger}_{T}(u)\right\}\right]
\end{align*}

with 
\begin{eqnarray*}
\phi_{t1}(F_{t}, \pi^\ast_{t})(O) & =&  I(T=t) I(Y \leq s)\left\{ 1+ \frac{\pi^\ast_{1-t}(X)}{\pi^\ast_t(X)} \frac{ \exp(\gamma_t)}{\{ 1-F_t(s|X) + F_t(s|X) \exp(\gamma_t)\}^2} \right\},  \\
\phi_{t2}(F_{t}, \pi^\ast_{t})(O) & =&  - I(T=t)  \left\{ \frac{\pi^\ast_{1-t}(X)}{\pi^\ast_t(X)} \frac{  F_t(s|X) \exp(\gamma_t)}{\{ 1-F_t(s|X) + F_t(s|X) \exp(\gamma_t)\}^2} \right\}, \\
\phi_{t3}(F_{t}, \pi^\ast_{t})(O) & =&  I(T=1-t) \left\{ \frac{F_t(s|X) \exp(\gamma_t)}{1-F_t(s|X) + F_t(s|X) \exp(\gamma_t)} \right\}.     
\end{eqnarray*}

Applying Lemma~\ref{inverse_weight}, we have 
\begin{eqnarray*}
&&E\left[g_t(G^\ast_{\cdot}, F_{t}, \pi^\ast_{t})(\widetilde{O})\right]\\
    &=& E\left[\frac{\Delta}{G^\ast_{T}(\widetilde{Y}^-| X)}\phi_{t1}(F_{t}, \pi^\ast_{t})(O)\right]\\
    &&+E\left[\left\{\frac{G_{T}(Y^-| X)  I(Y \leq s)}{G^\ast_{T}(Y^-| X)} + \frac{I( Y> s)G_{T}(s^-| X)}{G^\ast_{T}(s^-| X)}\right\} \left\{\phi_{t2}(F_{t}, \pi^\ast_{t})(O)+\phi_{t3}(F_{t}, \pi^\ast_{t})(O)\right\}\right]\\
    &&+E\left[\int_0^{\widetilde{Y}}\frac{l_t(F_{t}, \pi^\ast_{t})(u,T,X)}{G^\ast_{T}(u^-|X)}\left\{\exp\{\beta_{T}' X \}d\Lambda^{\dagger}_{T}(u)-\exp\{\beta^{\ast'}_{T} X \}d\Lambda^{\ast\dagger}_{T}(u)\right\}\right]\\
    &=& E\left[\frac{G_{T}(Y^-| X)}{G^\ast_{T}(Y^-| X)}\phi_{t1}(F_{t}, \pi^\ast_{t})(O)\right]\\
    &&+E\left[\frac{G_{T}(Y^-| X)  I(Y \leq s)}{G^\ast_{T}(Y^-| X)}\left\{\phi_{t2}(F_{t}, \pi^\ast_{t})(O)+\phi_{t3}((F_{t}, \pi^\ast_{t}))(O)\right\}\right]\\
    &&+E\left[\frac{I( Y> s)G_{T}(s^-| X)}{G^\ast_{T}(s^-| X)}\left\{\phi_{t2}(F_{t}, \pi^\ast_{t})(O)+\phi_{t3}(F_{t}, \pi^\ast_{t})(O)\right\}\right]\\
    &&+E\left[\int_0^{\widetilde{Y}}\frac{l_t(F_{t}, \pi^\ast_{t})(u,T,X)}{G^\ast_{T}(u^-|X)}\left\{\exp\{\beta_{T}' X \}d\Lambda^{\dagger}_{T}(u)-\exp\{{\beta^{\ast'}_{T}} X \}d\Lambda^{\ast\dagger}_{T}(u)\right\}\right]
\end{eqnarray*}

Let $E\left[g_t(G^\ast_{\cdot}, F_{t}, \pi^\ast_{t})(\widetilde{O})\right]=M_1+M_2+M_3+M_4$. Applying integration by parts to the first and second term, 
\begin{eqnarray*}
    M_1 &=& E\left[\frac{G_{T}(Y^-| X)}{G^\ast_{T}(Y^-| X)}\phi_{t1}(F_{t}, \pi^\ast_{t})(O)\right]\\
    &=& E\left[\int\frac{G_{T}(u^-| X)}{G^\ast_{T}(u^-| X)}\phi_{t1}(F_{t}, \pi^\ast_{t})(u, T, X)dF_T(u|X)\right]\\
    &=& E\left[\frac{G_{T}(u^-| X)}{G^\ast_{T}(u^-| X)}F_T(u|X)I(T=t) \left\{ 1+ \frac{\pi^\ast_{1-t}(X)}{\pi^\ast_t(X)} \frac{ \exp(\gamma_t)}{\{ 1-F_t(s|X) + F_t(s|X) \exp(\gamma_t)\}^2} \right\}\bigg|_0^s\right]\\
    &&-E\left[\int\phi_{t1}(F_{t}, \pi^\ast_{t})(u, T, X)F_T(u|X)\frac{dG_{T}(u^-| X)}{G^\ast_{T}(u^-| X)}\right]\\
    &&+E\left[\int\phi_{t1}(F_{t}, \pi^\ast_{t})(u, T, X)F_T(u|X)\frac{G_{T}(u^-| X)}{G^\ast_{T}(u^-| X)^2}dG^\ast_{T}(u^-| X)\right]\\
    &=& E\left[\frac{G_{t}(s^-| X)}{G^\ast_{t}(s^-| X)}F_t(s|X)I(T=t) \left\{ 1+ \frac{\pi^\ast_{1-t}(X)}{\pi^\ast_t(X)} \frac{ \exp(\gamma_t)}{\{ 1-F_t(s|X) + F_t(s|X) \exp(\gamma_t)\}^2} \right\}\right]\\
    &&-E\left[\int\phi_{t1}(F_{t}, \pi^\ast_{t})(u, T, X)F_t(u|X)\frac{dG_{T}(u^-| X)}{G^\ast_{T}(u^-| X)}\right]\\
    &&+E\left[\int\phi_{t1}(F_{t}, \pi^\ast_{t})(u, T, X)F_t(u|X)\frac{G_{T}(u^-| X)}{G^\ast_{T}(u^-| X)^2}dG^\ast_{T}(u^-| X)\right]
\end{eqnarray*}
with 
$$\phi_{t1}(F_{t}, \pi^\ast_{t})(u, T, X)=I(T=t) I(u \leq s)\left\{ 1+ \frac{\pi^\ast_{1-t}(X)}{\pi^\ast_t(X)} \frac{ \exp(\gamma_t)}{\{ 1-F_t(s|X) + F_t(s|X) \exp(\gamma_t)\}^2} \right\}.$$

\begin{eqnarray*}
    M_2 &=& E\left[\frac{G_{T}(Y^-| X)  I(Y \leq s)}{G^\ast_{T}(Y^-| X)}\left\{\phi_{t2}(F_{t}, \pi^\ast_{t})(O)+\phi_{t3}(F_{t}, \pi^\ast_{t})(O)\right\}\right]\\
    &=& E\left[\left\{\phi_{t2}(F_{t}, \pi^\ast_{t})(O)+\phi_{t3}(F_{t}, \pi^\ast_{t})(O)\right\}\int\frac{G_{T}(u^-| X)}{G^\ast_{T}(u^-| X)}I(u\leq s)dF_T(u|X)\right]\\
    &=& E\left[\left\{\phi_{t2}(F_{t}, \pi^\ast_{t})(O)+\phi_{t3}(F_{t}, \pi^\ast_{t})(O)\right\}\frac{G_{T}(u^-| X)}{G^\ast_{T}(u^-| X)}F_T(u|X)\bigg|_0^s\right]\\
    &&-E\left[\left\{\phi_{t2}(F_{t}, \pi^\ast_{t})(O)+\phi_{t3}(F_{t}, \pi^\ast_{t})(O)\right\}\int I(u\leq s)F_T(u|X)\frac{dG_{T}(u^-| X)}{G^\ast_{T}(u^-| X)}\right]\\
    &&+E\left[\left\{\phi_{t2}(F_{t}, \pi^\ast_{t})(O)+\phi_{t3}(F_{t}, \pi^\ast_{t})(O)\right\}\int I(u\leq s)F_T(u|X)\frac{G_{T}(u^-| X)}{G^\ast_{T}(u^-| X)^2}dG^\ast_{T}(u^-| X)\right]\\
    &=& E\left[\left\{\phi_{t2}(F_{t}, \pi^\ast_{t})(O)+\phi_{t3}(F_{t}, \pi^\ast_{t})(O)\right\}\frac{G_{T}(s^-| X)}{G^\ast_{T}(s^-| X)}F_T(s|X)\right]\\
    &&-E\left[\left\{\phi_{t2}(F_{t}, \pi^\ast_{t})(O)+\phi_{t3}(F_{t}, \pi^\ast_{t})(O)\right\}\int I(u\leq s)F_T(u|X)\frac{dG_{T}(u^-| X)}{G^\ast_{T}(u^-| X)}\right]\\
    &&+E\left[\left\{\phi_{t2}(F_{t}, \pi^\ast_{t})(O)+\phi_{t3}(F_{t}, \pi^\ast_{t})(O)\right\}\int I(u\leq s)F_T(u|X)\frac{G_{T}(u^-| X)}{G^\ast_{T}(u^-| X)^2}dG^\ast_{T}(u^-| X)\right]
\end{eqnarray*}
and
\begin{eqnarray*}
    M_3 &=& E\left[\frac{I( Y> s)G_{T}(s^-| X)}{G^\ast_{T}(s^-| X)}\left\{\phi_{t2}(F_{t}, \pi^\ast_{t})(O)+\phi_{t3}(F_{t}, \pi^\ast_{t})(O)\right\}\right]\\
    &=& E\left[\frac{G_{T}(s^-| X)}{G^\ast_{T}(s^-| X)}(1-F_T(s|X))\left\{\phi_{t2}(F_{t}, \pi^\ast_{t})(O)+\phi_{t3}(F_{t}, \pi^\ast_{t})(O)\right\}\right]
\end{eqnarray*}

Since $\exp\{\beta_{T}' X \}  d\Lambda^{\dagger}_{T}(u)=d\Lambda^{\dagger}_{T}(u|X)=-d\log G_{T}(u|X)=-\frac{dG_{T}(u|X)}{G_{T}(u|X)}$ and time is continuous, we have
\begin{align*}
    M_4 &=E\left[\int_0^{\widetilde{Y}}\frac{l_t(F_{t}, \pi^\ast_{t})(u,T,X)}{G^\ast_{T}(u^-|X)}\left\{\exp\{\beta_{T}' X \}d\Lambda^{\dagger}_{T}(u)-\exp\{\beta^{\ast'}_{T}X \}d\Lambda^{\ast\dagger}_{T}(u)\right\}\right]\\
    &=-E\left[\int_0^{\widetilde{Y}}\frac{l_t(F_{t}, \pi^\ast_{t})(u,T,X)}{G^\ast_{T}(u^-|X)}\left\{\frac{dG_{T}(u|X)}{G_{T}(u|X)}-\frac{dG^\ast_{T}(u|X)}{G^\ast_{T}(u|X)}\right\}\right]\\
    &=-E\left[\int_0^{\widetilde{Y}}\frac{l_t(F_{t}, \pi^\ast_{t})(u,T,X)}{G^\ast_{T}(u^-|X)}\left\{\frac{dG_{T}(u^-|X)}{G_{T}(u^-|X)}-\frac{dG^\ast_{T}(u^-|X)}{G^\ast_{T}(u^-|X)}\right\}\right]\\
    &=-E\left[\int I(u\leq \widetilde{Y})\frac{l_t(F_{t}, \pi^\ast_{t})(u,T,X)}{G^\ast_{T}(u^-|X)}\left\{\frac{dG_{T}(u^-|X)}{G_{T}(u^-|X)}-\frac{dG^\ast_{T}(u^-|X)}{G^\ast_{T}(u^-|X)}\right\}\right]\\
    &=-E\left[\int (1-F_T(u^-|X))\left\{l_t(F_{t}, \pi^\ast_{t})(u,T,X)\right\}\left\{\frac{dG_{T}(u^-|X)}{G^\ast_{T}(u^-|X)}-\frac{G_{T}(u^-|X)}{G^\ast_{T}(u^-|X)^2}dG^\ast_{T}(u^-|X)\right\}\right]\\
    &=-E\left[\int (F_t(s|X)-F_t(u^-|X))\phi_{t1}(F_{t}, \pi^\ast_{t})(u, T, X)\left\{\frac{dG_{T}(u^-|X)}{G^\ast_{T}(u^-|X)}-\frac{G_{T}(u^-|X)}{G^\ast_{T}(u^-|X)^2}dG^\ast_{T}(u^-|X)\right\}\right]\\
    &\hspace{1cm}-E\left[\int (1-F_T(u^-|X))I(u<s)\left\{\phi_{t2}(F_{t}, \pi^\ast_{t})(O)+\phi_{t3}(F_{t}, \pi^\ast_{t})(O)\right\}\right.\\
    &\hspace{3cm}\left.\times\left\{\frac{dG_{T}(u^-|X)}{G^\ast_{T}(u^-|X)}-\frac{G_{T}(u^-|X)}{G^\ast_{T}(u^-|X)^2}dG^\ast_{T}(u^-|X)\right\}\right]\\
    &=-E\left[\int (F_t(s|X)-F_t(u^-|X))\phi_{t1}(F_{t}, \pi^\ast_{t})(u, T, X)\left\{\frac{dG_{T}(u^-|X)}{G^\ast_{T}(u^-|X)}-\frac{G_{T}(u^-|X)}{G^\ast_{T}(u^-|X)^2}dG^\ast_{T}(u^-|X)\right\}\right]-\\
    &\hspace{1cm}-E\left[\int (1-F_T(u^-|X))I(u\leq s)\left\{\phi_{t2}(F_{t}, \pi^\ast_{t})(O)+\phi_{t3}(F_{t}, \pi^\ast_{t})(O)\right\}\right.\\
    &\hspace{3cm}\left.\times\left\{\frac{dG_{T}(u^-|X)}{G^\ast_{T}(u^-|X)}-\frac{G_{T}(u^-|X)}{G^\ast_{T}(u^-|X)^2}dG^\ast_{T}(u^-|X)\right\}\right]\\
\end{align*}

Putting everything together, and using the fact that $F(u^-|X)=F(u|X)$ (time is continuous), we have
\begin{eqnarray*}
    &&M_1+M_2+M_3+M_4\\
    &=& E\left[\frac{G_{t}(s^-| X)}{\widehat{G}_{t}(s^-| X)}F_t(s|X)I(T=t) \left\{ 1+ \frac{\pi^\ast_{1-t}(X)}{\pi^\ast_t(X)} \frac{ \exp(\gamma_t)}{\{ 1-F_t(s|X) + F_t(s|X) \exp(\gamma_t)\}^2} \right\}\right]\\
    &&+E\left[\left\{\phi_{t2}(F_{t}, \pi^\ast_{t})(O)+\phi_{t3}(F_{t}, \pi^\ast_{t})(O)\right\}\frac{G_{T}(s^-| X)}{G^\ast_{T}(s^-| X)}F_T(s|X)\right]\\
    &&+E\left[\frac{G_{T}(s^-| X)}{G^\ast_{T}(s^-| X)}(1-F_T(s|X))\left\{\phi_{t2}(F_{t}, \pi^\ast_{t})(O)+\phi_{t3}(F_{t}, \pi^\ast_{t})(O)\right\}\right]\\
    &&-E\left[F_t(s|X)\int \phi_{t1}(F_{t}, \pi^\ast_{t})(u, T, X)\left\{\frac{dG_{T}(u^-|X)}{G^\ast_{T}(u^-|X)}-\frac{G_{T}(u^-|X)}{G^\ast_{T}(u^-|X)^2}dG^\ast_{T}(u^-|X)\right\}\right]\\
    &&-E\left[\left\{\phi_{t2}(F_{t}, \pi^\ast_{t})(O)+\phi_{t3}(F_{t}, \pi^\ast_{t})(O)\right\}\right.\\
    &&\hspace{2cm}\left.\times\int I(u\leq s)\left\{\frac{dG_{T}(u^-|X)}{G^\ast_{T}(u^-|X)}-\frac{G_{T}(u^-|X)}{G^\ast_{T}(u^-|X)^2}dG^\ast_{T}(u^-|X)\right\}\right]
\end{eqnarray*}

Since 
\begin{eqnarray*}
    \int I(u\leq s)\frac{dG_{T}(u^-|X)}{G^\ast_{T}(u^-|X)} &=& \frac{G_{T}(u^-|X)}{G^\ast_{T}(u^-|X)}\bigg|_0^s+\int I(u\leq s)\frac{G_{T}(u^-|X)}{G^\ast_{T}(u^-|X)^2}dG^\ast_{T}(u^-|X)\\
    &=& \frac{G_{T}(s^-|X)}{G^\ast_{T}(s^-|X)}-1+\int I(u\leq s)\frac{G_{T}(u^-|X)}{G^\ast_{T}(u^-|X)^2}dG^\ast_{T}(u^-|X)
\end{eqnarray*}
we have
\begin{eqnarray*}
    &&M_1+M_2+M_3+M_4\\
    &=&E\left[F_t(s|X)I(T=t) \left\{ 1+ \frac{\pi^\ast_{1-t}(X)}{\pi^\ast_t(X)} \frac{ \exp(\gamma_t)}{\{ 1-F_t(s|X) + F_t(s|X) \exp(\gamma_t)\}^2} \right\}\right.\\
    &&\hspace{1cm}\left.+\phi_{t2}(F_{t}, \pi^\ast_{t})(O)+\phi_{t3}(F_{t}, \pi^\ast_{t})(O)\right]\\
    &=& E\left[F_t(s|X)I(T=t) \left\{ 1+ \frac{\pi^\ast_{1-t}(X)}{\pi^\ast_t(X)} \frac{ \exp(\gamma_t)}{\{ 1-F_t(s|X) + F_t(s|X) \exp(\gamma_t)\}^2} \right\}\right]\\
    &&-E\left[I(T=t)  \left\{ \frac{\pi^\ast_{1-t}(X)}{\pi^\ast_t(X)} \frac{  F_t(s|X) \exp(\gamma_t)}{\{ 1-F_t(s|X) + F_t(s|X) \exp(\gamma_t)\}^2} \right\}\right]\\
    &&+E\left[I(T=1-t) \left\{ \frac{F_t(s|X) \exp(\gamma_t)}{1-F_t(s|X) + F_t(s|X) \exp(\gamma_t)} \right\}\right]\\
    &=& E\left[F_t(s|X)I(T=t)+I(T=1-t) \left\{ \frac{F_t(s|X) \exp(\gamma_t)}{1-F_t(s|X) + F_t(s|X) \exp(\gamma_t)}\right\}\right]\\
    &=&E\left[F_t(s|X)\pi_t(X)+\pi_{1-t}(X) \left\{ \frac{F_t(s|X) \exp(\gamma_t)}{1-F_t(s|X) + F_t(s|X) \exp(\gamma_t)}\right\}\right]\\
    &=& \psi_t\\
\end{eqnarray*}
Applying Lemma~\ref{inverse_weight}, we have 
\begin{align*}
&E\left[g_t(\widetilde{P})(\widetilde{O})\right]\\
&=E\left[\frac{\Delta + (1-\Delta) I( \widetilde{Y} \geq s) }{G_{T}(\min(\widetilde{Y},s)^-| X)}\phi_{t}(P)(O)\right] + E\left[\frac{(1-\Delta)}{G_{T}(\widetilde{Y}^-| X)}l_t(P)(\widetilde{Y},T,X)\right]\\
    &\hspace{1cm}-E\left[\int_0^{\widetilde{Y}} \frac{l_t(P)(u,T,X)}{G_{T}(u^-|X)} \exp\{\beta_{T}' X \}d\Lambda^{\dagger}_{T}(u)\right]\\
    &=E\left[\phi_{t}(P)(O)\right]+ E\left[\int\frac{l_t(P)(u,T,X)}{G_{T}(u^-| X)}dN_c(u)\right]-E\left[\int_0^{\widetilde{Y}} \frac{l_t(P)(u,T,X)}{G_{T}(u^-|X)} \exp\{\beta_{T}' X \}d\Lambda^{\dagger}_{T}(u)\right]\\
    &= \psi_t+E\left[\int_0^{\widetilde{Y}}\frac{l_t(P)(u,T,X)}{G_{T}(u^-|X)}\exp\{\beta_{T}' X \}d\Lambda^{\dagger}_{T}(u)\right]-E\left[\int_0^{\widetilde{Y}}\frac{l_t(P)(u,T,X)}{G_{T}(u^-|X)}\exp\{\beta_{T}' X \}d\Lambda^{\dagger}_{T}(u)\right] \\
    &=\psi_t
\end{align*}
Thus, we conclude that $E\left[g_t(G^\ast_{\cdot}(\cdot|X), F_{t}(\cdot|X), \pi^\ast_{t}(X))(\widetilde{O})\right]=E\left[g_t(\widetilde{P})(\widetilde{O})\right]=\psi_t$. Applying similar strategies to $E\left[h_t(G^\ast_{\cdot})(\widetilde{O})\right]$, we have
\begin{align*}
    E\left[h_t(G^\ast_{\cdot})(\widetilde{O})\right] &=E\left[\frac{\Delta + (1-\Delta) I( \widetilde{Y} \geq s) }{G^\ast_{T}(\min(\widetilde{Y},s)^-| X)}\right] + E\left[\frac{(1-\Delta)I(\widetilde{Y}<s)}{G^\ast_{T}(\widetilde{Y}^-| X)}\right]\\
    &\hspace{1cm}-E\left[\int_0^{\widetilde{Y}} \frac{I(u<s)}{G^\ast_{T}(u^-|X)} \exp\{\beta^{\ast'}_{T} X \}d\Lambda^{\ast\dagger}_{T}(u)\right]\\
    &=E\left[\frac{\Delta + (1-\Delta) I( \widetilde{Y} \geq s) }{G^\ast_{T}(\min(\widetilde{Y},s)^-| X)}\right] + E\left[\frac{(1-\Delta)I(\widetilde{Y}<s)}{G^\ast_{T}(\min(\widetilde{Y},s)^-| X)}\right]\\
    &\hspace{1cm}-E\left[\int_0^{\widetilde{Y}} \frac{I(u<s)}{G^\ast_{T}(u^-|X)} \exp\{\beta^{\ast'}_{T} X \}d\Lambda^{\ast\dagger}_{T}(u)\right]\\
    &= E \left[  \frac{G_{T}(Y^-| X)  I(Y \leq s)}{G^\ast_{T}(Y^-| X)} + \frac{I( Y> s)G_{T}(s^-| X)}{G^\ast_{T}(s^-| X)} \right]\\
    &\hspace{1cm}-E\left[\int_0^{\widetilde{Y}}\frac{I(u<s)}{G^\ast_{T}(u^-|X)}\left\{\exp\{\beta_{T}' X \}d\Lambda^{\dagger}_{T}(u)-\exp\{\beta^{\ast'}_{T}X \}d\Lambda^{\ast\dagger}_{T}(u)\right\}\right]\\
    &= E \left[  \frac{G_{T}(Y^-| X)  I(Y \leq s)}{G^\ast_{T}(Y^-| X)} + \frac{I( Y> s)G_{T}(s^-| X)}{G^\ast_{T}(s^-| X)} \right]\\
    &\hspace{1cm}-E\left[\int \frac{I(u\leq \widetilde{Y}, u<s)}{G^\ast_{T}(u^-|X)}\left\{\frac{dG_{T}(u^-|X)}{G_{T}(u^-|X)}-\frac{dG^\ast_{T}(u^-|X)}{G^\ast_{T}(u^-|X)}\right\}\right]\\
    &= E \left[  \frac{G_{T}(Y^-| X)  I(Y \leq s)}{G^\ast_{T}(Y^-| X)} + \frac{I( Y> s)G_{T}(s^-| X)}{G^\ast_{T}(s^-| X)} \right]\\
    &\hspace{1cm}-E\left[\int I(u\leq Y, u<s)\left\{\frac{dG_{T}(u^-|X)}{G^\ast_{T}(u^-|X)}-\frac{G_{T}(u^-|X)}{G^\ast_{T}(u^-|X)^2}dG^\ast_{T}(u^-|X)\right\}\right]\\
    &= E \left[  \frac{G_{T}(Y^-| X)  I(Y \leq s)}{G^\ast_{T}(Y^-| X)} + \frac{I( Y> s)G_{T}(s^-| X)}{G^\ast_{T}(s^-| X)} \right]\\
    &\hspace{1cm}-E\left[\int I(u\leq Y\leq s)\left\{\frac{dG_{T}(u^-|X)}{G^\ast_{T}(u^-|X)}-\frac{G_{T}(u^-|X)}{G^\ast_{T}(u^-|X)^2}dG^\ast_{T}(u^-|X)\right\}\right]\\
    &\hspace{1cm}-E\left[\int I(u\leq s< Y)\left\{\frac{dG_{T}(u^-|X)}{G^\ast_{T}(u^-|X)}-\frac{G_{T}(u^-|X)}{G^\ast_{T}(u^-|X)^2}dG^\ast_{T}(u^-|X)\right\}\right]\\
    &= I(Y\leq s)+I(Y>s)=1
\end{align*}
And since 
\begin{align*}
    E\left[h_t(\widetilde{P})(\widetilde{O})\right] &=E\left[\frac{\Delta + (1-\Delta) I( \widetilde{Y} \geq s) }{G_{T}(\min(\widetilde{Y},s)^-| X)}\right] + E\left[\frac{(1-\Delta)I(\widetilde{Y}<s)}{G_{T}(\widetilde{Y}^-| X)}\right]\\
    &\hspace{1cm}-E\left[\int_0^{\widetilde{Y}} \frac{I(u<s)}{G_{T}(u^-|X)} \exp\{\beta_{T}' X \}d\Lambda^{\dagger}_{T}(u)\right]\\
    &=1,
\end{align*}
we conclude that $E\left[h_t(G^\ast_{\cdot})(\widetilde{O})\right]=E\left[h_t(\widetilde{P})(\widetilde{O})\right]$.

In the following proofs, we will use the following property: for any function $f(t)$ and any monotone function $G(t)$ defined on $[a, b]$, 
\begin{equation}\label{prop_int}
    \left|\int_a^b f(t)dG(t)\right|\leq \sup_{t\in[a, b]}\left|f(t)\right|\left|G(b)-G(a)\right|
\end{equation}

\begin{lemma}\label{rate}
$\vline \; E \left[  g_t(\widetilde{P}^*)(\widetilde{O}) - h_t(\widetilde{P}^*)(\widetilde{O})  \psi_t \right] \; \vline$
is bounded above by:
%the sum of eight terms, each of which is a second order product of differences between a component of $\widetilde{P}^*$ and the corresponding component of $\widetilde{P}$: 
\begin{align*}
& \sup_{u\in[0, \tau]}\left|\left|F^\ast_t(u|X)-F_t(u|X)\right|\right|_{L_2} \times \sup_{u\in[0, \tau]}\left|\left|G^\ast_{t}(u| X)-G_{t}(u| X)\right|\right|_{L_2} \\
& + \sup_{u\in[0, \tau]}\left|\left|F^\ast_t(u|X)-F_t(u|X)\right|\right|_{L_2} \times \sup_{u\in[0, \tau]}\left|\left|G^\ast_{1-t}(u| X)-G_{1-t}(u| X)\right|\right|_{L_2} \\
& + \sup_{u\in[0, \tau]}\left|\left|F^\ast_t(u|X)-F_t(u|X)\right|\right|_{L_2} \times \left|\left|\pi^\ast_t(X)-\pi_t(X)\right|\right|_{L_2} \\
& + \left\{ \sup_{u\in[0, \tau]}\left|\left|F^\ast_t(u|X)-F_t(u|X)\right|\right|_{L_2} \right\}^2 + E\left[\int_0^{\tau}(G^\ast_{t}(u|X)-G_{t}(u|X))^2du\right] \\
& + E\left[\int_0^{\tau}\left(F^\ast_t(u|X)-F_t(u|X)\right)^2du\right] + E\left[\int_0^{\tau}\left(\upsilon^{\dagger *}_{t}(u|X)-\upsilon^{\dagger}_{t}(u|X)\right)^2du\right],
\end{align*}
%\begin{enumerate}
%    \item  $\sup_{u\in[0, \tau]}\left|\left|F^\ast_t(u|X)-F_t(u|X)\right|\right|_{L_2} \times \sup_{u\in[0, \tau]}\left|\left|G^\ast_{t}(u| X)-G_{t}(u| X)\right|\right|_{L_2}$, 
%    \item $\sup_{u\in[0, \tau]}\left|\left|F^\ast_t(u|X)-F_t(u|X)\right|\right|_{L_2} \times \sup_{u\in[0, \tau]}\left|\left|G^\ast_{1-t}(u| X)-G_{1-t}(u| X)\right|\right|_{L_2}$
%    \item $\sup_{u\in[0, \tau]}\left|\left|F^\ast_t(u|X)-F_t(u|X)\right|\right|_{L_2} \times \left|\left|\pi^\ast_t(X)-\pi_t(X)\right|\right|_{L_2}$
%    \item $\left\{ \sup_{u\in[0, \tau]}\left|\left|F^\ast_t(u|X)-F_t(u|X)\right|\right|_{L_2} \right\}^2$
%    \item $E\left[\int_0^{\tau}(G^\ast_{t}(u|X)-G_{t}(u|X))^2du\right]$
%    \item $E\left[\int_0^{\tau}(G^\ast_{1-t}(u|X)-G_{1-t}(u|X))^2du\right]$
%    \item $E\left[\int_0^{\tau}\left(F^\ast_t(u|X)-F_t(u|X)\right)^2du\right]$
%    \item $E\left[\int_0^{\tau}\left(\upsilon^{\dagger *}_{t}(u|X)-\upsilon^{\dagger}_{t}(u|X)\right)^2du\right]$,
%\end{enumerate}
where $\widetilde{P}^*$ is any distribution for $\widetilde{O}$  and $\left| \left| c(X) \right| \right|_{L_2} = \sqrt{ \int c(x)^2 dF_X(x)}$. 
\end{lemma}

\textbf{Proof of Lemma~\ref{rate}}  As in Lemma~\ref{robust}, let $G^\ast_\cdot(\cdot|X)=(G^\ast_t(\cdot|X), G^\ast_{1-t}(\cdot|X))$ and \newline $G^\ast_\cdot(\cdot|X)=\exp\left(- \Lambda^{\ast\dagger}_\cdot(\cdot) \exp \{ {\beta^{\ast'}_\cdot} X \}\right)$.  Here, we let  $F^\ast_{t}(\cdot|X)=1-\exp\left(- \Upsilon^{\ast\dagger}_t(\cdot) \exp \{ {\alpha^\ast_t}' X \}\right)$. For simplicity, we will sometimes abbreviate the nuisance functions by $G^*_{\cdot}\coloneq G^*_{\cdot}(\cdot|X)$, $F^*_{t}\coloneq F^*_{t}(\cdot|X)$, $\pi^*_{t}\coloneq \pi^*_{t}(X)$, $G_{\cdot}\coloneq G_{\cdot}(\cdot|X)$, $F_{t}\coloneq F_{t}(\cdot|X)$, $\pi_{t}\coloneq \pi_{t}(X)$.  %In what follows $\widetilde{P}^*$ is notationally characterized by BLAH and $\widetilde{P}^*$ is notationally characterized.  
We will use $A \lesssim B$ to indicate $A$ is less than or equal to $B$ up to a constant factor. 

%For simplicity, we will abbreviate the notations for nuisance estimations inside of a function as $G_{T}\coloneq G_{T}(u|X)$, $F_{t}\coloneq F_{t}(u|X)$ and $\pi_{t}\coloneq \pi_{t}(X)$. We will use $X\lesssim Y$ to indicate $X$ is less than or equal to $Y$ up to a constant factor. 

Applying Lemma~\ref{robust}, we have 
$$ E \left[  g_t(\widetilde{P}^*)(\widetilde{O}) - h_t(\widetilde{P}^*)(\widetilde{O})  \psi_t \right] = H_1 + H_2$$
where
{\small
\begin{align*}
    H_1 &= \left\{E\left[g_t(\widetilde{P}^*)(\widetilde{O}) - h_t(\widetilde{P}^*)(\widetilde{O}) \psi_t\right]-E\left[g_t(G_{\cdot}, F^\ast_{t}, \pi_{t})(\widetilde{O}) - h_t(\widetilde{P})(\widetilde{O}) \psi_t\right]\right\}\\
    &\hspace{0.5cm}-\left\{E\left[g_t(G^\ast_\cdot, F_{t}, \pi^\ast_{t})(\widetilde{O}) - h_t(G^\ast_\cdot)(\widetilde{O}) \psi_t\right]-E\left[g_t(\widetilde{P})(\widetilde{O}) - h_t(\widetilde{P})(\widetilde{O}) \psi_t\right]\right\}\\
    &= \left\{E\left[\left(g_t(\widetilde{P}^*)(\widetilde{O}) - g_t(G_\cdot, F^\ast_{t}, \pi_{t})(\widetilde{O})\right)-\left(g_t(G^\ast_\cdot, F_{t}, \pi^\ast_{t})(\widetilde{O})-g_t(\widetilde{P})(\widetilde{O})\right)\right]\right\}\\
    H_2 &= E\left[g_t(G_\cdot, F^\ast_{t}, \pi_{t})(\widetilde{O}) - h_t(\widetilde{P})(\widetilde{O}) \psi_t\right]
\end{align*}
}

For any fixed $F^\ddagger_{t}(\cdot|X)$, 
{\small
\begin{align*}
 &g_t(G^\ast_\cdot, F^\ddagger_{t}, \pi^\ast_{t})(\widetilde{O}) - g_t(G_\cdot, F^\ddagger_{t}, \pi_{t})(\widetilde{O})\\
 & = \frac{\Delta + (1-\Delta)I( \widetilde{Y}\geq s)}{G^\ast_{T}(\min(\widetilde{Y}, s)^-| X)} \phi_t(F^\ddagger_{t}, \pi^\ast_{t})(O) + \frac{(1-\Delta)}{\widehat{G}_{T}(\widetilde{Y}^-| X)} l_t(F^\ddagger_{t}, \pi^\ast_{t})(\widetilde{Y},T,X) - \int_0^{\widetilde{Y}} \frac{l_t(F^\ddagger_{t}, \pi^\ast_{t})(u,T,X)}{\widehat{G}_{T}(u^-|X)} \exp\{\beta^{\ast'}_{T} X \}  d\Lambda^{\ast\dagger}_{T}(u)\\
 &\hspace{0.5cm}-\frac{\Delta + (1-\Delta)I( \widetilde{Y}\geq s)}{G_{T}(\min(\widetilde{Y}, s)^-| X)} \phi_t(F^\ddagger_{t}, \pi_{t})(O) - \frac{(1-\Delta)}{G_{T}(\widetilde{Y}^-| X)} l_t(F^\ddagger_{t}, \pi_{t})(\widetilde{Y},T,X) + \int_0^{\widetilde{Y}} \frac{l_t(F^\ddagger_{t}, \pi_{t})(u,T,X)}{G_{T}(u^-|X)} \exp\{\beta_{T}' X \}  d\Lambda^{\dagger}_{T}(u) \\
 & =  \frac{\Delta + (1-\Delta)I( \widetilde{Y}\geq s)}{G^\ast_{T}(\min(\widetilde{Y}, s)^-| X)} \left\{ \phi_t(F^\ddagger_{t}, \pi^\ast_{t})(O) - \phi_t(F^\ddagger_{t}, \pi_{t})(O) \right\} \\
 &\hspace{0.5cm}+ \{\Delta + (1-\Delta)I( \widetilde{Y}\geq s)\} \left\{ \frac{1}{G^\ast_{T}(\min(\widetilde{Y}, s)^-| X)} - \frac{1}{G_{T}(\min(\widetilde{Y}, s)^-| X)} \right\}  \phi_t(F^\ddagger_{t}, \pi_{t})(O) \\
 &\hspace{0.5cm}+\frac{(1-\Delta)}{G^\ast_{T}(\widetilde{Y}^-| X)} \left\{ l_t(F^\ddagger_{t}, \pi^\ast_{t})(\widetilde{Y},T,X) - l_t(F^\ddagger_{t}, \pi_{t})(\widetilde{Y},T,X) \right\} \\
 &\hspace{0.5cm}+(1-\Delta) \left\{ \frac{1}{G^\ast_{T}(\widetilde{Y}^-| X)} - \frac{1}{G_{T}(\widetilde{Y}^-| X)} \right\}  l_t(F^\ddagger_{t}, \pi_{t})(\widetilde{Y},T,X) \\
&\hspace{0.5cm} -\left\{ \int_0^{\widetilde{Y}} \frac{l_t(F^\ddagger_{t}, \pi^\ast_{t})(u,T,X) - l_t(F^\ddagger_{t}, \pi_{t})(u,T,X)}{G^\ast_{T}(u^-|X)} \exp\{\beta^{\ast'}_{T} X \}  d\Lambda^{\ast\dagger}_{T}(u)\right. \\
 &\hspace{0.5cm}+\left. \; \; \; \int_0^{\widetilde{Y}} l_t(F^\ddagger_{t}, \pi_{t})(u,T,X) \left\{ \frac{\exp\{{\beta^{\ast'}_{T}} X \}  d\Lambda^{\ast\dagger}_{T}(u)}{G^\ast_{T}(u^-|X)} - \frac{\exp\{\beta_{T}' X \}  d\Lambda^{\dagger}_{T}(u)}{G_{T}(u^-|X)} \right\} \right\} 
\end{align*}
}

Let $H_1=H_{11}+H_{12}+H_{13}+H_{14}-H_{15}-H_{16}$ where
{\small
\begin{align*}
H_{11} &= E\left[\frac{\Delta + (1-\Delta)I( \widetilde{Y}\geq s)}{G^\ast_{T}(\min(\widetilde{Y}, s)^-| X)} \left\{ \phi_t(F^\ast_{t}, \pi^\ast_{t})(O) - \phi_t(F^\ast_{t}, \pi_{t})(O)-\phi_t(F_{t}, \pi^\ast_{t})(O) + \phi_t(F_{t}, \pi_{t})(O)\right\}\right]\\
&= E\left[\frac{\Delta + (1-\Delta)I( \widetilde{Y}\geq s)}{G^\ast_{T}(\min(\widetilde{Y}, s)^-| X)} \left\{\frac{-(\pi^\ast_t(X)-\pi_t(X))}{\pi^\ast_t(X)\pi_t(X)}\right\}I(T=t)\exp(\gamma_t)\right.\\
& \hspace{2.5cm}\left.\times\left\{\frac{I(Y \leq s)-F^\ast_t(s|X)}{\{ 1-F^\ast_t(s|X) + F^\ast_t(s|X) \exp(\gamma_t)\}^2}-\frac{I(Y \leq s)-F_t(s|X)}{\{ 1-F_t(s|X) + F_t(s|X) \exp(\gamma_t)\}^2}\right\}\right]\\
&= E\left[\frac{\Delta I(Y \leq s)}{G^\ast_{T}(\min(\widetilde{Y}, s)^-| X)} \left\{\frac{-(\pi^\ast_t(X)-\pi_t(X))}{\pi^\ast_t(X)\pi_t(X)}\right\}I(T=t)\exp(\gamma_t)\right.\\
& \hspace{2.5cm}\left.\times\left\{\frac{1}{\{ 1-F^\ast_t(s|X) + F^\ast_t(s|X) \exp(\gamma_t)\}^2}-\frac{1}{\{ 1-F_t(s|X) + F_t(s|X) \exp(\gamma_t)\}^2}\right\}\right]-\\
&\hspace{0.5cm}-E\left[\frac{\Delta + (1-\Delta)I( \widetilde{Y}\geq s)}{G^\ast_{T}(\min(\widetilde{Y}, s)^-| X)} \left\{\frac{-(\pi^\ast_t(X)-\pi_t(X))}{\pi^\ast_t(X)\pi_t(X)}\right\}I(T=t)\exp(\gamma_t)\right.\\
& \hspace{2.5cm}\left.\times\left\{\frac{F^\ast_t(s|X)}{\{ 1-F^\ast_t(s|X) + \widehat{F}_t(s|X) \exp(\gamma_t)\}^2}-\frac{F_t(s|X)}{\{ 1-F_t(s|X) + F_t(s|X) \exp(\gamma_t)\}^2}\right\}\right]\\
&= E\left[\frac{\Delta I(Y \leq s)}{G^\ast_{T}(\min(\widetilde{Y}, s)^-| X)} \left\{\frac{-(\pi^\ast_t(X)-\pi_t(X))}{\pi^\ast_t(X)\pi_t(X)}\right\}I(T=t)\exp(\gamma_t)\right.\\
& \hspace{2.5cm}\left.\times\frac{2(1-\exp(\gamma_t))\{F^\ast_t(s|X)-F_t(s|X)\}-(1-\exp(\gamma_t))^2\{F^\ast_t(s|X)+F_t(s|X)\}\{F^\ast_t(s|X)-F_t(s|X)\}}{\{ 1-F^\ast_t(s|X) +F^\ast_t(s|X) \exp(\gamma_t)\}^2\{ 1-F_t(s|X) + F_t(s|X) \exp(\gamma_t)\}^2}\right]\\
&\hspace{0.5cm}-E\left[\frac{\Delta + (1-\Delta)I( \widetilde{Y}\geq s)}{G^\ast_{T}(\min(\widetilde{Y}, s)^-| X)} \left\{\frac{-(\pi^\ast_t(X)-\pi_t(X))}{\pi^\ast_t(X)\pi_t(X)}\right\}I(T=t)\exp(\gamma_t)\right.\\
& \hspace{2.5cm}\left.\times\frac{\{F^\ast_t(s|X)-F_t(s|X)\}-(1-\exp(\gamma_t))^2F^\ast_t(s|X)F_t(s|X)\{F^\ast_t(s|X)-F_t(s|X)\}}{\{ 1-F^\ast_t(s|X) + F^\ast_t(s|X) \exp(\gamma_t)\}^2\{ 1-F_t(s|X) + F_t(s|X) \exp(\gamma_t)\}^2}\right]
\end{align*}
}
Applying Jensen's inequality and Cauchy-Schwarz inequality, we have
\begin{align*}
|H_{11}| &\lesssim  \left|E\left[\left(\pi^\ast_t(X)-\pi_t(X)\right) \left(F^\ast_t(s|X)-F_t(s|X)\right)\right]\right|\\
&\leq  \left|\left|\left(\pi^\ast_t(X)-\pi_t(X)\right) \left(F^\ast_t(s|X)-F_t(s|X)\right)\right|\right|_{L_2}\\
&\leq  \left|\left|\pi^\ast_t(X)-\pi_t(X)\right|\right|_{L_2}\left|\left|F^\ast_t(s|X)-F_t(s|X)\right|\right|_{L_2}\\
&\leq  \left|\left|\pi^\ast_t(X)-\pi_t(X)\right|\right|_{L_2}\sup_{u\in[0, \tau]}\left|\left|F^\ast_t(u|X)-F_t(u|X)\right|\right|_{L_2}
\end{align*}

Since $\phi_t(F^\ast_{t}, \pi_{t})(O) - \phi_t(F_{t}, \pi_{t})(O)=\sum_{i=1}^3 \phi_{ti}(F^\ast_{t}, \pi_{t})(O) - \phi_{ti}(F_{t}, \pi_{t})(O)$, and 
{\small
\begin{eqnarray*}
\phi_{t1}(F^\ast_{t}, \pi_{t})(O) - \phi_{t1}(F_{t}, \pi_{t})(O)&=&I(T=t) I(Y \leq s)\frac{\pi_{1-t}(X)}{\pi_t(X)}\exp(\gamma_t)\\
&&\left\{ \frac{1}{\{ 1-F^\ast_t(s|X) + F^\ast_t(s|X) \exp(\gamma_t)\}^2}-\frac{1}{\{ 1-F_t(s|X) + F_t(s|X) \exp(\gamma_t)\}^2} \right\}\\
\phi_{t2}(F^\ast_{t}, \pi_{t})(O) - \phi_{t2}(F_{t}, \pi_{t})(O)&=& - I(T=t)\frac{\pi_{1-t}(X)}{\pi_t(X)}\exp(\gamma_t)\\
&&\left\{ \frac{  F^\ast_t(s|X)}{\{ 1-F^\ast_t(s|X) + F^\ast_t(s|X) \exp(\gamma_t)\}^2}-\frac{  F_t(s|X)}{\{ 1-F_t(s|X) + F_t(s|X) \exp(\gamma_t)\}^2}\right\}\\
\phi_{t3}(F^\ast_{t}, \pi_{t})(O) - \phi_{t3}(F_{t}, \pi_{t})(O)&=& I(T=1-t)\exp(\gamma_t)\\
&&\left\{ \frac{F^\ast_t(s|X) }{1-F^\ast_t(s|X) + F^\ast_t(s|X) \exp(\gamma_t)}-\frac{F_t(s|X) }{1-F_t(s|X) + F_t(s|X) \exp(\gamma_t)} \right\}
\end{eqnarray*}
}
we have
{\small
\begin{align*}
H_{12} &= E\left[\left\{\Delta + (1-\Delta)I( \widetilde{Y}\geq s)\right\} \left\{ \frac{1}{G^\ast_{T}(\min(\widetilde{Y}, s)^-| X)} - \frac{1}{G_{T}(\min(\widetilde{Y}, s)^-| X)} \right\}  \left\{\phi_t(F^\ast_{t}, \pi_{t})(O)-\phi_t(F_{t}, \pi_{t})(O)\right\}\right]\\
&= E\left[\left\{\Delta + (1-\Delta)I( \widetilde{Y}\geq s)\right\} \left\{ \frac{-\{G^\ast_{T}(\min(\widetilde{Y}, s)^-| X)-G_{T}(\min(\widetilde{Y}, s)^-| X)\}}{G^\ast_{T}(\min(\widetilde{Y}, s)^-| X)G_{T}(\min(\widetilde{Y}, s)^-| X)} \right\}\right.\\
&\left.\hspace{1.5cm}\times\left\{I(T=t) I(Y \leq s)\frac{\pi_{1-t}(X)}{\pi_t(X)}\exp(\gamma_t)\right.\right.\\
&\left.\left.\hspace{2cm}\times\frac{2(1-\exp(\gamma_t))\{F^\ast_t(s|X)-F_t(s|X)\}-(1-\exp(\gamma_t))^2\{F^\ast_t(s|X)+F_t(s|X)\}\{F^\ast_t(s|X)-F_t(s|X)\}}{\{ 1-F^\ast_t(s|X) + F^\ast_t(s|X) \exp(\gamma_t)\}^2\{ 1-F_t(s|X) + F_t(s|X) \exp(\gamma_t)\}^2}\right.\right.\\
&\left.\left.\hspace{1.5cm}- I(T=t)\frac{\pi_{1-t}(X)}{\pi_t(X)}\exp(\gamma_t)\right.\right.\\
&\left.\left.\hspace{2cm}\times\frac{\{F^\ast_t(s|X)-F_t(s|X)\}-(1-\exp(\gamma_t))^2F^\ast_t(s|X)F_t(s|X)\{F^\ast_t(s|X)-F_t(s|X)\}}{\{ 1-F^\ast_t(s|X) + F^\ast_t(s|X) \exp(\gamma_t)\}^2\{ 1-F_t(s|X) + F_t(s|X) \exp(\gamma_t)\}^2}\right.\right.\\
&\left.\left.\hspace{1.5cm}+\frac{I(T=1-t)\exp(\gamma_t)\{F^\ast_t(s|X)-F_t(s|X)\}}{\{ 1-F^\ast_t(s|X) + F^\ast_t(s|X) \exp(\gamma_t)\}\{ 1-F_t(s|X) + F_t(s|X) \exp(\gamma_t)\}}\right\}\right]
\end{align*}
}
Applying Jensen's inequality and Cauchy-Schwarz inequality, we have
\begin{align*}
|H_{12}| &\lesssim \sum_{t'\in\{0, 1\}}\left|E\left[\left(G^\ast_{t'}(\min(\widetilde{Y}, s)^-| X)-G_{t'}(\min(\widetilde{Y}, s)^-| X)\right) \left(F^\ast_t(s|X)-F_t(s|X)\right)\right]\right|\\
&\leq \sum_{t'\in\{0, 1\}}\left|\left|\left(G^\ast_{t'}(\min(\widetilde{Y}, s)^-| X)-G_{t'}(\min(\widetilde{Y}, s)^-| X)\right) \left(F^\ast_t(s|X)-F_t(s|X)\right)\right|\right|_{L_2}\\
&\leq \sum_{t'\in\{0, 1\}}\left|\left|G^\ast_{t'}(\min(\widetilde{Y}, s)^-| X)-G_{t'}(\min(\widetilde{Y}, s)^-| X)\right|\right|_{L_2}\left|\left|F^\ast_t(s|X)-F_t(s|X)\right|\right|_{L_2}\\
&\leq \sum_{t'\in\{0, 1\}}\sup_{u\in[0, s]}\left|\left|G^\ast_{t'}(u^-| X)-G_{t'}(u^-| X)\right|\right|_{L_2}\left|\left|F^\ast_t(s|X)-F_t(s|X)\right|\right|_{L_2}\\
&\leq \sum_{t'\in\{0, 1\}}\sup_{u\in[0, \tau]}\left|\left|G^\ast_{t'}(u^-| X)-G_{t'}(u^-| X)\right|\right|_{L_2}\sup_{u\in[0, \tau]}\left|\left|F^\ast_t(u|X)-F_t(u|X)\right|\right|_{L_2}
\end{align*}
Since for any fixed $F^\ddagger_{t}(\cdot|X)$, $l_t(F^\ddagger_{t}, \pi^\ast_{t})(\widetilde{Y},T,X) - l_t(F^\ddagger_{t}, \pi_{t})(\widetilde{Y},T,X)=\sum_{i=1}^{3}  l_{ti}(F^\ddagger_{t}, \pi^\ast_{t})(\widetilde{Y},T,X) - l_{ti}(F^\ddagger_{t}, \pi_{t})(\widetilde{Y},T,X)$, and 
{\small
\begin{eqnarray*}
    l_{t1}(F^\ddagger_{t}, \pi^\ast_{t})(\widetilde{Y},T,X) - l_{t1}(F^\ddagger_{t}, \pi_{t})(\widetilde{Y},T,X) &=& I(T=t)  I(\widetilde{Y} \leq s) \frac{ F^\ddagger_t(s|X) - F^\ddagger_t(\widetilde{Y}^-|X) }{1-F^\ddagger_t(\widetilde{Y}^-|X)}\times\\
    &&\frac{ \exp(\gamma_t)}{\{ 1-F^\ddagger_t(s|X) + F^\ddagger_t(s|X) \exp(\gamma_t)\}^2}\times\frac{-(\pi^\ast_t(X)-\pi_t(X))}{\pi^\ast_t(X)\pi_t(X)}\\
    l_{t2}(F^\ddagger_{t}, \pi^\ast_{t})(\widetilde{Y},T,X) - l_{t2}(F^\ddagger_{t}, \pi_{t})(\widetilde{Y},T,X) &=& I(\widetilde{Y}<s) \left\{\phi_{t2}(F^\ddagger_{t}, \pi^\ast_{t})(O)-\phi_{t2}(F^\ddagger_{t}, \pi_{t})(O)\right\}\\
    &=& - \frac{ I(T=t)I(\widetilde{Y}<s)F^\ddagger_t(s|X) \exp(\gamma_t)}{\{ 1-F^\ddagger_t(s|X) + F^\ddagger_t(s|X) \exp(\gamma_t)\}^2}\times  \frac{-(\pi^\ast_t(X)-\pi_t(X))}{\pi^\ast_t(X)\pi_t(X)}\\
    l_{t3}(F^\ddagger_{t}, \pi^\ast_{t})(\widetilde{Y},T,X) - l_{t3}(F^\ddagger_{t}, \pi_{t})(\widetilde{Y},T,X) &=& 0
\end{eqnarray*}
}
we have
{\small
\begin{align*}
    H_{13} &= E\left[\frac{(1-\Delta)}{G^\ast_{T}(\widetilde{Y}^-| X)} \left\{ l_t(F^\ast_{t}, \pi^\ast_{t})(\widetilde{Y},T,X)- l_t(F^\ast_{t}, \pi_{t})(\widetilde{Y},T,X)-l_t(F_{t}, \pi^\ast_{t})(\widetilde{Y},T,X)+ l_t(F_{t}, \pi_{t})(\widetilde{Y},T,X)\right\}\right]\\
    &= E\left[\frac{I(T=t)(1-\Delta)}{G^\ast_{T}(\widetilde{Y}^-| X)}\exp(\gamma_t)\times  \frac{-(\pi^\ast_t(X)-\pi_t(X))}{\pi^\ast_t(X)\pi_t(X)}\right.\\
    &\hspace{3cm}\left.\times\left\{I(\widetilde{Y} \leq s) \left\{\frac{F^\ast_t(s|X) - F^\ast_t(\widetilde{Y}^-|X)}{\{1-F^\ast_t(\widetilde{Y}^-|X)\}\{ 1-F^\ast_t(s|X) + F^\ast_t(s|X) \exp(\gamma_t)\}^2}\right.\right.\right.\\
    &\hspace{5cm}\left.\left.\left.-\frac{F_t(s|X) - F_t(\widetilde{Y}-|X)}{\{1-F_t(\widetilde{Y}-|X)\}\{ 1-F_t(s|X) + F_t(s|X) \exp(\gamma_t)\}^2}\right\}\right.\right.\\
    &\hspace{3cm}\left.\left.-I(\widetilde{Y}<s)\left\{\frac{ F^\ast_t(s|X)}{\{ 1-F^\ast_t(s|X) + F^\ast_t(s|X) \exp(\gamma_t)\}^2}-\frac{ F_t(s|X)}{\{ 1-F_t(s|X) + F_t(s|X) \exp(\gamma_t)\}^2}\right\}\right\}\right]
\end{align*}
}
Let $H_{13}=H_{131}-H_{132}$, and 
{\small
\begin{align*}
    H_{131} &= E\left[\frac{I(T=t)(1-\Delta)}{G^\ast_{T}(\widetilde{Y}^-| X)}\exp(\gamma_t)\times  \frac{-(\pi^\ast_t(X)-\pi_t(X))}{\pi^\ast_t(X)\pi_t(X)}\right.\\
    &\hspace{3cm}\left.\times I(\widetilde{Y} \leq s) \left\{\frac{F^\ast_t(s|X) - F^\ast_t(\widetilde{Y}^-|X)}{\{1-F^\ast_t(\widetilde{Y}^-|X)\}\{ 1-F^\ast_t(s|X) + F^\ast_t(s|X) \exp(\gamma_t)\}^2}\right.\right.\\
    &\hspace{5cm}\left.\left.-\frac{F_t(s|X) - F_t(\widetilde{Y}^-|X)}{\{1-F_t(\widetilde{Y}^-|X)\}\{ 1-F_t(s|X) + F_t(s|X) \exp(\gamma_t)\}^2}\right\}\right]\\
    &= E\left[\frac{I(T=t)(1-\Delta)}{G^\ast_{T}(\widetilde{Y}^-| X)}\exp(\gamma_t)\times  \frac{-(\pi^\ast_t(X)-\pi_t(X))}{\pi^\ast_t(X)\pi_t(X)}\times I(\widetilde{Y} \leq s)\right.\\
    &\hspace{2cm}\left.\times\left\{\frac{F^\ast_t(s|X)- F^\ast_t(\widetilde{Y}^-|X)}{1-F^\ast_t(\widetilde{Y}^-|X)}\left\{\frac{1}{\{ 1-F^\ast_t(s|X) + F^\ast_t(s|X) \exp(\gamma_t)\}^2}-\frac{1}{\{ 1-F_t(s|X) + F_t(s|X) \exp(\gamma_t)\}^2}\right\}\right.\right.\\
    &\hspace{3cm}\left.\left.+\frac{1}{\{ 1-F_t(s|X) + F_t(s|X) \exp(\gamma_t)\}^2}\left\{\frac{F^\ast_t(s|X) - F^\ast_t(\widetilde{Y}^-|X)}{1-F^\ast_t(\widetilde{Y}^-|X)}-\frac{F_t(s|X) - F_t(\widetilde{Y}^-|X)}{1-F_t(\widetilde{Y}^-|X)}\right\}\right\}\right]\\
    &= E\left[\frac{I(T=t)(1-\Delta)}{G^\ast_{T}(\widetilde{Y}^-| X)}\exp(\gamma_t)\times  \frac{-(\pi^\ast_t(X)-\pi_t(X))}{\pi^\ast_t(X)\pi_t(X)}\times I(\widetilde{Y} \leq s) \left\{\frac{F^\ast_t(s|X) - F^\ast_t(\widetilde{Y}^-|X)}{1-F^\ast_t(\widetilde{Y}^-|X)}\right.\right.\\
    &\hspace{1cm}\left.\left.\times\left\{\frac{2(1-\exp(\gamma_t))\{F^\ast_t(s|X)-F_t(s|X)\}-(1-\exp(\gamma_t))^2\{F^\ast_t(s|X)+F_t(s|X)\}\{F^\ast_t(s|X)-F_t(s|X)\}}{\{ 1-F^\ast_t(s|X) + F^\ast_t(s|X) \exp(\gamma_t)\}^2\{ 1-F_t(s|X) + F_t(s|X) \exp(\gamma_t)\}^2}\right\}\right.\right.\\
    &\hspace{2cm}\left.\left.+\frac{1}{\{ 1-F_t(s|X) + F_t(s|X) \exp(\gamma_t)\}^2}\right.\right.\\
    &\hspace{3cm}\left.\left.\times\left\{\frac{(1-F_t(\widetilde{Y}^-|X))\{F^\ast_t(s|X)-F_t(s|X)\}-(1-F_t(s|X))\{F^\ast_t(\widetilde{Y}^-|X)-F_t(\widetilde{Y}^-|X)\}}{\{1-F^\ast_t(\widetilde{Y}^-|X)\}\{1-F_t(\widetilde{Y}^-|X)\}}\right\}\right\}\right]
\end{align*}
}
Applying triangle inequality, Jensen's inequality and Cauchy-Schwarz inequality, we have
{\small
\begin{align*}
\left|H_{131}\right|\lesssim&\left|E\left[\left(\pi^\ast_t(X)-\pi_t(X)\right) \left(F^\ast_t(s|X)-F_t(s|X)\right)\right]+E\left[\left(\pi^\ast_t(X)-\pi_t(X)\right)\left(F^\ast_t(\widetilde{Y}^-|X)-F_t(\widetilde{Y}^-|X)\right)\right]\right|\\
\leq&\left\{\left|E\left[\left(\pi^\ast_t(X)-\pi_t(X)\right) \left(F^\ast_t(s|X)-F_t(s|X)\right)\right]\right|+\left|E\left[\left(\pi^\ast_t(X)-\pi_t(X)\right)\left(F^\ast_t(\widetilde{Y}^-|X)-F_t(\widetilde{Y}^-|X)\right)\right]\right|\right\}\\
\leq&\left\{\left|\left|\left(\pi^\ast_t(X)-\pi_t(X)\right) \left(F^\ast_t(s|X)-F_t(s|X)\right)\right|\right|_{L_2}+\left|\left|\left(\pi^\ast_t(X)-\pi_t(X)\right)\left(F^\ast_t(\widetilde{Y}^-|X)-F_t(\widetilde{Y}^-|X)\right)\right|\right|_{L_2}\right\}\\
\leq&\left\{\left|\left|\pi^\ast_t(X)-\pi_t(X)\right|\right|_{L_2} \left|\left|F^\ast_t(s|X)-F_t(s|X)\right|\right|_{L_2}+\left|\left|\pi^\ast_t(X)-\pi_t(X)\right|\right|_{L_2}\left|\left|F^\ast_t(\widetilde{Y}^-|X)-F_t(\widetilde{Y}^-|X)\right|\right|_{L_2}\right\}\\
\leq&\left|\left|\pi^\ast_t(X)-\pi_t(X)\right|\right|_{L_2} \sup_{u\in[0, s]}\left|\left|F^\ast_t(u|X)-F_t(u|X)\right|\right|_{L_2}\\
\leq&\left\{\left|\left|\pi^\ast_t(X)-\pi_t(X)\right|\right|_{L_2} \left|\left|F^\ast_t(s|X)-F_t(s|X)\right|\right|_{L_2}+\left|\left|\pi^\ast_t(X)-\pi_t(X)\right|\right|_{L_2}\left|\left|F^\ast_t(\widetilde{Y}^-|X)-F_t(\widetilde{Y}^-|X)\right|\right|_{L_2}\right\}\\
\lesssim&\left|\left|\pi^\ast_t(X)-\pi_t(X)\right|\right|_{L_2} \sup_{u\in[0, \tau]}\left|\left|F^\ast_t(u|X)-F_t(u|X)\right|\right|_{L_2}
\end{align*}
}

{\small
\begin{align*}
H_{132} &= E\left[\frac{I(T=t)(1-\Delta)}{G^\ast_{T}(\widetilde{Y}^-| X)}\exp(\gamma_t)\times  \frac{-(\pi^\ast_t(X)-\pi_t(X))}{\pi^\ast_t(X)\pi_t(X)}\times I(\widetilde{Y}<s)\right.\\
&\hspace{3cm}\left.\times\left\{\frac{ F^\ast_t(s|X)}{\{ 1-F^\ast_t(s|X) + F^\ast_t(s|X) \exp(\gamma_t)\}^2}-\frac{ F_t(s|X)}{\{ 1-F_t(s|X) + F_t(s|X) \exp(\gamma_t)\}^2}\right\}\right]\\
&= E\left[\frac{I(T=t)(1-\Delta)}{G^\ast_{T}(\widetilde{Y}^-| X)}\exp(\gamma_t)\times  \frac{-(\pi^\ast_t(X)-\pi_t(X))}{\pi^\ast_t(X)\pi_t(X)}\times I(\widetilde{Y}<s)\right.\\
&\hspace{3cm}\left.\left.\times\left\{\frac{\{F^\ast_t(s|X)-F_t(s|X)\}-(1-\exp(\gamma_t))^2F^\ast_t(s|X)F_t(s|X)\{F^\ast_t(s|X)-F_t(s|X)\}}{\{ 1-F^\ast_t(s|X) + F^\ast_t(s|X) \exp(\gamma_t)\}^2\{ 1-F_t(s|X) + F_t(s|X) \exp(\gamma_t)\}^2}\right\}\right\}\right]
\end{align*}
}
Applying Jensen's inequality and Cauchy-Schwarz inequality, we have
\begin{eqnarray*}
|H_{132}| \lesssim  \left|\left|\pi^\ast_t(X)-\pi_t(X)\right|\right|_{L_2} \sup_{u\in[0, \tau]}\left|\left|F^\ast_t(u|X)-F_t(u|X)\right|\right|_{L_2}
\end{eqnarray*}

Since $l_t(F^\ast_{t}, \pi_{t})(\widetilde{Y},T,X) - l_t(F_{t}, \pi_{t})(\widetilde{Y},T,X)=\sum_{i=1}^{3}  l_{ti}(F^\ast_{t}, \pi_{t})(\widetilde{Y},T,X) - l_{ti}(F_{t}, \pi_{t})(\widetilde{Y},T,X)$, and 
{\small
\begin{eqnarray*}
&&l_{t1}(F^\ast_{t}, \pi_{t})(\widetilde{Y},T,X) - l_{t1}(F_{t}, \pi_{t})(\widetilde{Y},T,X) \\
&& \hspace{1cm}=I(T=t)  I(\widetilde{Y} \leq s) \left\{\frac{ F^\ast_t(s|X) - F^\ast_t(\widetilde{Y}^-|X) }{1-F^\ast_t(\widetilde{Y}^-|X)}-\frac{ F_t(s|X) - F_t(\widetilde{Y}^-|X) }{1-F_t(\widetilde{Y}^-|X)}\right\} \\
& &\hspace{1.5cm}+I(T=t)  I(\widetilde{Y} \leq s)\frac{\pi_{1-t}(X)}{\pi_t(X)}\exp(\gamma_t)\\
&&\hspace{1.5cm}\times\left\{ \frac{F^\ast_t(s|X) - F^\ast_t(\widetilde{Y}^-|X)}{\{ 1-F^\ast_t(s|X) + F^\ast_t(s|X) \exp(\gamma_t)\}^2\{1-F^\ast_t(\widetilde{Y}^-|X)\}}\right.\\
&&\left.\hspace{3cm}-\frac{F_t(s|X)- F_t(\widetilde{Y}^-|X)}{\{ 1-F_t(s|X) + F_t(s|X) \exp(\gamma_t)\}^2\{1-F_t(\widetilde{Y}^-|X)\}} \right\}\\
&&l_{t2}(F^\ast_{t}, \pi_{t})(\widetilde{Y},T,X) - l_{t2}(F_{t}, \pi_{t})(\widetilde{Y},T,X)\\
&& \hspace{1cm}=- I(T=t)I(\widetilde{Y} < s)\frac{\pi_{1-t}(X)}{\pi_t(X)}\exp(\gamma_t)\\
&&\hspace{3cm}\times\left\{ \frac{  F^\ast_t(s|X)}{\{ 1-F^\ast_t(s|X) + F^\ast_t(s|X) \exp(\gamma_t)\}^2}-\frac{  F_t(s|X)}{\{ 1-F_t(s|X) + F_t(s|X) \exp(\gamma_t)\}^2}\right\}\\
&&l_{t3}(F^\ast_{t}, \pi_{t})(\widetilde{Y},T,X) - l_{t3}(F_{t}, \pi_{t})(\widetilde{Y},T,X)\\
&& \hspace{1cm}=I(T=1-t)I(\widetilde{Y} < s)\exp(\gamma_t)\times\left\{ \frac{F^\ast_t(s|X) }{1-F^\ast_t(s|X) + F^\ast_t(s|X) \exp(\gamma_t)}-\frac{F_t(s|X) }{1-F_t(s|X) + F_t(s|X) \exp(\gamma_t)} \right\}
\end{eqnarray*}
}
we have
{\small
\begin{eqnarray*}
    H_{14} &=& E\left[(1-\Delta) \left\{ \frac{1}{G^\ast_{T}(\widetilde{Y}^-| X)} - \frac{1}{G_{T}(\widetilde{Y}^-| X)} \right\}  \left\{l_t(F^\ast_{t}, \pi_{t})(\widetilde{Y},T,X)-l_t(F_{t}, \pi_{t})(\widetilde{Y},T,X)\right\}\right]\\
    &=& E\left[(1-\Delta) \left\{ \frac{-(G^\ast_{T}(\widetilde{Y}^-| X)-G_{T}(\widetilde{Y}^-| X))}{G^\ast_{T}(\widetilde{Y}^-| X)G_{T}(\widetilde{Y}^-| X)}\right\}  \left\{l_t(F^\ast_{t}, \pi_{t})(\widetilde{Y},T,X)-l_t(F_{t}, \pi_{t})(\widetilde{Y},T,X)\right\}\right]
\end{eqnarray*}
}
Applying the same strategies used for $H_{12}$ and $H_{13}$, we have 
\begin{eqnarray*}
    \left|H_{14}\right| \lesssim  \sum_{t'\in\{0, 1\}}\left|\left|G^\ast_{t'}(\widetilde{Y}^-| X)-G_{t'}(\widetilde{Y}^-| X)\right|\right|_{L_2}\sup_{u\in[0, s]}\left|\left|F^\ast_t(u|X)-F_t(u|X)\right|\right|_{L_2}
\end{eqnarray*}

Since each term in $l_t(F^\ast_{t}, \pi_{t})(\widetilde{Y},T,X)-l_t(F_{t}, \pi_{t})(\widetilde{Y},T,X)$ is multiplied by $I(\widetilde{Y}\leq s)$ or $I(\widetilde{Y}< s)$, we have 
\begin{eqnarray*}
    \left|H_{14}\right| &\lesssim&  \sum_{t'\in\{0, 1\}}\sup_{u\in[0, s]}\left|\left|G^\ast_{t'}(u^-| X)-G_{t'}(u^-| X)\right|\right|_{L_2}\sup_{u\in[0, s]}\left|\left|F^\ast_t(u|X)-F_t(u|X)\right|\right|_{L_2}\\
    &\leq&\sum_{t'\in\{0, 1\}}\sup_{u\in[0, \tau]}\left|\left|G^\ast_{t'}(u^-| X)-G_{t'}(u^-| X)\right|\right|_{L_2}\sup_{u\in[0, \tau]}\left|\left|F^\ast_t(u|X)-F_t(u|X)\right|\right|_{L_2}
\end{eqnarray*}
{\small
\begin{align*}
    H_{15} &= E\left[\int_0^{\widetilde{Y}} \frac{l_t(F^\ast_{t}, \pi^\ast_{t})(u,T,X) - l_t(F^\ast_{t}, \pi_{t})(u,T,X)-l_t(F_{t}, \pi^\ast_{t})(u,T,X)+ l_t(F_{t}, \pi_{t})(u,T,X)}{G^\ast_{T}(u^-|X)}\times\exp\{{\beta^{\ast'}_{T}} X \}  d\Lambda^{\ast\dagger}_{T}(u)\right]
\end{align*}
}
Using (\ref{prop_int}) and the fact that time is continuous, we have 
{\small
\begin{align*}
    \left|H_{15}\right| &\lesssim E\left[\left|\pi^\ast_t(X)-\pi_t(X)\right| \left|F^\ast_t(s|X)-F_t(s|X)\right|\sup_{u\in[0, s]}\left|\frac{1}{G^\ast_{T}(u^-|X)}\right|\left|\Lambda^{\ast\dagger}_{T}(s)-\Lambda^{\ast\dagger}_{T}(0)\right|\right]\\
    &\hspace{1cm}+E\left[\left|\pi^\ast_t(X)-\pi_t(X)\right|\sup_{u\in[0, s]}\left|\frac{F^\ast_t(u^-|X)-F_t(u^-|X)}{G^\ast_{T}(u^-|X)}\right|\left|\Lambda^{\ast\dagger}_{T}(s)-\Lambda^{\ast\dagger}_{T}(0)\right|\right]\\
    &\lesssim E\left[\left|\pi^\ast_t(X)-\pi_t(X)\right|\sup_{u\in[0, s]}\left|F^\ast_t(u^-|X)-F_t(u^-|X)\right|\right]\\
    &\leq \left|\left|\pi^\ast_t(X)-\pi_t(X)\right|\right|_{L_2}\sup_{u\in[0, s]}\left|\left|F^\ast_t(u^-|X)-F_t(u^-|X)\right|\right|_{L_2}\\
    &\leq \left|\left|\pi^\ast_t(X)-\pi_t(X)\right|\right|_{L_2}\sup_{u\in[0, \tau]}\left|\left|F^\ast_t(u^-|X)-F_t(u^-|X)\right|\right|_{L_2}
\end{align*}
}

{\small
\[
    H_{16} = E\left[\int_0^{\widetilde{Y}} \left\{l_t(F^\ast_{t}, \pi_{t})(u,T,X)-l_t(F_{t}, \pi_{t})(u,T,X)\right\} \left\{ \frac{\exp\{{\beta^{\ast'}_{T}} X \}  d\Lambda^{\ast\dagger}_{T}(u)}{G^\ast_{T}(u^-|X)} - \frac{\exp\{\beta_{T}' X \}  d\Lambda^{\dagger}_{T}(u)}{G_{T}(u^-|X)} \right\}\right]
\]
}

Using $\exp\{\beta^{\ddagger'}_{T} X \}  d\Lambda^{\ddagger\dagger}_{T}(u)=d\Lambda^{\ddagger\dagger}_{T}(u|X)=-d\log G^\ddagger_{T}(u|X)=-d\log G^\ddagger_{T}(u^-|X)=-\frac{dG^\ddagger_{T}(u^-|X)}{G^\ddagger_{T}(u^-|X)}$ for any fixed $\beta^{\ddagger'}_{T}$, $\Lambda^{\ddagger\dagger}_{T}(u)$, $\Lambda^{\ddagger\dagger}_{T}(u|X)$, $G^\ddagger_T(u|X)$ and integration by parts, we show that

\begin{eqnarray*}
    &&-\int_0^{\widetilde{Y}} \left\{l_t(F^\ast_{t}, \pi_{t})(u,T,X)-l_t(F_{t}, \pi_{t})(u,T,X)\right\} \left\{ \frac{dG^\ast_{T}(u^-|X)}{{G^\ast_{T}}(u^-|X)^2} - \frac{dG_{T}(u^-|X)}{G_{T}(u^-|X)^2} \right\}\\
    &=& -\left[\left\{ \frac{1}{G^\ast_{T}(u^-|X)} - \frac{1}{G_{T}(u^-|X)} \right\}\left\{l_t(F^\ast_{t}, \pi_{t})(u,T,X)-l_t(F_{t}, \pi_{t})(u,T,X)\right\}\right]\Bigg|_{0}^{\widetilde{Y}}\\
    &&- \int_0^{\widetilde{Y}}\left\{ \frac{1}{G^\ast_{T}(u^-|X)} - \frac{1}{G_{T}(u^-|X)} \right\}\frac{\partial}{\partial u}\left[l_t(F^\ast_{t}, \pi_{t})(u,T,X)-l_t(F_{t}, \pi_{t})(u,T,X)\right]du\\
    &=& -\left\{ \frac{1}{G^\ast_{T}(\widetilde{Y}^-|X)} - \frac{1}{G_{T}(\widetilde{Y}^-|X)} \right\}\left\{l_t(F^\ast_{t}, \pi_{t})(\widetilde{Y},T,X)-l_t(F_{t}, \pi_{t})(\widetilde{Y},T,X)\right\}\\
    && -\int_0^{\widetilde{Y}}\left\{ \frac{1}{G^\ast_{T}(u^-|X)} - \frac{1}{G_{T}(u^-|X)} \right\}\frac{\partial}{\partial u}\left[l_t(F^\ast_{t}, \pi_{t})(u,T,X)-l_t(F_{t}, \pi_{t})(u,T,X)\right]du
\end{eqnarray*}

Let $H_{16}=-H_{161}-H_{162}$, 
\begin{eqnarray*}
    H_{161} &=& E\left[\left\{ \frac{1}{G^\ast_{T}(\widetilde{Y}^-|X)} - \frac{1}{G_{T}(\widetilde{Y}^-|X)} \right\}\left\{l_t(F^\ast_{t}, \pi_{t})(\widetilde{Y},T,X)-l_t(F_{t}, \pi_{t})(\widetilde{Y},T,X)\right\}\right]\\
    &=& E\left[\left\{ \frac{-(G^\ast_{T}(\widetilde{Y}^-|X)-G_{T}(\widetilde{Y}^-|X))}{G^\ast_{T}(\widetilde{Y}^-|X)G_{T}(\widetilde{Y}^-|X)} \right\}\left\{l_t(F^\ast_{t}, \pi_{t})(\widetilde{Y},T,X)-l_t(F_{t}, \pi_{t})(\widetilde{Y},T,X)\right\}\right]
\end{eqnarray*}
Using the same strategies in $H_{12}$ and $H_{13}$ to deal with $l_t(F^\ast_{t}, \pi_{t})(\widetilde{Y},T,X)-l_t(F_{t}, \pi_{t})(\widetilde{Y},T,X)$, we have 
\begin{eqnarray*}
    \left|H_{161}\right| &\lesssim&  \sum_{t'\in\{0, 1\}}\left|\left|G^\ast_{t'}(\widetilde{Y}^-|X)-G_{t'}(\widetilde{Y}^-| X)\right|\right|_{L_2}\sup_{u\in[0, s]}\left|\left|F^\ast_t(u|X)-F_t(u|X)\right|\right|_{L_2}
\end{eqnarray*}
Since each term in $l_t(F^\ast_{t}, \pi_{t})(\widetilde{Y},T,X)-l_t(F_{t}, \pi_{t})(\widetilde{Y},T,X)$ is multiplied by $I(\widetilde{Y}\leq s)$ or $I(\widetilde{Y}< s)$, we have 
\begin{eqnarray*}
    \left|H_{161}\right| &\lesssim&  \sum_{t'\in\{0, 1\}}\sup_{u\in[0, s]}\left|\left|G^\ast_{t'}(u^-| X)-G_{t'}(u^-| X)\right|\right|_{L_2}\sup_{u\in[0, s]}\left|\left|F^\ast_t(u|X)-F_t(u|X)\right|\right|_{L_2}\\
    &\leq&  \sum_{t'\in\{0, 1\}}\sup_{u\in[0, \tau]}\left|\left|G^\ast_{t'}(u^-| X)-G_{t'}(u^-| X)\right|\right|_{L_2}\sup_{u\in[0, \tau]}\left|\left|F^\ast_t(u|X)-F_t(u|X)\right|\right|_{L_2}
\end{eqnarray*}
Since 
\begin{eqnarray*}
    \frac{\partial}{\partial u}\left[l_t(F^\ast_{t}, \pi_{t})(u,T,X)-l_t(F_{t}, \pi_{t})(u,T,X)\right]&=&\sum_{i=1}^3\frac{\partial}{\partial u}\left[l_{ti}(F^\ast_{t}, \pi_{t})(u,T,X)-l_{ti}(F_{t}, \pi_{t})(u,T,X)\right]\\
    &=& \sum_{i=1}^3\left\{\frac{\partial}{\partial u}l_{ti}(F^\ast_{t}, \pi_{t})(u,T,X)-\frac{\partial}{\partial u}l_{ti}(F_{t}, \pi_{t})(u,T,X)\right\}
\end{eqnarray*}
and for fixed $F^\ddagger_{t}(u|X)$, 
\begin{eqnarray*}
    \frac{\partial}{\partial u}l_{t1}(F^\ddagger_{t}, \pi_{t})(u,T,X)&=&I(T=t)  I(u \leq s) \left\{1+  \frac{\pi_{1-t}(X)}{\pi_t(X)} \frac{ \exp(\gamma_t)}{\{ 1-F^\ast_t(s|X) + F^\ast_t(s|X) \exp(\gamma_t)\}^2} \right\}\times\\
    &&-\frac{ (1-F^\ast_t(s|X))f^\ast_t(u^-|X) }{(1-F^\ast_t(u^-|X))^2}\\
    \frac{\partial}{\partial u}l_{t2}(F^\ddagger_{t}, \pi_{t})(u,T,X)&=&0\\
    \frac{\partial}{\partial u}l_{t3}(F^\ddagger_{t}, \pi_{t})(u,T,X)&=&0
\end{eqnarray*}

So 
{\small
\begin{eqnarray*}
&&\frac{\partial}{\partial u}\left[l_t(F^\ast_{t}, \pi_{t})(u,T,X)-l_t(F_{t}, \pi_{t})(u,T,X)\right]=\frac{\partial}{\partial u}l_{t1}(F^\ast_{t}, \pi_{t})(u,T,X)-\frac{\partial}{\partial u}l_{t1}(F_{t}, \pi_{t})(u,T,X)\\
&=& -I(T=t)I(u \leq s)\left\{\frac{ (1-F^\ast_t(s|X))F^\ast_t(u^-|X) }{(1-F^\ast_t(u^-|X))^2}-\frac{ (1-F_t(s|X))f_t(u^-|X) }{(1-F_t(u^-|X))^2}\right\}\\
&&-I(T=t)I(u \leq s)\frac{\pi_{1-t}(X)}{\pi_t(X)}\frac{ (1-F^\ast_t(s|X))F^\ast_t(u^-|X) }{(1-F^\ast_t(u^-|X))^2}\exp(\gamma_t)\times\\
&&\hspace{1.5cm}\times\left\{\frac{1}{\{ 1-F^\ast_t(s|X) + F^\ast_t(s|X) \exp(\gamma_t)\}^2}-\frac{1}{\{ 1-F_t(s|X) + F_t(s|X) \exp(\gamma_t)\}^2}\right\}\\
&&-I(T=t)I(u \leq s)\frac{\pi_{1-t}(X)}{\pi_t(X)}\frac{ \exp(\gamma_t)}{\{ 1-F_t(s|X) + F_t(s|X) \exp(\gamma_t)\}^2}\\
&&\hspace{3cm}\times\left\{\frac{ (1-F^\ast_t(s|X))F^\ast_t(u^-|X) }{(1-F^\ast_t(u^-|X))^2}-\frac{ (1-F_t(s|X))f_t(u^-|X) }{(1-F_t(u^-|X))^2}\right\}
\end{eqnarray*}
}

\begin{eqnarray*}
    H_{162} &=& E\left[\int_0^{\widetilde{Y}}\left\{ \frac{1}{G^\ast_{T}(u^-|X)} - \frac{1}{G_{T}(u^-|X)} \right\}\frac{\partial}{\partial u}\left[l_t(F^\ast_{t}, \pi_{t})(u,T,X)-l_t(F_{t}, \pi_{t})(u,T,X)\right]du\right]
\end{eqnarray*}

Let $H_{162}=-H_{1621}-H_{1622}-H_{1623}$, where
{\small
\begin{align*}
    H_{1621}&=E\left[\int_0^{\widetilde{Y}}I(T=t)I(u \leq s)\left\{ \frac{1}{G^\ast_{T}(u^-|X)} - \frac{1}{G_{T}(u^-|X)} \right\}\right.\\
    &\hspace{5cm}\left.\times\left\{\frac{ (1-F^\ast_t(s|X))F^\ast_t(u^-|X) }{(1-F^\ast_t(u^-|X))^2}-\frac{ (1-F_t(s|X))f_t(u^-|X) }{(1-F_t(u^-|X))^2}\right\}du\right]\\
    &= E\left[\int_0^{\widetilde{Y}}\frac{-I(T=t)I(u \leq s)(G^\ast_{T}(u^-|X)-G_{T}(u^-|X))}{G^\ast_{T}(u^-|X)G_{T}(u^-|X)}\right.\\
    &\hspace{5cm}\left.\times\left\{\frac{ (1-F^\ast_t(s|X)) }{(1-F^\ast_t(u^-|X))^2}dF^\ast_t(u^-|X)-\frac{ (1-F_t(s|X)) }{(1-F_t(u^-|X))^2}dF_t(u^-|X)\right\}\right]\\
    &=E\left[\int_0^{\widetilde{Y}}\frac{-I(T=t)I(u \leq s)(G^\ast_{T}(u^-|X)-G_{T}(u^-|X))}{G^\ast_{T}(u^-|X)G_{T}(u^-|X)}\right.\\
    &\hspace{5cm}\left.\times\left\{ (1-F^\ast_t(s|X))d\left(\frac{1}{1-F^\ast_t(u^-|X)}\right)-(1-F_t(s|X))d\left(\frac{1}{1-F_t(u^-|X)}\right)\right\}\right]\\
    &= E\left[\int_0^{\min(\widetilde{Y}, s)}\frac{-I(T=t)(G^\ast_{T}(u^-|X)-G_{T}(u^-|X))}{G^\ast_{T}(u^-|X)G_{T}(u^-|X)}\right.\\
    &\hspace{5cm}\left.\times\left\{ (1-F^\ast_t(s|X))d\left(\frac{1}{1-F^\ast_t(u^-|X)}\right)-(1-F_t(s|X))d\left(\frac{1}{1-F_t(u^-|X)}\right)\right\}\right]\\
    &= E\left[\int_0^{\min(\widetilde{Y}, s)}\frac{-I(T=t)(G^\ast_{T}(u^-|X)-G_{T}(u^-|X))}{G^\ast_{T}(u^-|X)G_{T}(u^-|X)}\right.\\
    &\hspace{5cm}\left.\times\left\{ (1-F^\ast_t(s|X))\left\{d\left(\frac{1}{1-F^\ast_t(u^-|X)}\right)-d\left(\frac{1}{1-F_t(u^-|X)}\right)\right\}\right.\right.\\
    &\hspace{6cm}\left.\left.-\left\{F^\ast_t(s|X)-F_t(s|X)\right\}d\left(\frac{1}{1-F_t(u^-|X)}\right)\right\}\right]
\end{align*}
}

Let $H_{1621}=-H_{16211}-H_{16212}$, where
{\small
\begin{align*}
    H_{16211} &= E\left[\int_0^{\min(\widetilde{Y}, s)}\frac{I(T=t)(G^\ast_{T}(u^-|X)-G_{T}(u^-|X))}{G^\ast_{T}(u^-|X)G_{T}(u^-|X)}(1-F^\ast_t(s|X))\times\left\{d\left(\frac{1}{1-F^\ast_t(u^-|X)}\right)-d\left(\frac{1}{1-F_t(u^-|X)}\right)\right\}\right]\\
    &= E\left[\int_0^{\min(\widetilde{Y}, s)}\frac{I(T=t)(G^\ast_{T}(u^-|X)-G_{T}(u^-|X))}{G^\ast_{T}(u^-|X)G_{T}(u^-|X)}(1-F^\ast_t(s|X))\left\{\frac{\upsilon_t^{\ast\dagger}(u^-|X)}{S^\ast_t(u^-|X)}du-\frac{\upsilon_t^{\dagger}(u^-|X)}{S_t(u^-|X)}du\right\}\right]\\
    &= E\left[\int_0^{\min(\widetilde{Y}, s)}\frac{I(T=t)(G^\ast_{T}(u^-|X)-G_{T}(u^-|X))}{G^\ast_{T}(u^-|X)G_{T}(u^-|X)}(1-F^\ast_t(s|X))\left\{\frac{\upsilon_t^{\ast\dagger}(u^-|X)}{S^\ast_t(u^-|X)}-\frac{\upsilon_t^{\dagger}(u^-|X)}{S_t(u^-|X)}\right\}du\right]
\end{align*}
}

By the Cauchy-Schwarz inequality,
{\small
\begin{align*}
    H_{16211}^2&\leq nE\left[\int_0^{\min(\widetilde{Y}, s)}\frac{I(T=t)(G^\ast_{T}(u^-|X)-G_{T}(u^-|X))}{G^\ast_{T}(u^-|X)G_{T}(u^-|X)}(1-F^\ast_t(s|X))\left\{\frac{\upsilon_t^{\ast\dagger}(u^-|X)}{S^\ast_t(u^-|X)}-\frac{\upsilon_t^{\dagger}(u^-|X)}{S_t(u^-|X)}\right\}du\right]^2\\
    &\leq nE\left[\int_0^{\min(\widetilde{Y}, s)}\frac{I(T=t)(G^\ast_{T}(u^-|X)-G_{T}(u^-|X))^2}{G^{\ast 2}_{T}(u^-|X)G_{T}(u^-|X)^2}du\right]\\
    &\hspace{3cm}\times E\left[\int_0^{\min(\widetilde{Y}, s)}(1-F^\ast_t(s|X))^2\left(\frac{\upsilon_t^{\ast\dagger}(u^-|X)}{S^\ast_t(u^-|X)}-\frac{\upsilon_t^{\dagger}(u^-|X)}{S_t(u^-|X)}\right)^2du\right]
\end{align*}
}
Since 
\begin{eqnarray*}
    \frac{\upsilon_t^{\ast\dagger}(u^-|X)}{S^\ast_t(u^-|X)}-\frac{\upsilon_t^{\dagger}(u^-|X)}{S_t(u^-|X)} &=& -\frac{\upsilon_t^{\ast\dagger}(u^-|X)}{S^\ast_t(u^-|X)S_t(u^-|X)}\left(S^\ast_t(u^-|X)-S_t(u^-|X)\right)\\
    &&+\frac{1}{S_t(u^-|X)}\left(\upsilon_t^{\ast\dagger}(u^-|X)-\upsilon_t^{\dagger}(u^-|X)\right)
\end{eqnarray*}

and $ab\leq\frac{a^2+b^2}{2}$, we have
\begin{comment}
\begin{align*}
    H_{16211}^2&\lesssim n\sum_{k=1}^K E\left[\int_0^{\min(\widetilde{Y}, s)}(\widehat{G}_{T}(u^-|X)-G_{T}(u^-|X))^2du\right]\\
    &\hspace{1.5cm}\times \left\{E\left[\int_0^{\min(\widetilde{Y}, s)}\left(F^\ast_t(u^-|X)-F_t(u^-|X)\right)^2du\right]+E\left[\int_0^{\min(\widetilde{Y}, s)}\left(\exp\{\widehat{\alpha}_{t}' X \} -\exp\{\alpha_{t}' X \}\right)^2du\right]\right.\\
    &\hspace{5cm}\left.+E\left[\int_0^{\min(\widetilde{Y}, s)}\left(\widehat{\upsilon}^{\dagger}_{t}(u^-)-\upsilon^{\dagger}_{t}(u^-)\right)^2du\right]\right\}\\
    &\leq n\sum_{k=1}^K E\left[\int_0^{s}(\widehat{G}_{T}(u^-|X)-G_{T}(u^-|X))^2du\right]\\
    &\hspace{1.5cm}\times \left\{E\left[\int_0^{s}\left(F^\ast_t(u^-|X)-F_t(u^-|X)\right)^2du\right]+E\left[\int_0^{s}\left(\exp\{\widehat{\alpha}_{t}' X \} -\exp\{\alpha_{t}' X \}\right)^2du\right]\right.\\
    &\hspace{5cm}\left.+E\left[\int_0^{\min(\widetilde{Y}, s)}\left(\widehat{\upsilon}^{\dagger}_{t}(u^-)-\upsilon^{\dagger}_{t}(u^-)\right)^2du\right]\right\}
\end{align*}
\end{comment}
\begin{align*}
    H_{16211}^2&\lesssim nE\left[\int_0^{\min(\widetilde{Y}, s)}(G^\ast_{t}(u^-|X)-G_{t}(u^-|X))^2du\right]\\
    &\hspace{1.5cm}\times \left\{E\left[\int_0^{\min(\widetilde{Y}, s)}\left(F^\ast_t(u^-|X)-F_t(u^-|X)\right)^2du\right]+E\left[\int_0^{\min(\widetilde{Y}, s)}\left(\upsilon^{\ast\dagger}_{t}(u^-|X)-\upsilon^{\dagger}_{t}(u^-|X)\right)^2du\right]\right\}\\
    &\leq n E\left[\int_0^{s}(G^\ast_{t}(u^-|X)-G_{t}(u^-|X))^2du\right]\\
    &\hspace{1.5cm}\times \left\{E\left[\int_0^{s}\left(F^\ast_t(u^-|X)-F_t(u^-|X)\right)^2du\right]+E\left[\int_0^{s}\left(\upsilon^{\ast\dagger}_{t}(u^-|X)-\upsilon^{\dagger}_{t}(u^-|X)\right)^2du\right]\right\}\\
    &\leq n E\left[\int_0^{\tau}(G^\ast_{t}(u^-|X)-G_{t}(u^-|X))^2du\right]\\
    &\hspace{1.5cm}\times \left\{E\left[\int_0^{\tau}\left(F^\ast_t(u^-|X)-F_t(u^-|X)\right)^2du\right]+E\left[\int_0^{\tau}\left(\upsilon^{\ast\dagger}_{t}(u^-|X)-\upsilon^{\dagger}_{t}(u^-|X)\right)^2du\right]\right\}\\
    &\lesssim n \left\{E\left[\int_0^{\tau}(G^\ast_{t}(u^-|X)-G_{t}(u^-|X))^2du\right]^2+E\left[\int_0^{\tau}\left(F^\ast_t(u^-|X)-F_t(u^-|X)\right)^2du\right]^2\right.\\
    &\hspace{1.5cm}\left.+E\left[\int_0^{\tau}\left(\upsilon^{\ast\dagger}_{t}(u^-|X)-\upsilon^{\dagger}_{t}(u^-|X)\right)^2du\right]^2\right\}
\end{align*}
Since $E\left[\int_0^{\tau}(G^\ast_{t}(u^-|X)-G_{t}(u^-|X))^2du\right]\geq0$, $E\left[\int_0^{\tau}\left(F^\ast_t(u^-|X)-F_t(u^-|X)\right)^2du\right]\geq0$ and\\
$E\left[\int_0^{\tau}\left(\upsilon^{\ast\dagger}_{t}(u^-|X)-\upsilon^{\dagger}_{t}(u^-|X)\right)^2du\right]\geq0$, we have
\begin{align*}
    H_{16211}^2&\lesssim \left\{E\left[\int_0^{\tau}(G^\ast_{t}(u^-|X)-G_{t}(u^-|X))^2du\right]^2+E\left[\int_0^{\tau}\left(F^\ast_t(u^-|X)-F_t(u^-|X)\right)^2du\right]^2\right.\\
    &\hspace{1.5cm}\left.+E\left[\int_0^{\tau}\left(\upsilon^{\ast\dagger}_{t}(u^-|X)-\upsilon^{\dagger}_{t}(u^-|X)\right)du\right]^2\right\}\\
    &\leq \left\{E\left[\int_0^{\tau}(G^\ast_{t}(u^-|X)-G_{t}(u^-|X))^2du\right]+E\left[\int_0^{\tau}\left(F^\ast_t(u^-|X)-F_t(u^-|X)\right)^2du\right]\right.\\
    &\hspace{1.5cm}\left.+E\left[\int_0^{\tau}\left(\upsilon^{\ast\dagger}_{t}(u^-|X)-\upsilon^{\dagger}_{t}(u^-|X)\right)^2du\right]\right\}^2\\
    &\leq \left\{E\left[\int_0^{\tau}(G^\ast_t(u^-|X)-G_t(u^-|X))^2du\right]\right.\\
    &\hspace{1.5cm}\left.+E\left[\int_0^{\tau}\left(F^\ast_t(u^-|X)-F_t(u^-|X)\right)^2du\right]+E\left[\int_0^{\tau}\left(\upsilon^{\ast\dagger}_{t}(u^-|X)-\upsilon^{\dagger}_{t}(u^-|X)\right)^2du\right]\right\}^2
\end{align*}

Thus, 
\begin{align*}
    |H_{16211}|&\leq E\left[\int_0^{\tau}(G^\ast_t(u^-|X)-G_t(u^-|X))^2du\right]\\
    &\hspace{1.5cm}+E\left[\int_0^{\tau}\left(F^\ast_t(u^-|X)-F_t(u^-|X)\right)^2du\right]+E\left[\int_0^{\tau}\left(\upsilon^{\ast\dagger}_{t}(u^-|X)-\upsilon^{\dagger}_{t}(u^-|X)\right)^2du\right]
\end{align*}

{\small
\begin{align*}
H_{16212} &=E\left[\int_0^{\min(\widetilde{Y}, s)}\frac{I(T=t)(G^\ast_{T}(u^-|X)-G_{T}(u^-|X))}{G^\ast_{T}(u^-|X)G_{T}(u^-|X)}(1-F^\ast_t(s|X))\times\left\{F^\ast_t(s|X)-F_t(s|X)\right\}d\left(\frac{1}{1-F_t(u^-|X)}\right)\right]
\end{align*}
}

So applying (\ref{prop_int}), 
\begin{eqnarray*}
    \left|H_{16212}\right|&\lesssim& E\left[\sup_{u\in[0, \min(\widetilde{Y}, s)]}\left|G^\ast_{t}(u^-|X)-G_{t}(u^-|X)\right|\left|F^\ast_t(s|X)-F_t(s|X)\right|\right]\\
    &\leq& E\left[\sup_{u\in[0, s]}\left|G^\ast_{t}(u^-|X)-G_{t}(u^-|X)\right|\left|F^\ast_t(s|X)-F_t(s|X)\right|\right]\\
    &\leq&\sup_{u\in[0, s]}\left|\left|G^\ast_{t}(u^-|X)-G_{t}(u^-|X)\right|\right|_{L_2}\left|\left|F^\ast_t(s|X)-F_t(s|X)\right|\right|_{L_2}\\
    &\leq&\sup_{u\in[0, \tau]}\left|\left|G^\ast_{t}(u^-|X)-G_{t}(u^-|X)\right|\right|_{L_2}\sup_{u\in[0, \tau]}\left|\left|F^\ast_t(u|X)-F_t(u|X)\right|\right|_{L_2}
\end{eqnarray*}
Since 
{\small
\begin{align*}
    H_{1622} &= E\left[\int_0^{\widetilde{Y}}I(T=t)I(u \leq s)\left\{ \frac{1}{G^\ast_{T}(u^-|X)} - \frac{1}{G_{T}(u^-|X)} \right\}\frac{\pi_{1-t}(X)}{\pi_t(X)}\frac{ (1-F^\ast_t(s|X))F^\ast_t(u^-|X) }{(1-F^\ast_t(u^-|X))^2}\right.\\
    &\hspace{2.5cm}\left.\times\exp(\gamma_t)\left\{\frac{1}{\{ 1-F^\ast_t(s|X) + F^\ast_t(s|X) \exp(\gamma_t)\}^2}-\frac{1}{\{ 1-F_t(s|X) + F_t(s|X) \exp(\gamma_t)\}^2}\right\}du\right]\\
    &= E\left[\int_0^{\widetilde{Y}}I(T=t)I(u \leq s)\left\{ \frac{-(G^\ast_{T}(u^-|X)-G_{T}(u^-|X))}{G^\ast_{T}(u^-|X)G_{T}(u^-|X)} \right\}\frac{\pi_{1-t}(X)}{\pi_t(X)}\frac{ (1-F^\ast_t(s|X))}{(1-F^\ast_t(u^-|X))^2}\right.\\
    &\hspace{2.5cm}\left.\times\exp(\gamma_t)\left\{\frac{1}{\{ 1-F^\ast_t(s|X) + F^\ast_t(s|X) \exp(\gamma_t)\}^2}-\frac{1}{\{ 1-F_t(s|X) + F_t(s|X) \exp(\gamma_t)\}^2}\right\}du\right]\\
\end{align*}
}
Applying (\ref{prop_int}), we have
\begin{eqnarray*}
    \left| H_{1622}\right| &\lesssim& E\left[\sup_{u\in[0, \min(\widetilde{Y}, s)]}\left|G^\ast_{t}(u^-|X)-G_{t}(u^-|X)\right|\left|F^\ast_t(s|X)-F_t(s|X)\right|\right]\\
    &\leq& E\left[\sup_{u\in[0, s]}\left|G^\ast_{t}(u^-|X)-G_{t}(u^-|X)\right|\left|F^\ast_t(s|X)-F_t(s|X)\right|\right]\\
    &\leq& \sup_{u\in[0, s]}\left|\left|G^\ast_{t}(u^-|X)-G_{t}(u^-|X)\right|\right|_{L_2}\left|\left|F^\ast_t(s|X)-F_t(s|X)\right|\right|_{L_2}\\
    &\leq& \sup_{u\in[0, \tau]}\left|\left|G^\ast_{t}(u^-|X)-G_{t}(u^-|X)\right|\right|_{L_2}\sup_{u\in[0, \tau]}\left|\left|F^\ast_t(u|X)-F_t(u|X)\right|\right|_{L_2}
\end{eqnarray*}
{\small
\begin{align*}
H_{1623}&=E\left[\int_0^{\widetilde{Y}}I(T=t)I(u \leq s)\left\{ \frac{1}{G^\ast_{T}(u^-|X)} - \frac{1}{G_{T}(u^-|X)} \right\}\frac{\pi_{1-t}(X)}{\pi_t(X)}\frac{ \exp(\gamma_t)}{\{ 1-F_t(s|X) + F_t(s|X) \exp(\gamma_t)\}^2}\right.\\
&\hspace{3cm}\times\left.\left\{\frac{ (1-F^\ast_t(s|X))F^\ast_t(u^-|X)}{(1-F^\ast_t(u^-|X))^2}-\frac{ (1-F_t(s|X))f_t(u^-|X) }{(1-F_t(u^-|X))^2}\right\}du\right]\\
&= E\left[\int_0^{\widetilde{Y}}I(T=t)I(u \leq s)\left\{ \frac{-(G^\ast_{T}(u^-|X)-G_{T}(u^-|X))}{G^\ast_{T}(u^-|X)G_{T}(u^-|X)} \right\}\frac{\pi_{1-t}(X)}{\pi_t(X)}\frac{ \exp(\gamma_t)}{\{ 1-F_t(s|X) + F_t(s|X) \exp(\gamma_t)\}^2}\right.\\
&\hspace{3cm}\left.\times\left\{\frac{ (1-F^\ast_t(s|X)) }{(1-F^\ast_t(u^-|X))^2}dF^\ast_t(u^-|X)-\frac{ (1-F_t(s|X)) }{(1-F_t(u^-|X))^2}dF_t(u^-|X)\right\}\right]\\
&= E\left[\int_0^{\widetilde{Y}}I(T=t)I(u \leq s)\left\{ \frac{-(G^\ast_{T}(u^-|X)-G_{T}(u^-|X))}{G^\ast_{T}(u^-|X)G_{T}(u^-|X)} \right\}\frac{\pi_{1-t}(X)}{\pi_t(X)}\frac{ \exp(\gamma_t)}{\{ 1-F_t(s|X) + F_t(s|X) \exp(\gamma_t)\}^2}\right.\\
&\hspace{2cm}\left.\times\left\{ (1-F^\ast_t(s|X))\left\{d\left(\frac{1}{1-F^\ast_t(u^-|X)}\right)-d\left(\frac{1}{1-F_t(u^-|X)}\right)\right\}\right.\right.\\
&\hspace{5cm}-\left.\left.\left\{F^\ast_t(s|X)-F_t(s|X)\right\}d\left(\frac{1}{1-F_t(u^-|X)}\right)\right\}\right]
\end{align*}
}
Let $H_{1623}=-H_{16231}-H_{16232}$, 
{\small
\begin{align*}
    H_{16231} &= \sqrt{n}E\left[\int_0^{\widetilde{Y}}I(T=t)I(u \leq s)\left\{ \frac{(G^\ast_{T}(u^-|X)-G_{T}(u^-|X))}{G^\ast_{T}(u^-|X)G_{T}(u^-|X)} \right\}\frac{\pi_{1-t}(X)}{\pi_t(X)}\frac{ \exp(\gamma_t)}{\{ 1-F_t(s|X) + F_t(s|X) \exp(\gamma_t)\}^2}\right.\\
    &\hspace{3cm}\times\left.(1-F^\ast_t(s|X))\left\{d\left(\frac{1}{1-F^\ast_t(u^-|X)}\right)-d\left(\frac{1}{1-F_t(u^-|X)}\right)\right\}\right]
\end{align*}
}
Applying the same strategies used for $H_{16212}$, we have 
\begin{comment}
\begin{align*}
    H_{16231}^2&\lesssim nE\left[\int_0^{\min(\widetilde{Y}, s)}(G^\ast_{T}(u^-|X)-G_{T}(u^-|X))^2du\right]\\
    &\hspace{1.5cm}\times \left\{E\left[\int_0^{\min(\widetilde{Y}, s)}\left(F^\ast_t(u^-|X)-F_t(u^-|X)\right)^2du\right]+E\left[\int_0^{\min(\widetilde{Y}, s)}\left(\exp\{\widehat{\alpha}_{t}' X \} -\exp\{\alpha_{t}' X \}\right)^2du\right]\right.\\
    &\hspace{5cm}\left.+E\left[\int_0^{\min(\widetilde{Y}, s)}\left(\widehat{\upsilon}^{\dagger}_{t}(u^-)-\upsilon^{\dagger}_{t}(u^-)\right)^2du\right]\right\}
\end{align*}
\end{comment}
\begin{align*}
    H_{16231}^2&\lesssim E\left[\int_0^{\min(\widetilde{Y}, s)}(G^\ast_{t}(u^-|X)-G_{t}(u^-|X))^2du\right]\\
    &\hspace{1.5cm}\times \left\{E\left[\int_0^{\min(\widetilde{Y}, s)}\left(F^\ast_t(u^-|X)-F_t(u^-|X)\right)^2du\right]+E\left[\int_0^{\min(\widetilde{Y}, s)}\left(\upsilon^{\ast\dagger}_{t}(u^-|X)-\upsilon^{\dagger}_{t}(u^-|X)\right)^2du\right]\right\}\\
    &\leq E\left[\int_0^{s}(G^\ast_{t}(u^-|X)-G_{t}(u^-|X))^2du\right]\\
    &\hspace{1.5cm}\times \left\{E\left[\int_0^{s}\left(F^\ast_t(u^-|X)-F_t(u^-|X)\right)^2du\right]+E\left[\int_0^{s}\left(\upsilon^{\ast\dagger}_{t}(u^-|X)-\upsilon^{\dagger}_{t}(u^-|X)\right)^2du\right]\right\}\\
    &\leq E\left[\int_0^{\tau}(G^\ast_{t}(u^-|X)-G_{t}(u^-|X))^2du\right]\\
    &\hspace{1.5cm}\times \left\{E\left[\int_0^{\tau}\left(F^\ast_t(u^-|X)-F_t(u^-|X)\right)^2du\right]+E\left[\int_0^{\tau}\left(\upsilon^{\ast\dagger}_{t}(u^-|X)-\upsilon^{\dagger}_{t}(u^-|X)\right)^2du\right]\right\}\\
    &\leq \left\{E\left[\int_0^{\tau}(G^\ast_{t}(u^-|X)-G_{t}(u^-|X))^2du\right]+E\left[\int_0^{\tau}\left(F^\ast_t(u^-|X)-F_t(u^-|X)\right)^2du\right]\right.\\
    &\hspace{1.5cm}\left.+E\left[\int_0^{\tau}\left(\upsilon^{\ast\dagger}_{t}(u^-|X)-\upsilon^{\dagger}_{t}(u^-|X)\right)^2du\right]\right\}^2\\
    &\leq \left\{E\left[\int_0^{\tau}(G^\ast_{t}(u^-|X)-G_{t}(u^-|X))^2du\right]\right.\\
    &\hspace{1.5cm}\left.+E\left[\int_0^{\tau}\left(F^\ast_t(u^-|X)-F_t(u^-|X)\right)^2du\right]+E\left[\int_0^{\tau}\left(\upsilon^{\ast\dagger}_{t}(u^-|X)-\upsilon^{\dagger}_{t}(u^-|X)\right)^2du\right]\right\}^2
\end{align*}

Thus, 
\begin{align*}
    |H_{16231}|&\leq E\left[\int_0^{\tau}(G^\ast_{t}(u^-|X)-G_{t}(u^-|X))^2du\right]\\
    &\hspace{1.5cm}+E\left[\int_0^{\tau}\left(F^\ast_t(u^-|X)-F_t(u^-|X)\right)^2du\right]+E\left[\int_0^{\tau}\left(\upsilon^{\ast\dagger}_{t}(u^-|X)-\upsilon^{\dagger}_{t}(u^-|X)\right)^2du\right]
\end{align*}

\begin{align*}
    H_{16232} &= E\left[\int_0^{\widetilde{Y}}I(T=t)I(u \leq s)\left\{ \frac{(G^\ast_{T}(u^-|X)-G_{T}(u^-|X))}{G^\ast_{T}(u^-|X)G_{T}(u^-|X)} \right\}\frac{ \exp(\gamma_t)}{\{ 1-F_t(s|X) + F_t(s|X) \exp(\gamma_t)\}^2}\right.\\
    &\hspace{4cm}\left.\times\frac{\pi_{1-t}(X)}{\pi_t(X)}(1-F^\ast_t(s|X))\left\{F^\ast_t(s|X)-F_t(s|X)\right\}d\left(\frac{1}{1-F_t(u^-|X)}\right)\right]
\end{align*}

Applying (\ref{prop_int}), we have
\begin{eqnarray*}
    \left| H_{16232}\right| &\lesssim& E\left[\sup_{u\in[0, \min(\widetilde{Y}, s)]}\left|G^\ast_{t}(u^-|X)-G_{t}(u^-|X)\right|\left|F^\ast_t(s|X)-F_t(s|X)\right|\right]\\
    &\leq& E\left[\sup_{u\in[0, s]}\left|G^\ast_{t}(u^-|X)-G_{t}(u^-|X)\right|\left|F^\ast_t(s|X)-F_t(s|X)\right|\right]\\
    &\leq& \sup_{u\in[0, s]}\left|\left|G^\ast_{t}(u^-|X)-G_{t}(u^-|X)\right|\right|_{L_2}\left|\left|F^\ast_t(s|X)-F_t(s|X)\right|\right|_{L_2}\\
    &\leq& \sup_{u\in[0, \tau]}\left|\left|G^\ast_{t}(u^-|X)-G_{t}(u^-|X)\right|\right|_{L_2}\sup_{u\in[0, \tau]}\left|\left|F^\ast_t(u|X)-F_t(u|X)\right|\right|_{L_2}
\end{eqnarray*}

\begin{eqnarray*}
    H_2 &=& E\left[g_t(G_\cdot, F^\ast_{t}, \pi_{t})(\widetilde{O}) - h_t(\widetilde{P})(\widetilde{O}) \psi_t\right]\\
    &=& E\left[\frac{\Delta + (1-\Delta)I( \widetilde{Y}\geq s)}{G_{T}(\min(\widetilde{Y}, s)^-| X)} \left\{\phi_t(F^\ast_{t}, \pi_{t})(O)-\psi_t\right\}\right] \\
    &&+E\left[\frac{(1-\Delta)}{G_{T}(\widetilde{Y}^-| X)} \left\{l_t(F^\ast_{t}, \pi_{t})(\widetilde{Y},T,X)-I(\widetilde{Y}<s)\psi_t\right\}\right]\\
    &&-E\left[\int_0^{\widetilde{Y}} \frac{l_t(F^\ast_{t}, \pi_{t})(u,T,X)-I(u<s)\psi_t}{G_{T}(u^-|X)} \exp\{\beta_{T}' X \}d\Lambda^{\dagger}_{T}(u)\right]\\
    &=& E\left[\frac{\Delta + (1-\Delta)I( \widetilde{Y}\geq s)}{G_{T}(\min(\widetilde{Y}, s)^-| X)} \left\{\phi_t(F^\ast_{t}, \pi_{t})(O)-\psi_t\right\}\right] \\
     &&+E\left[\int \frac{l_t(F^\ast_{t}, \pi_{t})(u,T,X)-I(u<s)\psi_t}{G_{T}(u^-|X)} dN_c(u)\right]\\
    &&-E\left[\int_0^{\widetilde{Y}} \frac{l_t(F^\ast_{t}, \pi_{t})(u,T,X)-I(u<s)\psi_t}{G_{T}(u^-|X)} \exp\{\beta_{T}' X \}d\Lambda^{\dagger}_{T}(u)\right]\\
    &=& E\left[\frac{\Delta + (1-\Delta)I( \widetilde{Y}\geq s)}{G_{T}(\min(\widetilde{Y}, s)^-| X)} \left\{\phi_t(F^\ast_{t}, \pi_{t})(O)-\psi_t\right\}\right] \\
    &=& E\left[\phi_t(F^\ast_{t}, \pi_{t})(O)\right]-\psi_t
\end{eqnarray*}

Since 
\begin{align*}
    E\left[\phi_t(F^\ast_{t}, \pi_{t})(O)\right]&-\psi_t \\
    = E&\left[\pi_t(X)F_t(s|X)\left\{ 1+ \frac{\pi_{1-t}(X)}{\pi_t(X)} \frac{ \exp(\gamma_t)}{\{ 1-F^\ast_t(s|X) + F^\ast_t(s|X) \exp(\gamma_t)\}^2} \right\}\right]\\
    &-E\left[\pi_t(X)\left\{ \frac{\pi_{1-t}(X)}{\pi_t(X)} \frac{  F^\ast_t(s|X) \exp(\gamma_t)}{\{ 1-F^\ast_t(s|X) + F^\ast_t(s|X) \exp(\gamma_t)\}^2} \right\}\right]\\
    &+E\left[\pi_{1-t}(X)\left\{ \frac{F^\ast_t(s|X) \exp(\gamma_t)}{1-F^\ast_t(s|X) + F^\ast_t(s|X) \exp(\gamma_t)} \right\}\right]\\
    &-E\left[\pi_t(X)F_t(s|X)+\pi_{1-t}(X)\frac{F_t(s|X) \exp(\gamma_t)}{1-F_t(s|X) + F_t(s|X) \exp(\gamma_t)}\right]\\
    = E&\left[\pi_{1-t}(X)\exp(\gamma_t)\frac{  F_t(s|X)-F^\ast_t(s|X)}{\{ 1-F^\ast_t(s|X) + F^\ast_t(s|X) \exp(\gamma_t)\}^2}\right]\\
    &+E\left[\pi_{1-t}(X)\exp(\gamma_t)\left\{ \frac{F^\ast_t(s|X)}{1-F^\ast_t(s|X) + F^\ast_t(s|X) \exp(\gamma_t)}-\frac{F_t(s|X)}{1-F_t(s|X) + F_t(s|X) \exp(\gamma_t)} \right\}\right]\\
    = E&\left[\pi_{1-t}(X)\exp(\gamma_t)\frac{  F_t(s|X)-F^\ast_t(s|X)}{\{ 1-F^\ast_t(s|X) + F^\ast_t(s|X) \exp(\gamma_t)\}^2}\right]\\
    &+E\left[\pi_{1-t}(X)\exp(\gamma_t)\left\{ \frac{F^\ast_t(s|X)-F_t(s|X)}{\{1-F^\ast_t(s|X) + F^\ast_t(s|X) \exp(\gamma_t)\}\{1-F_t(s|X) + F_t(s|X) \exp(\gamma_t)\}} \right\}\right]\\
    = E&\left[\frac{ \pi_{1-t}(X)\exp(\gamma_t)(\exp(\gamma_t)-1) \left\{F^\ast_t(s|X)-F_t(s|X)\right\}^2}{\{ 1-F^\ast_t(s|X) + F^\ast_t(s|X) \exp(\gamma_t)\}^2\{1-F_t(s|X) + F_t(s|X) \exp(\gamma_t)\}}\right]
\end{align*}
we have 
\begin{eqnarray*}
    H_2 &=& E\left[\phi_t(F^\ast_{t}, \pi_{t})(O)\right]-\psi_t\\
    &=& E\left[\frac{ \pi_{1-t}(X)\exp(\gamma_t)(\exp(\gamma_t)-1) \left\{F^\ast_t(s|X)-F_t(s|X)\right\}^2}{\{ 1-F^\ast_t(s|X) + F^\ast_t(s|X) \exp(\gamma_t)\}^2\{1-F_t(s|X) + F_t(s|X) \exp(\gamma_t)\}}\right]
\end{eqnarray*}
Thus,
\begin{eqnarray*}
    \left| H_{2}\right| &\lesssim& \left|\left|F^\ast_t(s|X)-F_t(s|X)\right|\right|_{L_2}^2\\
    &\leq&\sup_{u\in[0, \tau]}\left|\left|F^\ast_t(u|X)-F_t(u|X)\right|\right|_{L_2}^2
\end{eqnarray*}
Combining all the upper bounds, we have that $\Big|E \left[  g_t(\widetilde{P}^*)(\widetilde{O}) - h_t(\widetilde{P}^*)(\widetilde{O})  \psi_t \right]\Big|$ is bounded above by

\begin{align*}
& \sup_{u\in[0, \tau]}\left|\left|F^\ast_t(u|X)-F_t(u|X)\right|\right|_{L_2} \times \sup_{u\in[0, \tau]}\left|\left|G^\ast_{t}(u| X)-G_{t}(u| X)\right|\right|_{L_2} \\
& + \sup_{u\in[0, \tau]}\left|\left|F^\ast_t(u|X)-F_t(u|X)\right|\right|_{L_2} \times \sup_{u\in[0, \tau]}\left|\left|G^\ast_{1-t}(u| X)-G_{1-t}(u| X)\right|\right|_{L_2} \\
& + \sup_{u\in[0, \tau]}\left|\left|F^\ast_t(u|X)-F_t(u|X)\right|\right|_{L_2} \times \left|\left|\pi^\ast_t(X)-\pi_t(X)\right|\right|_{L_2} \\
& + \left\{ \sup_{u\in[0, \tau]}\left|\left|F^\ast_t(u|X)-F_t(u|X)\right|\right|_{L_2} \right\}^2 + E\left[\int_0^{\tau}(G^\ast_{t}(u|X)-G_{t}(u|X))^2du\right] \\
& + E\left[\int_0^{\tau}\left(F^\ast_t(u|X)-F_t(u|X)\right)^2du\right] + E\left[\int_0^{\tau}\left(\upsilon^{\dagger *}_{t}(u|X)-\upsilon^{\dagger}_{t}(u|X)\right)^2du\right]
\end{align*}

This concludes Lemma~\ref{rate}.

\begin{theorem}\label{robust_est}
    $\widehat{\psi}_t$ in (11) is a consistent estimator of $\psi_t$ if $F_{t}(\cdot|X)$ is correctly specified. 
\end{theorem}
\textbf{Proof of Theorem~\ref{robust_est}} Let $\widehat{G}_\cdot(\cdot|X)\xrightarrow{P}G^\ast_\cdot(\cdot|X)$, $\widehat{\pi}_{t}(X)\xrightarrow{P}\pi^\ast_{t}(X)$ for any $G^\ast_\cdot(\cdot|X)$ and $\pi^\ast_{t}(X)$, where $G^\ast_\cdot(\cdot|X)=\exp\left(- \Lambda^{\ast\dagger}_\cdot(\cdot) \exp \{ {\beta^{\ast'}_\cdot} X \}\right)$. And let $\widehat{F}_t(\cdot|X)\xrightarrow{P}F_{t}(\cdot|X)$ where $F_{t}(\cdot|X)$ is the true conditional distribution of $Y$ given $X$ and $T=t$. By the weak law of large numbers and the continuous mapping theorem, 
$$
\widehat{\psi}_t = \frac{\sum_{k=1}^K \sum_{i \in S_k} g_t\left(\widehat{\widetilde{P}}^{(-k)}\right)(\widetilde{O}_i) }{\sum_{k=1}^K \sum_{i \in S_k} h_t\left(\widehat{\widetilde{P}}^{(-k)}\right)(\widetilde{O}_i)}\xrightarrow{P}\frac{E\left[g_t(G^\ast_\cdot(\cdot|X), F_{t}(\cdot|X), \pi^\ast_{t}(X))(\widetilde{O})\right]}{E\left[h_t(G^\ast_\cdot(\cdot|X))(\widetilde{O})\right]}
$$

Using Lemma~\ref{robust}, we know that $E\left[g_t(G^\ast_\cdot(\cdot|X), F_{t}(\cdot|X), \pi^\ast_{t}(X))(\widetilde{O})\right]=E\left[g_t(\widetilde{P})(\widetilde{O})\right]$ and  $E\left[h_t(G^\ast_\cdot(\cdot|X))(\widetilde{O})\right]=E\left[h_t(\widetilde{P})(\widetilde{O})\right]$. Thus, $\widehat{\psi}_t\xrightarrow{P}\psi_t$. 

\begin{theorem}\label{little_op}
    Given $\widehat{\widetilde{P}}$ is an consistent estimator of $\widetilde{P}$, and the upper bound in Lemma~\ref{rate} is $o_p(n^{-1/2})$ for $\widetilde{P}^\ast=\widehat{\widetilde{P}}$, then the observed data influence function under $K$-fold sample-splitting satisfies the following:
    \begin{eqnarray*}  
        \frac{1}{\sqrt{n}}\sum_{k=1}^K \sum_{i \in S_k}\left\{g_t(\widehat{\widetilde{P}}^{(-k)})(\widetilde{O}_i) - h_t(\widehat{\widetilde{P}}^{(-k)})(\widetilde{O}_i) \psi_t\right\} =\frac{1}{\sqrt{n}}\sum_{k=1}^K \sum_{i \in S_k}\left\{g_t(\widetilde{P})(\widetilde{O}_i) - h_t(\widetilde{P})(\widetilde{O}_i) \psi_t\right\}+o_p(1)
    \end{eqnarray*}
\end{theorem}
\textbf{Proof of Theorem~\ref{little_op}} 

Since $E\left[g_t(\widetilde{P})(\widetilde{O}) - h_t(\widetilde{P})(\widetilde{O}) \psi_t\right]=0$, we have
\begin{eqnarray*}
    &&\frac{1}{\sqrt{n}}\sum_{k=1}^K \sum_{i \in S_k}\left\{g_t(\widehat{\widetilde{P}}^{(-k)})(\widetilde{O}_i) - h_t(\widehat{\widetilde{P}}^{(-k)})(\widetilde{O}_i) \psi_t\right\}\\
    &=&\frac{1}{\sqrt{n}}\sum_{k=1}^K \sum_{i \in S_k}\left\{g_t(\widetilde{P})(\widetilde{O}) - h_t(\widetilde{P})(\widetilde{O}) \psi_t\right\}+R_{n,1}+R_{n,2}
\end{eqnarray*}
where
{\small
\begin{align*}
R_{n,1} &= \sqrt{n}\sum_{k=1}^K\left(\frac{n_k}{n}\right)(\frac{1}{n_k}\sum_{i\in S_k}-E) \left[  \left\{ g_t(\widehat{\widetilde{P}}^{(-k)})(\widetilde{O}) - h_t(\widehat{\widetilde{P}}^{(-k)})(\widetilde{O}) \psi_t \right\}-\left\{ g_t(\widetilde{P})(\widetilde{O}) - h_t(\widetilde{P})(\widetilde{O}) \psi_t \right\} \right]\\
R_{n,2} &=\sqrt{n}\sum_{k=1}^K\left(\frac{n_k}{n}\right) E \left[  g_t(\widehat{\widetilde{P}}^{(-k)})(\widetilde{O}) - h_t(\widehat{\widetilde{P}}^{(-k)})(\widetilde{O}) \psi_t \right]
\end{align*}
}

By the sample splitting proposition in~\cite{kennedy2023semiparametricdoublyrobusttargeted}, we have $R_{n,1}=o_P(1)$. Since we apply the same nuisance estimation methods for each fold $k$, the upper bound in Lemma~\ref{rate} is $o_P(n^{-1/2})$ for $\widetilde{P}^\ast=\widehat{\widetilde{P}}^{(-k)}$ for $k=1, ..., K$. And since $n_k=O(n)$, we have $R_{n,2}=o_P(1)$. This concludes Theorem~\ref{little_op}.

\begin{theorem}\label{F_distribute}
    Suppose the following:
    \begin{enumerate}[label=(\roman*)]
        \item Assumptions in Section 2.2 and correct specifications of models for $\widetilde{P}$
        \item Upper bound in Lemma~\ref{rate} is $o_P(n^{-1/2})$ for $\widetilde{P}^\ast=\widehat{\widetilde{P}}$
    \end{enumerate}
    Then the split-sample estimator of $\psi_t$ is $\sqrt{n}$-consistent and is asymptotically normal. 
\end{theorem}

\textbf{Proof of Theorem~\ref{F_distribute}} 

Applying the mean value theorem to $\frac{1}{n}\sum_{k=1}^K\sum_{i\in S_k}\left\{g_t(\widehat{\widetilde{P}}^{(-k)})(\widetilde{O}_i) - h_t(\widehat{\widetilde{P}}^{(-k)})(\widetilde{O}_i) \widehat{\psi}_t\right\}$ around $\widehat{\psi}_t$, 
\begin{eqnarray*}
    \sqrt{n}(\widehat{\psi}_t-\psi_t) &=& -\frac{\frac{1}{\sqrt{n}}\sum_{k=1}^K\sum_{i\in S_k}\left\{g_t(\widehat{\widetilde{P}}^{(-k)})(\widetilde{O}_i) - h_t(\widehat{\widetilde{P}}^{(-k)})(\widetilde{O}_i) \psi_t\right\}}{\frac{\partial}{\partial \widehat{\psi}_t}\frac{1}{n}\sum_{k=1}^K\sum_{i\in S_k}\left\{g_t(\widehat{\widetilde{P}}^{(-k)})(\widetilde{O}_i) - h_t(\widehat{\widetilde{P}}^{(-k)})(\widetilde{O}_i)\widehat{\psi}_t\right\}\bigg|_{\widehat{\psi}_t=\widetilde{\psi}_t}}\\
    &=& \frac{\frac{1}{\sqrt{n}}\sum_{k=1}^K\sum_{i\in S_k}\left\{g_t(\widehat{\widetilde{P}}^{(-k)})(\widetilde{O}_i) - h_t(\widehat{\widetilde{P}}^{(-k)})(\widetilde{O}_i) \psi_t\right\}}{\frac{1}{n}\sum_{k=1}^K\sum_{i\in S_k}\left\{h_t(\widehat{\widetilde{P}}^{(-k)})(\widetilde{O}_i)\right\}}
\end{eqnarray*}
where $\widetilde{\psi}_t$ is some value between $\widehat{\psi}_t$ and $\psi_t$. Under Theorem~\ref{little_op}, 
    \begin{eqnarray*}  
        \frac{1}{\sqrt{n}}\sum_{k=1}^K \sum_{i \in S_k}\left\{g_t(\widehat{\widetilde{P}}^{(-k)})(\widetilde{O}_i) - h_t(\widehat{\widetilde{P}}^{(-k)})(\widetilde{O}_i) \psi_t\right\} =\frac{1}{\sqrt{n}}\sum_{k=1}^K \sum_{i \in S_k}\left\{g_t(\widetilde{P})(\widetilde{O}_i) - h_t(\widetilde{P})(\widetilde{O}_i) \psi_t\right\}+o_P(1)
    \end{eqnarray*}

By the central limit theorem and the fact that $E\left[g_t(\widetilde{P})(\widetilde{O}) - h_t(\widetilde{P})(\widetilde{O}) \psi_t\right]=0$, 
\[
\frac{1}{\sqrt{n}}\sum_{k=1}^K \sum_{i \in S_k}\left\{g_t(\widetilde{P})(\widetilde{O}_i) - h_t(\widetilde{P})(\widetilde{O}_i) \psi_t\right\}\xrightarrow{D}N\left(0, E\left[(g_t(\widetilde{P})(\widetilde{O}) - h_t(\widetilde{P})(\widetilde{O}) \psi_t)^2\right]\right)
\]

Since $\frac{1}{n}\sum_{k=1}^K\sum_{i\in S_k}\left\{h_t(\widehat{\widetilde{P}}^{(-k)})(\widetilde{O}_i)\right\}\xrightarrow{P}E\left[h_t(\widetilde{P})(\widetilde{O})\right]$, using Slutsky's theorem, 
$$\sqrt{n}(\widehat{\psi}_t-\psi_t)\xrightarrow{D}N\left(0, \frac{E\left[(g_t(\widetilde{P})(\widetilde{O}) - h_t(\widetilde{P})(\widetilde{O}) \psi_t)^2\right]}{E\left[h_t(\widetilde{P})(\widetilde{O})\right]^2}\right)$$

We can estimate the variance of $\psi_t$ by
$$\left\{\frac{1}{n}\displaystyle\sum_{k=1}^K\sum_{i\in\mathcal{S}_k}h_t(\widehat{\widetilde{P}}^{(-k)})(\widetilde{O}_i)\right\}^{-2}\left\{\frac{1}{n^2}\displaystyle\sum_{k=1}^K\sum_{i\in\mathcal{S}_k}\left(g_t(\widehat{\widetilde{P}}^{(-k)})(\widetilde{O}_i) - h_t(\widehat{\widetilde{P}}^{(-k)})(\widetilde{O}_i) \widehat{\psi}_t\right)^2\right\}$$

To prove consistency of this variance estimator, we show
$$\frac{1}{n}\displaystyle\sum_{k=1}^K\sum_{i\in\mathcal{S}_k}\left(g_t(\widehat{\widetilde{P}}^{(-k)})(\widetilde{O}_i) - h_t(\widehat{\widetilde{P}}^{(-k)})(O_i) \widehat{\psi}_t\right)^2\xrightarrow{P}E\left[(g_t(\widetilde{P})(\widetilde{O}) - h_t(\widetilde{P})(\widetilde{O}) \psi_t)^2\right]$$
$$\frac{1}{n}\displaystyle\sum_{k=1}^K\sum_{i\in\mathcal{S}_k}h_t(\widehat{\widetilde{P}}^{(-k)})(\widetilde{O}_i)\xrightarrow{P}E\left[h_t(\widetilde{P})(\widetilde{O})\right]$$

\end{document}